\documentclass[prd,aps,preprint,showpacs]{revtex4}
\usepackage{graphicx}
\usepackage{pslatex}
\usepackage{makeidx}
\usepackage[dvips]{color}

\newcommand{\be}{\begin{equation}}
\newcommand{\ee}{\end{equation}}
\newcommand{\bi}{\begin{itemize}}
\newcommand{\ei}{\end{itemize}}
\newcommand{\bea}{\begin{eqnarray}}
\newcommand{\eea}{\end{eqnarray}}

\newcommand{\etal}{{\em et~al.}}

\newcommand{\lsim}{\mathrel{\lower4pt\hbox{$\sim$}}
\hskip-12.5pt\raise1.6pt\hbox{$<$}\;}
\newcommand{\Dslash}{{D}\!\!\!\!\!/\,}

\newcommand{\Tr}{\mbox{Tr}}
\newcommand{\VEV}[1]{\left\langle #1\right\rangle}
\renewcommand{\Re}{\mbox{Re}}
\renewcommand{\Im}{\mbox{Im}}
\newcommand{\pbp}{\bar\psi\psi}

\newcommand{\ie}{{\it i.e.}}
\newcommand{\eg}{{\it e.g.}}
\newcommand{\vs}{{\it vs.~}}

\begin{document}

\title{QCD Thermodynamics from the Lattice}

\author{C.E. DeTar} \affiliation{Physics Department, University of Utah,
Salt Lake City, UT 84112, USA}

\author{U.M. Heller}
\affiliation{American Physical Society, One Research Road,
Ridge, NY 11961, USA}

\date{\today}

\begin{abstract}
We review the current methods and results of lattice simulations of
quantum chromodynamics at nonzero temperatures and densities.  The
review is intended to introduce the subject to interested
nonspecialists and beginners.  It includes a brief overview of lattice
gauge theory, a discussion of the determination of the crossover
temperature, the QCD phase diagram at zero and nonzero densities, the
equation of state, some in-medium properties of hadrons including
charmonium, and some plasma transport coefficients.
\end{abstract}

\maketitle

\section{Introduction}

Quantum chromodynamics is the well-established theory of interacting
quarks and gluons.  Although its Lagrangian is simple and elegant,
except for high energy processes where perturbation theory is
applicable, it is very difficult to solve.  Over the past three
decades {\it ab initio} numerical and computational methods have been
devised for obtaining nonperturbative solutions.  They have become
refined to the point that a few dozen calculated quantities (decay
constants, mass splittings, etc.) agree with known experimental values
to a precision of a couple percent \cite{Bazavov:2009bb}.  These
successes provide the opportunity to push the calculations with some
confidence into new regimes that have not been thoroughly explored
experimentally.  In this review we will be interested in numerical
simulations of strongly interacting matter under the extreme
conditions of high temperatures and/or high baryon number densities.

Shortly after the big bang the universe was very likely dominated by a
high temperature plasma of quarks, antiquarks, and gluons.  As the
universe expanded and cooled, hadrons emerged that make up today's
universe. Knowing the characteristics of the plasma and the nature of
the transition to hadrons is clearly important for understanding these
stages in the development of the universe.  In the cores of some dense
stars it is conceivable that the baryon number density is sufficiently
high that hadrons lose their identities and merge into a plasma of
quarks and gluons.  The equation of state of such a dense plasma, for
example, is important for understanding conditions leading to a
collapse to a black hole.  In heavy ion collisions at RHIC, FAIR, and
soon at the LHC we seek to produce a quark-gluon plasma and study its
properties.  Since so little is known about the plasma, we turn to
numerical simulation of high-temperature and moderate-density QCD to
predict its properties and to guide the experimental investigation.
Apart from the phenomenological interest in such simulations, there is
also intrinsic theoretical interest in understanding the behavior of
confining field theories under extreme conditions.  In particular,
there are tantalizing predictions of still new states of matter at
very high densities \cite{Wilczek:1999ym}. Lattice QCD thermodynamics
is understandably a popular and vigorous field of research.

Certainly, present day lattice simulations can't answer all of our
questions.  The current standard methodology assumes thermal
equilibrium.  Moreover, simulations at nonzero densities are still in
their infancy, so much of what we know is restricted to zero or very
small baryon number densities.  To apply lattice results to the
phenomenology of heavy ion collisions requires an intermediate model,
such as hydrodynamics, which takes input from lattice simulations,
adds model assumptions, and makes predictions about the rapidly
evolving, emerging matter.  For this purpose the most important
quantities obtained from lattice simulations are the phase diagram as
a function of temperature and baryon number density, the equation of
state, speed of sound, and transport properties, such as the
viscosities and thermal conductivity.

In this review, intended for nonspecialists and beginners, we give a
brief overview of the lattice methodology and discuss a variety of
numerical results.  We discuss challenges and potential sources of
systematic error.  In Sec.~\ref{sec:lgt_thermo} we give a brief
introduction to lattice gauge theory and discuss the advantages and
disadvantages of various fermion formulations.  We discuss a variety
of observables used to determine the transition temperature $T_c$ in
Sec.~\ref{sec:Tc} and comment on some disparate results.  In
Sec.~\ref{sec:phase_diag_zero_dens} we review our current
understanding of the phase diagram at zero baryon density, and in
Sec.~\ref{sec:phase_diag_nonzero_dens} we do the same for nonzero
baryon number densities.  We discuss the variety of methods in current
use for simulating at nonzero densities.  We review the equation of
state in Sec.~\ref{sec:eos}. In Sec.~\ref{sec:in_medium} we discuss
some properties of hadrons in the high temperature medium, and in
Sec.~\ref{sec:transport} some results for transport coefficients.
Finally, in Sec.~\ref{sec:outlook} we summarize briefly the current
state of the field, list outstanding problems, and list some prospects
for resolving them.

\section{Thermodynamics in Lattice Gauge Theory}
\label{sec:lgt_thermo}

\subsection{Quantum partition function}

Quantum thermodynamics at a fixed, large volume is based on the
partition function in the quantum canonical ensemble
%
\be
Z = \Tr [\exp(-H/T)],
\label{eq:Censemble}
\ee
where $H$ is the quantum Hamiltonian operator, $T$ is the temperature,
and the trace is taken over the physical Hilbert space.  At nonzero
densities the grand canonical ensemble is appropriate:
\be
Z = \Tr \left[\exp\left(-H/T + \sum_i \mu_i N_i/T\right)\right],
\label{eq:GCensemble}
\ee
where $\mu_i$ is the chemical potential for the $i$th species and
$N_i$ is the corresponding conserved flavor number.  For example, in
QCD we may introduce a separate chemical potential for each quark
flavor.  Zero chemical potential for a given flavor implies equal
numbers of quarks and antiquarks of that flavor, so zero baryon number
density, zero strangeness, etc.

The expectation value of an observable ${\cal O}$ at temperature $T$
is computed with respect to this ensemble through
\be
\VEV{\cal O} =  \Tr \left[{\cal O} \exp\left(-H/T + 
  \sum_i \mu_i N_i/T\right)\right]/Z .
\ee

\subsection{Feynman path integral partition function}

The Feynman path integral formalism provides a practical basis for the
computation of thermodynamic quantities, especially in quantum field
theory, where there are many degrees of freedom.  It converts the
trace over quantum states into a multidimensional integration over
classical variables \cite{FeynmanHibbs}.  It is beyond the scope of
this review to give a detailed derivation of the path integral
formulation, particularly for a gauge theory with fermion fields.
There are standard references \cite{MontvayBook, CreutzBook, DeGrandDeTarBook}.

The classical variables in the Feynman path integral are the path
``histories'' of the fundamental fields in Euclidean (imaginary) time
$\tau$.  (Imaginary, because the Boltzmann weight factor $\exp(-H/T)$ is, in
effect, a time evolution operator $\exp(-iHt)$ for an imaginary time
$-i/T$.)  For computational purposes the histories are discretized in
$\tau$.  The quantum fields at any given time are also discretized in
three-dimensional coordinate space ${\bf x}$.  The resulting path
integral is then a multidimensional integral over variables defined on
a four-dimensional space-time lattice ($x = {\bf x}, \tau)$.  The
discretization of space and time introduces an error, but the error
vanishes as the lattice spacing is taken to zero (continuum limit).

\subsection{Scalar field example}

For a concrete example, consider a scalar field theory described by
the Lagrange density
\be
   L(\phi) = \frac{1}{2} \sum_\mu \left[\frac{\partial}{dx_\mu}\phi(x) \right]^2
      + V[\phi(x)] ,
\ee
where $V$ describes the mass and self-interaction.  On a hypercubic
lattice with point separation $a$ and a central-difference
discretization of the derivatives, we can write a lattice
approximation
\be
  L[\phi(x)] = \frac{1}{8a^2} \sum_\mu \left[\phi(x + a \hat\mu) - 
    \phi(x - a \hat\mu)\right]^2 + V[\phi(x)].
\ee
where $\hat\mu$ is a unit coordinate vector in the $\mu$ direction.  A
Euclidean time history is then specified simply by giving the values
of the field $\phi({\bf x},\tau)$ on all the lattice points $({\bf
  x},\tau)$.  Each such history corresponds to a classical Euclidean
action $S(\phi)$, which is computed by summing its Lagrange density
over the lattice points
\be
  S(\phi) = \sum_{{\bf x},\tau}L[\phi({\bf x}, \tau)].
\ee

The partition function then becomes a multidimensional integral over
the values of the field $\phi({\bf x},\tau)$ at each point, weighted by
the exponential of the classical Euclidean action:
\be
   Z = \int\prod_{\bf x, \tau}d\phi({\bf x},\tau)\, \exp[-S(\phi)] .
\ee

Two important conditions on the Euclidean time history are inherited
from the definition [Eq.~(\ref{eq:Censemble})] of the partition function:
First, the time history $\tau$ ranges over a finite interval from $0$
to $a(N_\tau-1)$ where
\be
    1/T = aN_\tau,
\ee
which establishes the relation between the temperature and the
Euclidean time extent of the lattice.  Second, to reproduce the trace
over quantum states, the bosonic field $\phi$ must be periodic under
$\tau \rightarrow \tau + aN_\tau$.

Similarly, the expectation value of an operator ${\cal O}(\phi)$,
which depends on the field $\phi$, is given by
\be
   \VEV{\cal O} = \int\prod_{\bf x, \tau} d\phi({\bf x},\tau)\,
           {\cal O}(\phi) \exp[-S(\phi)]/Z,
\ee
where we replace the field operator $\phi$ with its classical value 
when we insert ${\cal O}(\phi)$ in the integrand.

\subsection{QCD on the Lattice}

For a renormalizable, asymptotically free theory, such as QCD, the
lattice formulation takes on a larger significance than just a
convenient computational device.  The lattice regulates ultraviolet
divergences.  The lattice constant $a$ provides an upper bound or
``cutoff'' scale $\pi/a$ for momenta.  From this point of view the
lattice formulation of the theory is every bit as respectable as other
regularization schemes.  Of course, as usual, we define the theory in
the limit in which the cutoff is removed, \ie, $a \rightarrow 0$.
Before this is done all quantities we calculate have cutoff errors
that vanish in the continuum limit.

With the lattice regulator we apply the usual renormalization process:
We select a few experimental values and use them to fix the bare
parameters of the theory (quark masses and gauge coupling).  In this
way the bare parameters depend on the cutoff (lattice spacing), but
the physical predictions should approach a cutoff-independent value in
the limit of zero lattice spacing.  In principle all regularization
schemes should agree in the limit that their cutoffs are taken away.

For QCD the fields are fermions and gluons. The groundwork for the
lattice formulation of QCD with fermions was laid down by
Wilson~\cite{Wilson:1974sk} in 1974, although there was other seminal
work on lattice theories with local gauge invariance by
Wegner~\cite{Wegner:1971rs}, Smit, and
Polyakov~\cite{Polyakov:1975rs}. To preserve gauge invariance, gluon
variables are introduced as $SU(3)$ matrices on the links between
nearest neighbors of the lattice. There are four forward links per
site, corresponding to the four components of the color vector
potential $A^c_\mu(x)$.  The matrix for the link joining the site $x$
with the site $x + a\hat\mu$ is then
\be
   U_\mu(x) = \exp[igaA^c_\mu(x)\lambda^c/2] ,
\ee
where $\lambda^c$ are the Gell-Mann generators of $SU(3)$. 

For the pure Yang-Mills theory of gluons a simple lattice form of the
classical action is constructed from the plaquettes $U_{P,\mu\nu}(x)$,
\ie, the product of the link matrices surrounding the unit square in
the forward $\mu\nu$ direction at site $x$.  The single-plaquette
Wilson action is simply the sum over all such plaquettes:
\be
  S_G(U) = \sum_{x,\mu,\nu} {\beta \over 6} \Re \Tr [1 - U_{P,\mu\nu}(x)],
  \label{eq:Wilson_1plaq}
\ee
The gauge coupling $\alpha_s = g^2/4\pi$ appears in the coefficient
\be
  \beta = 6/g^2.
\ee
In the continuum limit the plaquette reduces to the familiar 
square of the field strength tensor summed over eight colors $c$:
\be
  \Re \Tr [1 - U_{P,\mu\nu}(x)] \rightarrow \frac{g^2 a^4}{4}
        \sum_c(F^c_{\mu\nu})^2 + {\cal O}(a^6) .
\ee
In fact any closed planar loop, normalized by the area in lattice
units, has the same continuum limit, but with a different ${\cal
  O}(a^6)$ cutoff error.  For example a $2\times 1$ rectangular
version of the plaquette could also be used.  If the two components
are combined with the proper choice of coefficients, one can construct
an improved gluon action that eliminates the leading cutoff
correction, leaving errors at the next order ${\cal O}(a^8)$.
Relative to the leading continuum contribution, which carries the
volume factor $a^4$, such actions are called ``tree-level ${\cal
  O}(a^2)$'' improved.  Further improvements can even eliminate
quantum cutoff corrections of the type ${\cal O}(a^2 \alpha_s)$.  The
``tadpole L\"uscher-Weisz'' actions
\cite{Luscher:1984xn,Luscher:1985zq} are in this category.  Improving
actions in this way is desirable, since it brings a calculation closer
to the continuum limit at a given lattice spacing
\cite{Symanzik:1983gh}.

The quark fields $\psi(x)$, one for each flavor, have values on each
lattice site.  Since they are fermions, they require special treatment
in the functional integration: their classical values are
anticommuting Grassmann numbers.  The fermion contribution to the
action for each flavor can be written generally as
\be
   S_F(U,\psi) = \sum_{x,y} \bar \psi(x) M(U,x,y) \psi(y) ,
\ee
where $M(U,x,y)$ is the Dirac matrix -- essentially a lattice rendering
of the familiar Dirac operator $\Dslash + m$.  The functional
integral for the partition function then becomes
\be
  Z = \int [dU][d\psi][d\bar \psi]\,\exp[-S_G(U) - S_F(U,\psi)].
\label{eq:partitionQCD}
\ee
Since the dependence on the quark fields is simply bilinear, and
computing numerically with anticommuting numbers is nontrivial, it is
standard to integrate out the quark fields immediately, following the
rules of Grassmann integration, leaving only an integration over the
gauge fields, weighted by the determinant of the Dirac matrix:
\be
  Z  = \int [dU]\,\exp[-S_G(U)]\det[M(U)].
\label{eq:det_action}
\ee

There are many ways to formulate a lattice fermion action, each with
its advantages and disadvantages. A great deal of effort over the past
couple of decades has been devoted to improving the lattice treatment
of fermions.  We sketch the formulations here.  For more detail, see
\cite{MontvayBook, CreutzBook, DeGrandDeTarBook}.

\subsubsection{Wilson fermions}

The original Wilson rendering of the Dirac operator $D_\nu \gamma_\nu + m$
starts from a simple central difference approximation to the derivative:
\be
   \nabla_\nu \psi(x) = \frac{1}{2a}[U_\nu(x) \psi(x + \hat \nu a) - 
       U_\nu^\dagger(x-\hat\nu a)\psi(x - \hat \nu a)] ,
\label{eq:cov_deriv}
\ee
where the link matrices $U_\nu(x)$ provide the gauge-covariance.  The
action constructed from this operator is
\be
  S_{F,\rm naive} = \sum_{x,\nu} \bar \psi(x)(\nabla_\nu \gamma_\nu + m)
       \psi(x) .
\label{eq:naive}
\ee
It describes sixteen degenerate particles where only one is desired.
Wilson remedied this undesirable ``fermion doubling'' problem by
adding an irrelevant term to the action
\be
  S_{F,\rm Wilson} =  S_{F,\rm naive} - 
     \frac{ar}{2} \sum_{x,\nu} \bar \psi(x) \Delta_\nu \psi(x),
\label{eq:Wils_ferm1}
\ee
where $r$ is usually set to 1 and $\Delta_\nu \psi(x)$ is the covariant
Laplacian,
\be
  \Delta_\nu \psi(x) =
     \frac{1}{a^2} [U_\nu(x) \psi(x+\hat\nu a)
       + U_\nu^\dagger(x-\hat\nu a) \psi(x-\hat\nu a) - 2 \psi(x)]. 
\label{eq:lapl}
\ee
The added term gives fifteen of the
doublers masses of order of the cutoff scale $1/a$, leaving only one
light state.  The unwanted doublers thus become inaccessibly heavy in
the continuum limit.

It is customary to rearrange the terms in the Wilson action and
multiply the field $\psi(x)$ by a constant to give
\be
  S_{F,\rm Wilson} =  \bar \psi(x) \psi(x) +
    \kappa \sum_{x,\nu} \bar \psi(x) \left[(1 + \gamma_\nu) U_\nu(x)\psi(x+\hat\nu a)
    + (1 - \gamma_\nu) U_\nu^\dagger(x-\hat\nu a) \psi(x-\hat\nu a) \right] ,
\ee
where the ``hopping parameter'' $\kappa = 1/(8 + 2 ma)$ controls the
quark mass.  Improvements to the Wilson formalism include removing
tree-level ${\cal O}(a)$ errors by introducing a ``clover'' term in
the action \cite{Sheikholeslami:1985ij} and, for two flavors,
introducing a ``twisted mass''
\cite{Frezzotti:2000nk,Frezzotti:2003ni}.

For thermodynamics applications the chief drawback of Wilson fermions
has been (1) an explicit breaking of chiral symmetry at nonzero
lattice spacing, (2) a difficulty reaching low quark masses, and (3) a
relatively poor representation of the quark dispersion relation.  None
of these difficulties is insurmountable.  Chiral symmetry is restored
in the continuum limit.  

\begin{figure}
\begin{center}
\includegraphics[width=0.6\textwidth]{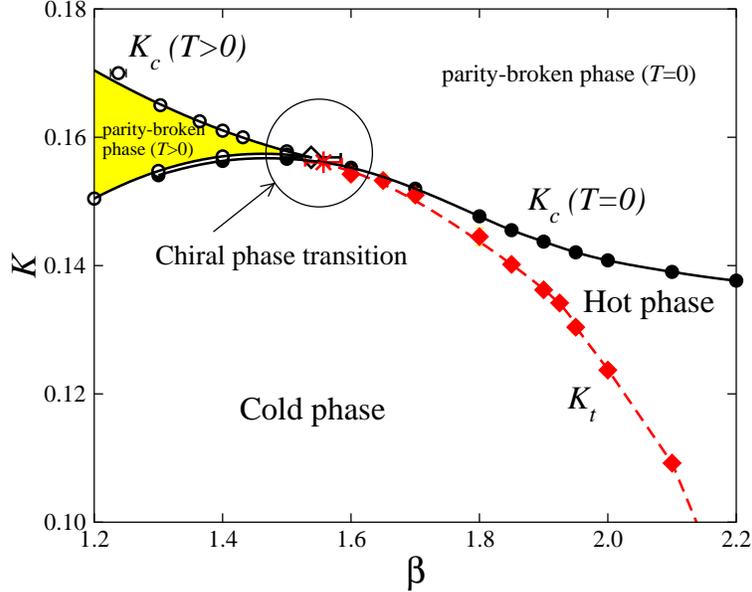}\\
\end{center}
\caption{The bare-parameter phase diagram for two flavors of
  clover-improved Wilson fermions and an improved gauge action for
  zero and nonzero temperatures, illustrating the mapping necessary
  for thermodynamics studies with Wilson fermions.  In this plot the
  hopping parameter $\kappa$ is denoted by $K$.  The line of chiral
  critical hopping parameters $\kappa_c(T=0)$ was determined from the
  vanishing of the pion mass.  The line $\kappa_t$ indicates the high
  temperature crossover at $N_\tau=4$.  It was determined from the
  Polyakov line (see section \ref{sec:Polyakov}).  The region ``chiral
  phase transition'' shows where the thermal crossover happens for
  small pion masses. The parity broken phases come from lattice
  artifacts of Wilson fermions. The data are from the CP-PACS
  collaboration, \cite{AliKhan:2000iz}, as shown in
  \cite{Ejiri:2007qk}.  }
\label{fig:phasediag_cppacs}
\end{figure}

It is necessary to search for the value $\kappa = \kappa_c$ where the
pion mass vanishes.  Since this value depends on the inverse gauge
coupling $\beta$, one gets a curve $\kappa_c(\beta)$ in bare parameter
$\kappa-\beta$ space as shown in Fig.~\ref{fig:phasediag_cppacs}.
Lines of constant pion mass form a family of such curves (not shown)
with the pion mass increasing as $\kappa$ decreases.  Also shown is a
high temperature crossover line $\kappa_t(\beta)$.  Its location
depends on $N_\tau$.  Where it intersects the $\kappa_c$ line, we
expect a true chiral phase transition.  Pushing to stronger coupling
(smaller $\beta$) or negative quark masses (higher $\kappa$) from
there takes us into the realm of lattice artifacts: the theory has a
parity broken phase at unphysical values of the bare parameters, as
indicated.


\subsubsection{Staggered fermions}

The staggered fermion approach starts from the naive action in
Eq.~(\ref{eq:naive}).  Through a field transformation, the Dirac matrices
can be diagonalized exactly giving four identical actions, each of
them with one spin per site.  If we keep only one of the actions, we
reduce the lattice fermion degrees of freedom by a factor of four,
which still leaves us with four fermion doublers.  These residual
degrees of freedom are called ``tastes.''  Without further
intervention, they would overcount the sea quark effects by a factor of
four.  To get approximately the correct counting, we replace the
fermion determinant by its fourth root for each of the desired
flavors:
\be
  Z_{\rm stagg}  = \int [dU]\,\exp[-S_G(U)]\prod_i \det[M_i(U)]^{1/4}.
\label{eq:root_det_action}
\ee
In the continuum limit at nonzero quark masses, the eigenvalues of the
determinant cluster in increasingly tighter quartets as expected from
fermion doubling \cite{Follana:2005km}.  Then we have an $SU(4)$ taste
symmetry, so taking the fourth root is equivalent to using only one of
them as a sea quark species.  This procedure has generated
considerable controversy.  Although there is no rigorous proof that
the method is valid, all indications so far are that the approximation
is under control as long as we take the continuum limit before we take
the quark masses to zero or fit data to a chiral model with taste
symmetry breaking properly included~\cite{Sharpe:2006re}, in which
case the limits are completely under control.

At nonzero lattice spacing the taste symmetry is broken, which
introduces lattice artifacts.  For example, mesons composed of a
valence quark and antiquark come in nondegenerate taste multiplets of
sixteen tastes.  In the continuum limit they are degenerate.

\begin{figure}
\begin{center}
\includegraphics[width=0.47\textwidth]{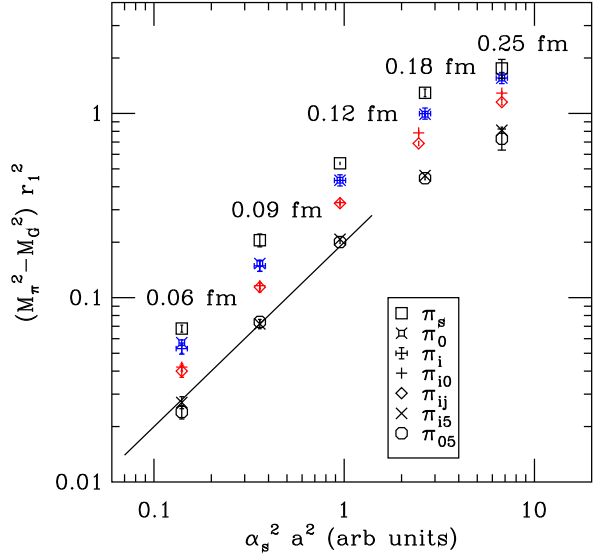}\\
\end{center}
\caption{ Plot showing that the lattice artifact taste splitting of
  pion masses vanishes as $\alpha_s^2 a^2$ in the continuum limit.
  The splitting is measured as the difference of the squared masses of
  the multiplet member and the Goldstone pion member.  It is given in
  units of $r_1 \approx 0.318$ fm. The plot symbols distinguish the
  members of the multiplet.  (The subscripts in the legend denote
  the Dirac-gamma-matrix-style classification of the pion tastes,
  ranging from singlet ($s$) and $\gamma_0$ to $\gamma_0\gamma_5$.)
  The line is drawn with unit log-log slope to test proportionality to
  $\alpha_s^2 a^2$. }
\label{fig:taste_split}
\end{figure}

The asqtad
\cite{Lepage:1997id,Bernard:1997mz,Lagae:1997he,Lagae:1998pe,Orginos:1998ue,Toussaint:1998sa,Lepage:1998vj,Orginos:1999cr,Bernard:1999xx}
and p4fat3 \cite{Heller:1999xz,Karsch:2000ps} improvements of the staggered fermion
formalism eliminate errors of ${\cal O}(a^2)$ in the quark dispersion
relation and suppress taste splitting significantly.  The asqtad
suppression is somewhat better, presumably because it eliminates all
tree level ${\cal O}(a^2)$ errors.  The recently proposed HISQ action
does still better~\cite{Follana:2006rc}. In Fig.~\ref{fig:taste_split}
we compare the predicted and measured scaling of the splitting in the
asqtad pion taste multiplet.

Taste splitting can also be reduced simply by replacing the gauge link
matrices in the action by smoothed gauge links -- for example the
Dublin ``stout links''~\cite{Aoki:2006br}.  Unlike the asqtad approach, this
method does not eliminate ${\cal O}(a^2)$ errors systematically.  Thus
the free quark dispersion relation is still unimproved.

Is taste-symmetry breaking really a problem for thermodynamics?  It is
believed to be most dramatic for the pion and less noticeable for more
massive states \cite{Ishizuka:1993mt}.  One could argue that close to
the crossover temperature and away from the critical point, so many
excited states participate, as in the hadron resonance gas model, that
pions do not matter much.  But if we approach the critical point at
fixed lattice spacing, taste splitting is likely to have a strong
effect on the critical behavior: we may even get a chiral symmetry
restoring transition in the wrong universality class.  And certainly
at quite low temperatures where pions dominate the statistical
ensemble, taste splitting makes a difference.

Taste-symmetry breaking also complicates the definition of the
``physical'' quark mass in a thermodynamics simulation.  At zero
temperature it is traditional to adjust the up and down quark masses
so that the Goldstone pion (the lightest one in the taste multiplet)
has the physical pion mass.  This is legitimate, because we may
restrict our attention to Green's functions whose external legs are
the Goldstone pion.  In a thermodynamics simulation, however, all
members of the taste multiplet participate in the thermal ensemble.
Thus it is more appropriate to tune the average multiplet mass, \eg,
the rms pion mass to the physical pion mass.  At a nonzero lattice
spacing, the multiplet splitting may be so large, that goal is
unreachable.  In that case the physical point is reached only by
reducing the lattice spacing together with the light quark mass.  It
is simply incorrect to claim a thermodynamics calculation is done at a
physical pion mass when the rms mass is still much higher.

\subsubsection{Domain-wall fermions}

Neither the Wilson fermion formulation, including the clover-improved
and twisted-mass version, nor the staggered fermion formulation are
entirely satisfactory discretizations of fermions. Wilson fermions
explicitly break chiral symmetry and its recovery requires a fine
tuning. Staggered fermions, while preserving a remnant of chiral
symmetry, have a remaining doubling problem, requiring the fourth-root
trick, which is still somewhat controversial.

A more sophisticated, somewhat indirect and more costly discretization
of fermions goes under the name of ``domain-wall fermions'' and was
developed by Kaplan \cite{Kaplan:1992bt} and by Furman and Shamir
\cite{Furman:1994ky}. Furman and Shamir's construction has become
standard. An additional, fifth dimension of length $L_s$ is introduced
and one considers 5-d Wilson fermions with no gauge links in the
fifth direction, and the 4-d gauge links independent of the fifth
coordinate, $s$,
\begin{equation}
S_{DW} = \sum_{s=0}^{L_s-1} \sum_x \bar\psi(x,s) \left\{ 
 \sum_\mu \left( \gamma_\mu \nabla_\mu - \frac{1}{2} \Delta_\mu \right)
 \psi(x,s) - M \psi(x,s) - P_- \psi(x,s+1) - P_+ \psi(x,s-1) \right\} ~,
\label{eq:S_DW}
\end{equation}
where $P_\pm = \frac{1}{2} (1 \pm \gamma_5)$ are chiral projectors and
we have set $r = a = 1$. The parameter $M$, often referred to as
domain-wall height, is introduced here with a sign opposite that of
the usual mass term for Wilson fermions
[Eq.~(\ref{eq:Wils_ferm1})]. It needs to be chosen in the interval $0
< M < 2$.  For free fermions the optimal choice is $M=1$, while in the
interacting case $M$ should be somewhat larger. The fermion fields
satisfy the boundary condition in the fifth direction,
\begin{equation}
P_- \psi(x,L_s) = - m_f P_- \psi(x,0) ~,~~~
 P_+ \psi(x,-1) = 
 - m_f P_+ \psi(x,L_s-1) ~,
\label{eq:DW_bc}
\end{equation}
where $m_f$ is a bare quark mass.

\begin{figure}
\begin{center}
\includegraphics[width=0.6\textwidth]{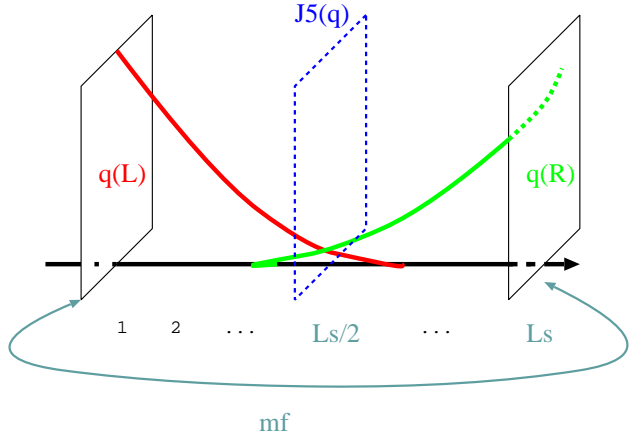}\\
\end{center}
\caption{Sketch, courtesy of Taku Izubuchi, of the domain-wall fermion
setup. Left and right handed modes are exponentially bound to the left
and right domain walls. The residual mass $m_{res}$ is determined from
an axial Ward identity applied in the center slice.
}
\label{fig:dwf_sketch}
\end{figure}

\begin{figure}
\begin{center}
\includegraphics[clip=true,width=0.47\textwidth]{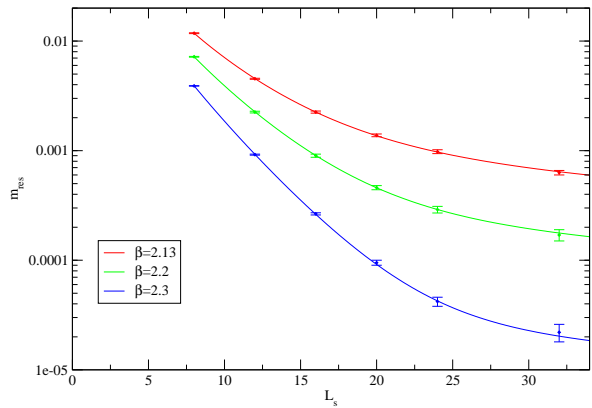}\\
\end{center}
\caption{Plot of the residual mass $m_{res}$ as a function of $L_s$
  showing its desired suppression with increasing $L_s$ and increasing
  inverse gauge coupling $\beta$.  Also shown are fits to an
  exponential fall-off plus a $1/L_s$ contribution, from a recent
  $2+1$ flavor dynamical domain-wall fermion simulation
  \cite{Antonio:2008zz}.  }
\label{fig:mres_2+1}
\end{figure}

The domain-wall fermion action, Eq.~(\ref{eq:S_DW}), has 5-d chiral
zero modes $\Psi$ bound exponentially to the boundaries at $s=0$ and
$s=L_s-1$, which are identified with the chiral modes of 4-d fermions
as,
\begin{equation}
q^R(x) = P_+ \psi(x,L_s-1) ~,~~~ q^L(x) = P_- \psi(x,0) ~,~~~
 \bar q^R(x) = \bar\psi(x,L_s-1) P_- ~,~~~ \bar q^L(x) = \bar\psi(x,0)
 P_+ ~.
\label{eq:DW_4d_q}
\end{equation}
The left- and right-handed modes $q^L$ and $q^R$ do not interact for
$m_f=0$ when $L_s \to \infty$ and the domain-wall action has a chiral
symmetry. At finite $L_s$ the chiral symmetry is slightly broken.  A
popular measure of the chiral symmetry breaking is called ``residual
mass'', $m_{res}$. It is determined from the axial Ward identity
applied at the midpoint between the two domain walls, as sketched in
Fig.~\ref{fig:dwf_sketch}. This residual mass was expected to fall off
exponentially in $L_s$. But, due to lattice artifacts of Wilson
fermions with large negative mass, there is a contribution to
$m_{res}$ that decreases only like $1/L_s$
\cite{Edwards:1998sh,Golterman:2005fe}.  An example from a recent
dynamical domain-wall fermion simulation \cite{Antonio:2008zz} is
shown in Fig.~\ref{fig:mres_2+1}.  Nevertheless, often $L_s =
\mathcal{O}(10-20)$ is large enough to keep the chiral symmetry
breaking negligibly small, especially at smaller lattice spacing
(weaker coupling).

Domain-wall fermions, therefore, solve, or at least substantially
alleviate, explicit chiral symmetry breaking without a doubling
problem. The price is a computational cost roughly a factor of $L_s$
larger than that for Wilson type fermions.

\begin{figure}
\begin{center}
\includegraphics[width=0.6\textwidth]{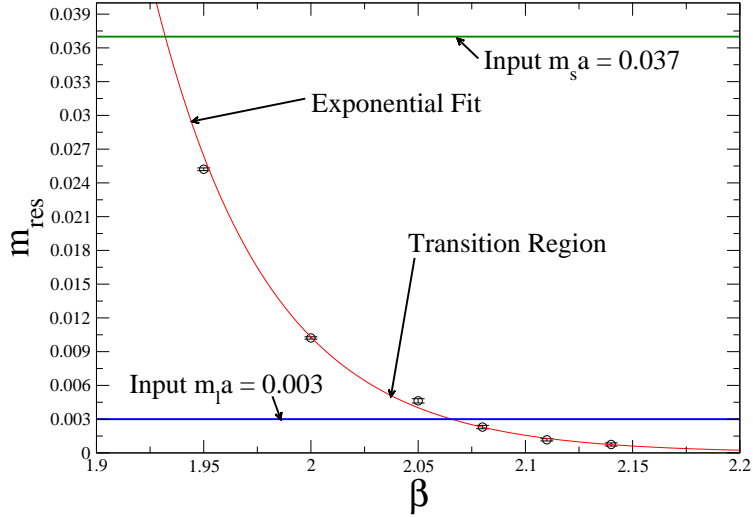}\\
\end{center}
\caption{Residual mass $m_{res}$ for the recent nonzero-temperature
  simulations on $N_\tau=8$ lattices with $L_s=32$
  \cite{Cheng:2008ge}.  At lower $\beta$, corresponding to lower
  temperatures, $m_{res}$ increases rapidly, and is larger than the
  input light quark mass already in the transition region.  }
\label{fig:mres_ft08}
\end{figure}

Early, $N_\tau=4$ nonzero-temperature domain-wall fermion simulations
suffered from large residual mass, since the lattice spacing in the
transition/crossover region is large, leading to much heavier quarks
than desired \cite{Chen:2000zu}. More recent simulations with $N_\tau=8$,
still using $L_S=32$ and even 96 are described in \cite{Cheng:2008ge}.
Even for the $L_s=32$ simulations with $N_t=8$ the residual mass is
uncomfortably large in the transition region, and getting worse at
lower temperatures, corresponding to smaller $\beta$ as shown in
Fig.~\ref{fig:mres_ft08}, since one would like $m_{res}$ to be small
compared with the input light quark masses.

\subsubsection{Overlap fermions}

Related to the domain-wall fermions of the previous subsection are the
so-called overlap fermions developed by Narayanan and Neuberger
\cite{Narayanan:1994gw, Neuberger:1997fp}. They retain a complete
chiral symmetry without the doubling problem, albeit again at
substantial additional computational cost.

The overlap Dirac operator for massless fermions can be written as
\cite{Neuberger:1997fp},
\begin{equation}
a D_{ov} = M \left[ 1 + \gamma_5 \varepsilon\left( \gamma_5 D_W(-M) \right)
\right] ~,
\label{eq:D_ov}
\end{equation}
where $D_W(-M)$ is the usual Wilson Dirac operator with negative mass
$m = -M$. As with domain-wall fermions $0 < M < 2$ should be used.
For a hermitian matrix $X$, $\varepsilon(X)$ is the matrix sign function,
that can be defined as
\begin{equation}
\varepsilon(X) = \frac{X}{\sqrt{X^2}} ~.
\label{eq:eps_X}
\end{equation}

Using the fact that $\varepsilon^2(X) = 1$ it is easy to see that the
Neuberger-Dirac operator satisfies the so-called Ginsparg-Wilson
relation \cite{Ginsparg:1981bj},
\begin{equation}
\left\{ \gamma_5, D_{ov} \right\} = a D_{ov} \gamma_5 R D_{ov} ~,
\label{eq:G_W}
\end{equation}
with $R = 1/M$. Equivalently, when the inverse of $D_{ov}$ is
well defined, it satisfies
\begin{equation}
\left\{ \gamma_5, D^{-1}_{ov} \right\} = a \gamma_5 R ~.
\label{eq:G_W_inv}
\end{equation}
Chiral symmetry, in the continuum, implies that the massless fermion
propagator anticommutes with $\gamma_5$. As seen above, the massless
overlap propagator violates this only by a local term that vanishes
in the continuum limit. According to Ginsparg and Wilson this is the
mildest violation of the continuum chiral symmetry on the lattice
possible. L\"uscher \cite{Luscher:1998pqa} has shown that any Dirac
operator satisfying the Ginsparg-Wilson relation (\ref{eq:G_W}) has a
modified chiral symmetry at finite lattice spacing,
\begin{equation}
\delta \psi = i \epsilon \gamma_5 \left( 1 - \frac{a}{2M} D \right) \psi ~,~~~
 \delta \bar\psi = i \epsilon \bar\psi \left( 1 - \frac{a}{2M} D \right)
 \gamma_5 ~.
\label{eq:chi_sym}
\end{equation}
or
\begin{equation}
\delta \psi = i \epsilon \gamma_5 \left( 1 - \frac{a}{M} D \right) \psi
 = i \epsilon \hat\gamma_5 \psi ~,~~~
 \delta \bar\psi = i \epsilon \bar\psi \gamma_5 ~,
\label{eq:chi_sym2}
\end{equation}
with $\hat\gamma_5 = \gamma_5 \left( 1 - \frac{a}{M} D \right)$
satisfying $\hat\gamma_5^\dagger = \hat\gamma_5$ and, using the
G-W relation, Eq.~(\ref{eq:G_W}), $\hat\gamma_5^2 = 1$.

So far only one exploratory study, on a $6^3 \times 4$ lattice, of
nonzero-temperature overlap fermions has been done \cite{Fodor:2003bh}.
The main difficulty and computational cost for overlap fermions comes
from the numerical implementation of the matrix sign function,
Eq.~(\ref{eq:eps_X}).

\subsection{Cutoff effects}

In selecting a fermion formalism for a thermodynamics study, it is
important to be aware of possible lattice artifacts (cutoff effects).
There are two important categories of artifacts.  One comes from an
imperfect rendering of chiral symmetry.  The other, from the free quark
dispersion relation.

It is obviously important to get the chiral symmetry right if we are
simulating close to a chiral phase transition.  Each action has its
problems with chiral symmetry.  For staggered fermions the taste
splitting interferes.  For Wilson fermions, the chiral symmetry is
explicitly broken at nonzero lattice spacing.  For these actions the
obvious remedy is to reduce the lattice spacing.  For domain-wall
fermions, chiral symmetry is broken to the extent the fifth dimension
is not infinite, and, for overlap fermions, chiral symmetry is broken
to the extent the matrix sign function is only approximated in
numerical simulations.  For the latter two chiral actions, this type
of error can be reduced without also reducing the lattice spacing.

At high temperatures where quarks are effectively deconfined, it would
seem important to have a good quark dispersion relation, so, for
example, we get an accurate value for the energy density and pressure.
This artifact can be studied analytically for free fermions.  Recently
Hegde {\it et al.}  \cite{Hegde:2008nx} looked at deviations from the
expected free-fermion Stefan-Boltzmann relation for the pressure $p$
as a function of $1/N_\tau^2$ (equivalently $a^2$) and chemical
potential $\mu/T$:
\be
       \frac{p}{T^4} = 
           \sum_{k=0}^\infty A_{2k} P_{2k}(\mu/\pi T) 
                  \left(\frac{\pi}{N_\tau}\right)^{2k}
\label{eq:Pexpansion}
\ee
where $P_{2k}(\mu/\pi T)$ is a polynomial normalized so that
$P_{2k}(0) = 1$.  The leading term $A_0$ is the Stefan-Boltzmann term.
The ratios of higher coefficients $A_{2k}/A_0$ measure the strength of
the cutoff effects.  These terms determine the ability of the action
to approximate the continuum free fermion dispersion relation,
and they are useful in comparing actions to the extent
free quarks are relevant in an interacting plasma.  Table
\ref{tab:SB} reproduces their results for a variety of actions.  We
see that the hypercube action \cite{Bietenholz:1996pf} has pleasingly
small coefficients.  The Naik (asqtad) and p4 (p4fat3) actions remove
the second order term as designed, but the p4 action is better at
sixth order.  The standard (unimproved) staggered action (regardless
of gauge-link smearing) does as poorly as does the standard (and
clover-improved) Wilson actions.  The overlap and domain-wall actions
constructed from the standard Wilson kernel unfortunately inherit its
poor behavior. Improving the kernel fermion action would help to
reduce these cutoff effects.

\begin{table}
\begin{center}
\begin{tabular}{|c|c|c|c|}
\hline
action & $A_{2}/A_{0}$ & $A_{4}/A_{0}$ & $A_{6}/A_{0}$\tabularnewline
\hline
\hline
standard staggered & $248/147$ & $635/147$ & $3796/189$\tabularnewline
\hline
Naik & $0$ & $-1143/980$ & $-365/77$\tabularnewline
\hline
p4 & $0$ & $-1143/980$ & $73/2079$\tabularnewline
\hline
\hline
standard Wilson & $248/147$ & $635/147$ & $13351/8316$\tabularnewline
\hline
hypercube & $-0.242381$ & $0.114366$ & $-0.0436614$\tabularnewline
\hline
\hline
overlap/ & $248/147$ & $635/147$ & $3796/189$\tabularnewline
domain wall &~&~&\tabularnewline
\hline
\end{tabular}
\caption{Continuum limit scaling behavior of free massless quarks in
  various lattice formulations, based on an expansion
  [Eq.~(\protect\ref{eq:Pexpansion})] of the pressure in powers of
  $1/N_\tau^2$ from \protect\cite{Hegde:2008nx}.  Shown are ratios of
  the expansion coefficients to the ideal, leading Stefan-Boltzmann
  coefficient.  A small ratio indicates good scaling.
\label{tab:SB}
}
\end{center}
\end{table}

\section{Determining the Transition Temperature}
\label{sec:Tc}

We want to know the temperature of the transition from confined
hadronic matter to a quark-gluon plasma for two obvious reasons: to
interpret experimental data and to understand QCD as a field theory.
If the transition is only a crossover, a likely possibility for QCD at
the physical value of the quark masses as discussed below, and a true
phase transition occurs only at unphysical values of the quark masses,
then these two purposes diverge.  A crossover temperature is
imprecise, so its meaning could vary with the observable, but one can
at least speak of a range of temperatures over which
phenomenologically interesting changes take place, or one could choose
one observable to identify a temperature.  A true phase
transition has a precise temperature defined by the singularity of the
partition function, and all observables capable of producing a signal
should agree about the temperature.

In this section we discuss a variety of observables commonly used to
detect the transition.  In the following sections we discuss what we
have learned from them about the phase structure of QCD.

Two observables are traditionally used to determine the temperature of
the transition: the Polyakov loop and the chiral condensate.  The
Polyakov loop is a natural indicator of deconfinement.  The chiral
condensate is an indicator of chiral symmetry restoration.

\subsection{Polyakov loop and the free energy of color screening}
\label{sec:Polyakov}

The Polyakov loop is an order parameter for a high-temperature,
deconfining phase transition in QCD in the limit of infinite quark
masses.  At finite quark masses it is no longer an order parameter,
but it is still used to locate the transition. It is built from the
product of time-like gauge-link matrices.  It is the expectation value
of the color trace of that product:


\be
   L({\bf x},a,T) = \VEV{\Tr \prod_{\tau=0}^{N_\tau-1}U_0({\bf x},\tau)}.
\label{eq:Polyakov}
\ee
This quantity is gauge invariant because the combined boundary
conditions for gluon and fermion fields require that gauge
transformations be periodic under $t \rightarrow t + N_\tau$.
Translational invariance insures that it is independent of ${\bf x}$.
It can be shown that the Polyakov loop measures the change in the free
energy of the ensemble under the introduction of a static quark
(excluding its mass).
\be
  L(a,T) = \exp[-F_L(a,T)/T].
\ee
In that sense the Polyakov loop is a useful phenomenological quantity
as we now explain. When a static quark is introduced it must be
screened so that the ensemble remains a color singlet.  At low
temperatures, screening is achieved by binding to it the lightest
antiquark, forming a static-light meson.  The free energy cost then
consists of the self-energy of the static charge, the binding energy,
and the self-energy of the light quark.  In the quark plasma, color
neutrality is achieved through a collective shift of the plasma
charges, as in Debye screening in an ordinary electrical plasma.
Aside from the self-energy of the static quark, which is the same at
all temperatures, the additional free energy cost is small.  So we
expect $F_q(a,T)$ to decrease abruptly in the transition from the
confining regime to the plasma regime.

\begin{figure}
\begin{center}
\includegraphics[width=0.6\textwidth]{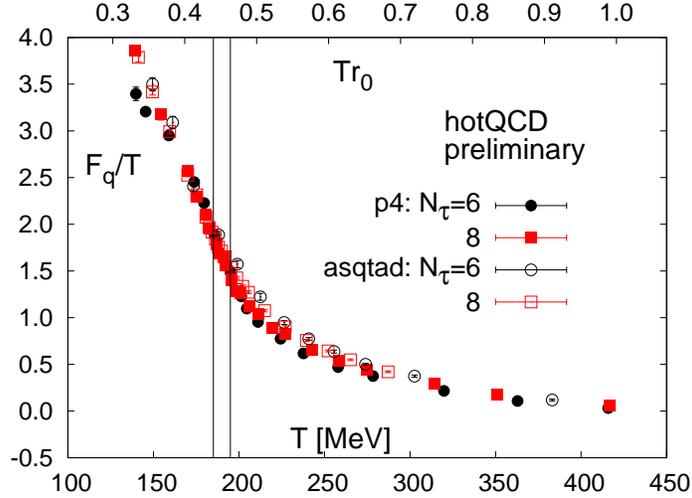}\\
\end{center}
\caption{Renormalized screening free energy of a static quark (from
  the renormalized Polyakov loop) \vs temperature in MeV units (bottom
  scale) and $r_0$ units (top scale) for $N_\tau = 6$ and $8$ from a
  HotQCD study comparing p4fat3 and asqtad staggered fermion
  formulations
  \protect\cite{Gupta:Lat2008,DeTar:2007as,Bazavov:2009zn}. Measurements
  are taken along a line of constant physics with $m_{ud} = 0.1 m_s$.
  The vertical bands here and in HotQCD figures below indicate a
  temperature range 185 - 195 MeV and serve to facilitate comparison.}
  \label{fig:renormPloop}
\end{figure}

The static-quark self energy diverges as $1/a$ in the limit of small
lattice spacing, so it is convenient to remove it from the definitions
of the free energy and the Polyakov loop:
\be
  F_L(a,T) = F_{\rm static}(a) + F_q(T) \quad \quad L_{\rm renorm}(T) = \exp[-F_q(T)/T].
\ee
Figure \ref{fig:renormPloop} illustrates the free energy from a recent
lattice simulation.  (Here and elsewhere, the temperature scale is
given in MeV and in units of the Sommer parameter
\cite{Sommer:1993ce}, $r_0 \approx 0.467$ fm.  The latter is defined
in terms of the potential $V(r)$ between a heavy quark and antiquark.
It is the distance where $r^2dV(r)/dr = 1.65$.)  The renormalized free
energy behaves as expected.

If we take the masses of all the quarks to infinity, we arrive at the
pure $SU(3)$ Yang-Mills ensemble, which has a first-order deconfining
phase transition with zero $L(a,T)$ at low temperatures and nonzero at
high temperatures.  The free energy is correspondingly infinite for $T
< T_c$ and finite above.  In this limit the Polyakov loop is a true
order parameter for the transition.

\subsection{Chiral condensate}

\subsubsection{Chiral symmetry restoration}

The second traditional observable is the chiral condensate.  It is the
order parameter for a high-temperature, chiral-symmetry restoring
phase transition at zero up and down quark masses.  At nonzero quark
masses, it is no longer an order parameter, but, like the Polyakov
loop, it is used as an indicator of the transition.  It is defined
for each quark flavor $i$ as the derivative of the thermodynamic
potential $\ln Z$ with respect to the quark mass,
\be
  \VEV{\bar\psi_i(x)\psi_i(x)} = \frac{T}{V}\frac{\partial \ln Z}{\partial m_i}
       = \frac{T}{V}\VEV{\Tr M_i^{-1}},
\ee
or the expectation value of the trace of the inverse of the fermion
matrix.  When the $u$ and $d$ quark masses both vanish, QCD has a
$U(1)\times SU(2)\times SU(2)$ chiral symmetry, which is spontaneously
broken at low temperatures to $U(1)\times SU(2)$, \ie, the familiar
baryon number and isospin symmetries.  At high temperatures the full
chiral symmetry is restored.  The chiral condensate $\VEV{\pbp}$ is
the order parameter of the broken symmetry.  It is nonzero at low
temperatures and zero at high temperatures.  With only two flavors,
the phase transition is expected to be second order, so the chiral
condensate is continuous at the transition. When the quark masses are
small, but nonzero, as they are in nature, the symmetry is explicitly
broken and the chiral condensate does not vanish at high temperatures,
but it is small.

\begin{figure}
\begin{center}
\includegraphics[width=0.6\textwidth]{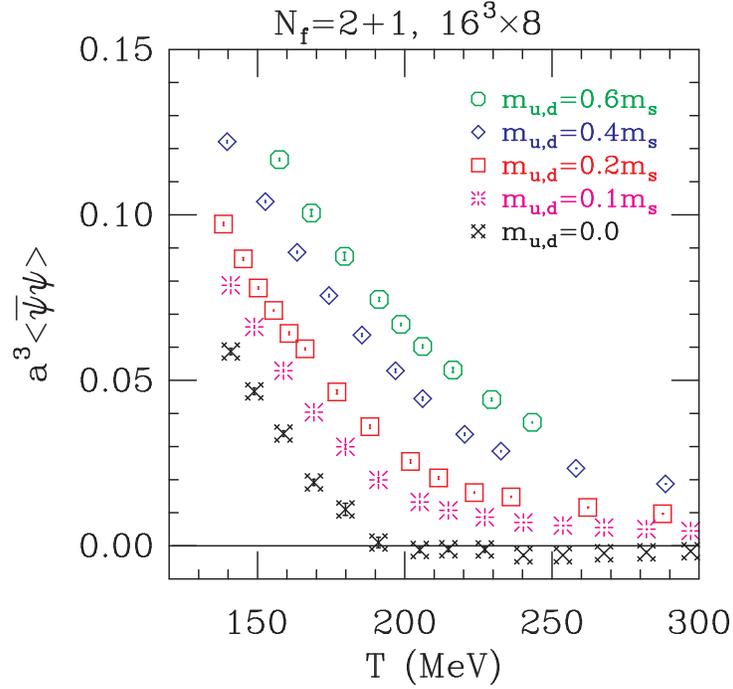}\\
\end{center}
\caption{Chiral condensate \vs temperature in MeV units ($r_0$ scale)
  for $N_\tau = 8$  from \cite{Bernard:2004je}
  using the asqtad fermion formulation. Measurements were taken along
  lines of constant physics with a range of light, degenerate
  up and down quark masses $m_{ud}$ specified in the legend as a 
  fraction of the strange quark mass $m_s$.  An extrapolation to
  zero quark mass is also shown.}
\label{fig:pbp_ud}
\end{figure}

Figure \ref{fig:pbp_ud} illustrates the behavior of the chiral
condensate from a recent lattice simulation with two light quark
flavors and one massive strange quark.  Measurements were taken
along ``lines of constant physics'', \ie, curves in the space of the
bare parameters (gauge coupling and quark masses) along which the pion
and kaon masses are approximately constant, whether or not they have
their correct experimental values.  The extrapolation to zero light
quark mass appears to be consistent with the expected behavior of this
chiral order parameter.

\subsubsection{Chiral multiplets}

The restoration of chiral symmetry leads to symmetry multiplets in the
hadronic spectrum.  At low temperatures, where the symmetry is
spontaneously broken, the spectrum consists of the familiar hadrons.
At high temperatures, where the symmetry is restored, they may have
analogs as resonant plasma excitations, at least not too far above the
crossover temperature.  They also control screening in the plasma in
analogy with the Yukawa interaction.  (See
Sec.~\ref{subsec:screening}.)

When the light quarks are massless, spontaneous symmetry breaking
requires that the pion be massless.  If the symmetry is restored at
high temperatures, the pion, suitably defined as a state, acquires a
mass.  Of course, in nature, the light quarks are not massless, so the
symmetry is only approximate, and the pion has a small mass at low
temperatures.

Another consequence of restoring the chiral symmetry with massless $u$
and $d$ quarks is that all hadronic states involving those quarks
would fall into larger symmetry multiplets.  Thus, for example, the
three pions become degenerate with the $f_0$, the three $a_0$'s become
degenerate with the $\eta$, and nucleons become degenerate with
parity-partner nucleons.

The classical QCD Lagrangian suggests a further $U(1)$ chiral
symmetry, which would conserve a flavor-singlet axial charge.  This
symmetry is broken at the quantum level.  This quantum phenomenon is
called the Adler-Bell-Jackiw axial anomaly~\cite{PeskinSchroederBook}.
Whether the strength of the anomaly decreases in conjunction with the
high temperature transition is an open question.

If the anomaly also vanishes, the eight meson states listed above fall
into a single degenerate supermultiplet.  Again, if the light quarks
are not precisely massless or the anomaly does not completely vanish,
these statements are only approximate.

Whether or not hadron-like resonances are observable in experiments,
the multiplets appear, nonetheless, in calculations, most notably in
simulations of hadronic screening.

\subsubsection{Singularities of the chiral condensate}

Although we require a numerical simulation to determine the chiral
condensate, from general considerations we can predict some of its
singularities at small quark mass $m$ and small lattice spacing $a$:
\bea
\VEV{\pbp(a,m,T)} \sim \left\{
\begin{array}{ll}
  c_{1/2}(a,T) \sqrt{m} + c_1 m/a^2 + {\rm analytic}
      & \mbox{$T < T_c$} , \\
  c_1 m/a^2 + c_\delta m^{1/\delta} + {\rm analytic} & \mbox{$T = T_c$} ,
  \label{eq:pbp_sing} \\
  c_1 m/a^2 + {\rm analytic}
      & \mbox{$T > T_c$} .
\end{array}
\right. 
\eea
Knowing the behavior of the condensate, and in particular its
singularities, is important for locating the phase transition.
The $m/a^2$ singularity is easily derived in perturbation theory from
a one-quark-loop diagram.  The $\sqrt{m}$ singularity at low
temperatures arises in chiral perturbation theory at one-loop order.
In this case the pion makes the loop.  It is an infrared singularity
caused by the vanishing of the pion mass at zero quark mass
\cite{Karsch:2008ch}.  Thus it appears only in the confined phase
where the pion is massless.  If we take $T \rightarrow 0$ before
$m\rightarrow 0$ the square root singularity is replaced by the usual
chiral $\log(m)$.  The term $m^{1/\delta}$ is the expected critical
behavior at the transition temperature.  [For the expected 3-d $O(4)$
universality class, $\delta = 0.56$.]  The RBC-Bielefeld group discusses
evidence for the expected mass dependence \cite{Karsch:2008ch}.

\begin{figure}
\begin{center}
\hspace*{15mm}
\includegraphics[width=0.6\textwidth]{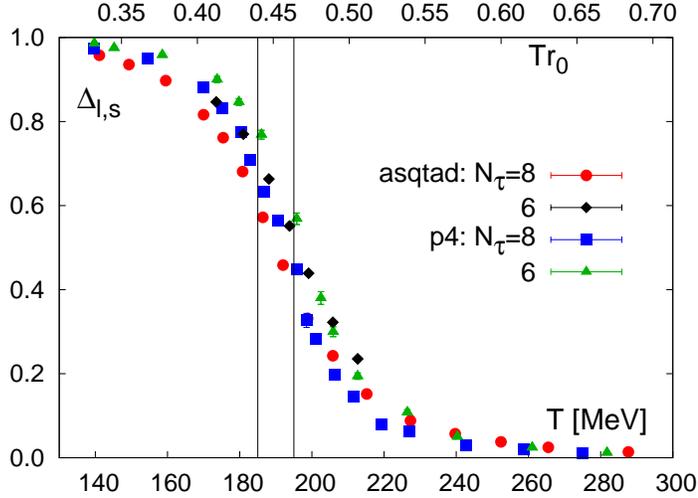}\\
\end{center}
\caption{To give an indication of its variation with lattice spacing,
  we plot the chiral condensate difference ratio \vs temperature in
  MeV units (bottom scale) and $r_0$ units (top scale) for $N_\tau =
  6$ and $8$ from a HotQCD study.  Results are given for both the
  p4fat3 and asqtad staggered fermion formulations
  \protect\cite{Bazavov:2009zn}. Measurements are taken along a line
  of constant physics with $m_{ud} = 0.1 m_s$.}
\label{fig:diff}
\end{figure}

In a calculation with three quarks with masses $m_u = m_d = m_\ell$
and $m_s$, it is convenient for comparing results of different
calculations to eliminate the ultraviolet divergence by taking a
linear combination of the light-quark and strange-quark chiral
condensates
\bea
 D_{\ell,s}(T) &=& \langle \bar \psi \psi \rangle|_\ell 
     - \frac{m_\ell}{m_s} \langle \psi \psi \rangle|_s , \nonumber \\
 \Delta_{\ell,s}(T) &=& D_{\ell,s}(T)/D_{\ell,s}(T=0) .
\eea
The ratio $\Delta_{\ell,s}$ of the high-temperature and
zero-temperature value also eliminates a common scalar-density
renormalization factor $Z_S$.  This is the quantity plotted in
Fig.~\ref{fig:diff} from a recent simulation.  It shows the expected
dramatic fall-off at the crossover.

\subsection{Other Observables}

Susceptibilities are often used as indicators of a phase transition.
They measure fluctuations in the related observables.  Since a
transition or crossover is usually accompanied by fluctuations in an
order parameter, the related susceptibilities tend to peak there.

\subsubsection{Quark number susceptibility}

In the low temperature phase, fluctuations in quark number are
suppressed by confinement for the same reason that the free energy of
screening of a static quark is large there.  At high temperatures,
fluctuations are common.  There can also be cross-correlations.  The
relevant observable for a quark of flavor $i$ is the expectation
$\VEV{N_i^2/V}$ for spatial volume $V$ and total quark number $N_i$.
This is the quark number susceptibility.  It controls event-by-event
fluctuations in the associated flavor in heavy ion collisions.  For
flavors $i$ and $j$ the generalized susceptibility (including cross
correlations) is
\be
 \chi_{ij} = \VEV{N_iN_j/V} 
     =  \frac{T}{V}\frac{\partial^2 \ln Z}{\partial \mu_i \partial \mu_j}.
\label{eq:general_susc}
\ee
We discuss the Taylor expansion of this observable in $\mu_i$ in
Sec.~\ref{subsec:eos_nzmu}.

\begin{figure}
    \hspace*{-5mm}
    \includegraphics[width=0.6\textwidth]{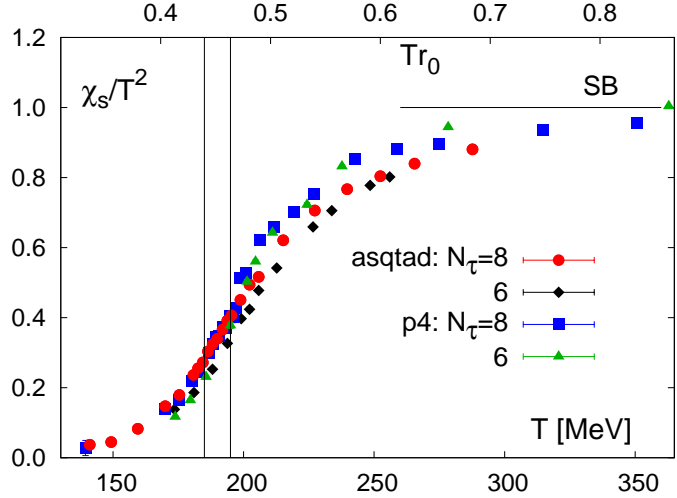}
\caption{Strange quark number susceptibility divided by the square of
  the temperature \vs temperature in MeV units (bottom scale) and
  $r_0$ units (top scale) for $N_\tau = 6$ and $8$.  Measurements are
  taken along a line of constant physics with $m_{ud} = 0.1
  m_s$. Results are from a HotQCD study comparing p4fat3 and asqtad
  staggered fermion formulations \protect\cite{Bazavov:2009zn}.}
\label{fig:strange_susc}
\end{figure}

Figure \ref{fig:strange_susc} illustrates the behavior of the strange
quark number susceptibility $\chi_{ss}$.  It shows an abrupt rise at
the crossover.  Because it has a relatively high signal to noise
ratio, this quantity is often used to define the crossover
temperature.

\begin{figure}
    \hspace*{-5mm}
    \includegraphics[width=0.6\textwidth]{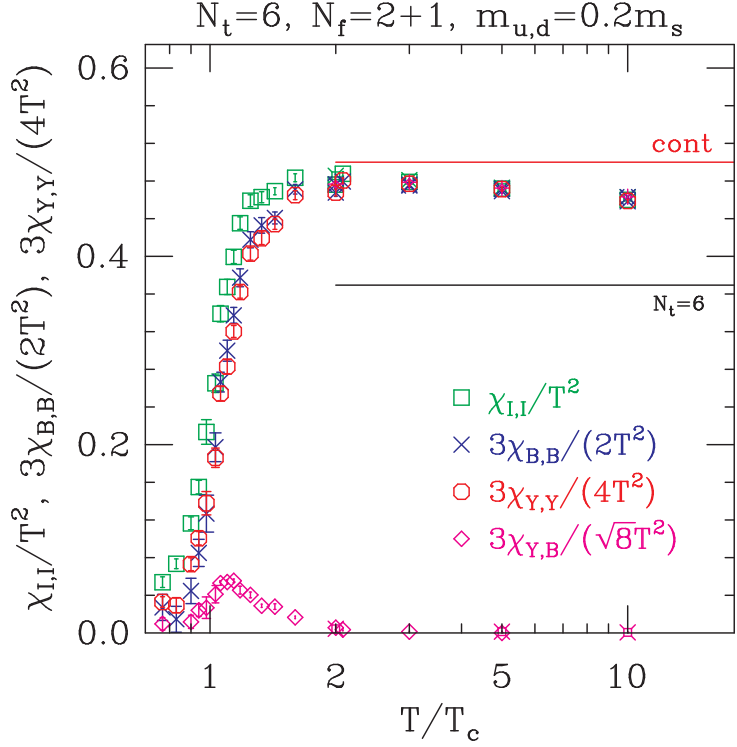}
\caption{Chiral susceptibility matrix in the $I$, $Y$, $B$ basis.
  divided by the square of the temperature \vs temperature in units of
  the crossover temperature $T_c$ for $N_\tau = 6$.  Measurements are
  taken along a line of constant physics with $m_{ud} = 0.2 m_s$ from
  \protect\cite{Bernard:2004je}.
}
\label{fig:full_susc}
\end{figure}

We can transform the generalized quark number susceptibility
$\chi_{ij}$ from the flavor basis to the basis in which the isospin
$I$, hypercharge $Y$, and baryon number $B$ are diagonal.  The
resulting quantities are shown in Fig.~\ref{fig:full_susc}.  The
diagonal susceptibilities all show the expected abrupt rise at the
crossover temperature.  The offdiagonal susceptibility $\chi_{Y,B}$
shows a small nonzero value above the crossover.  The positive
correlation between hypercharge and baryon number at these
temperatures can either be understood in terms of fluctuations in
light quark degrees of freedom or in terms of persistent three-quark
baryon states: Light up and down quarks have positive baryon number
(1/3) and hypercharge (1/3) and their antiquarks have the opposite
values.  In both cases their fluctuations lead to a positive
correlation.  Strange quarks have positive baryon number (1/3) but
negative hypercharge (-2/3).  They would lead to a negative
correlation, but because of their higher mass, they are less
prevalent.  So we are left with a net positive correlation.  At higher
temperatures the mass difference is irrelevant and the correlations
cancel.  Similar arguments can be made for three-quark baryonic
states, where nonstrange baryons are more prevalent than strange
baryons.

\begin{figure}
\begin{center}
\includegraphics[width=0.5\textwidth]{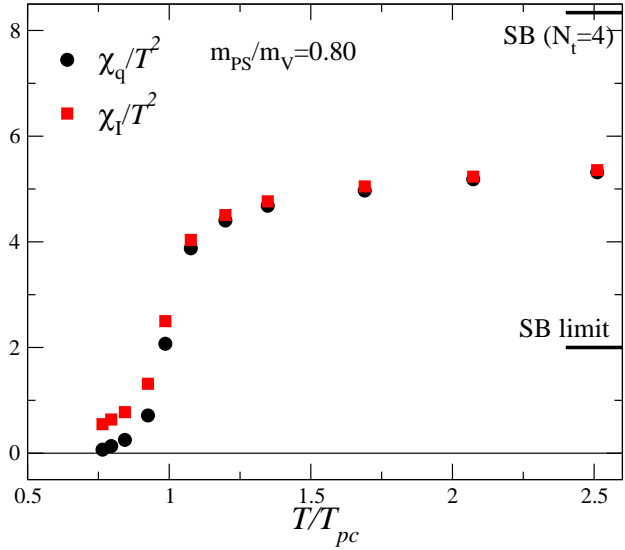}\\
\end{center}
\caption{Quark number susceptibilities with Wilson fermions on $N_\tau=4$
lattices along a line of constant physics with pseudoscalar to
vector meson mass ratio $m_{PS}/m_V=0.8$ from \cite{Maezawa:2007ew}.
}
\label{fig:Wils_qno}
\end{figure}

In Fig.~\ref{fig:Wils_qno} we show a computation of baryon ($\chi_q$)
and isospin ($\chi_I$) quark number susceptibilities from a recent
computation with two flavors of clover-improved Wilson fermions on
$16^3 \times 4$ lattices \cite{Maezawa:2007ew}, using a different
normalization from that of Fig.~\ref{fig:full_susc}.  The cutoff
effects for this Wilson fermion simulation are seen to be
significantly larger than with improved staggered fermions, as
expected from Table~\ref{tab:SB}.

\subsubsection{Chiral susceptibility}

The various chiral susceptibilities are based on the second derivative
of the thermodynamic potential with respect to the quark masses
\be
  \chi_{ij} = \frac{T}{V}\frac{\partial^2 \ln Z}{\partial m_i \partial m_j}.
\ee
For two equal mass light quarks $u$ and $d$, the derivatives can be
converted to expectation values of products of the inverse of the
fermion Dirac matrix $M$ for those species.  The commonly reported
susceptibilities are the ``disconnected'' chiral susceptibility
\be
 \chi_{\rm disc} = \frac{T}{V}\left[
   \left\langle (\Tr M^{-1})^2\right\rangle - 
           \left\langle \Tr M^{-1} \right\rangle^2\right],
\ee
the connected chiral susceptibility
\be
 \chi_{\rm conn} = \frac{T}{V}
   \left\langle\Tr M^{-2}\right\rangle,
\ee
the isosinglet susceptibility
\be
 \chi_{\rm sing} = \chi_{\rm conn} + 2\chi_{\rm disc} ,
\ee
and the isotriplet susceptibility
\be
  \chi_{\rm trip} = 2\chi_{\rm conn}.
\ee

\begin{figure}
  \begin{tabular}{cc}
  \includegraphics[width=0.6\textwidth]{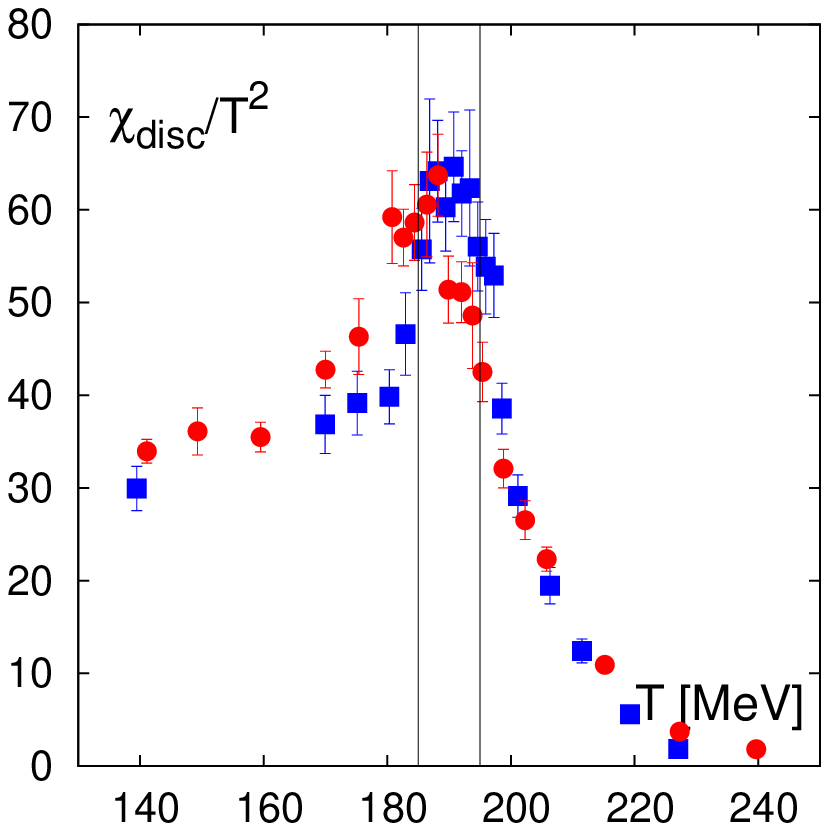}
  &
  \hspace*{-25mm}
  \includegraphics[width=0.6\textwidth]{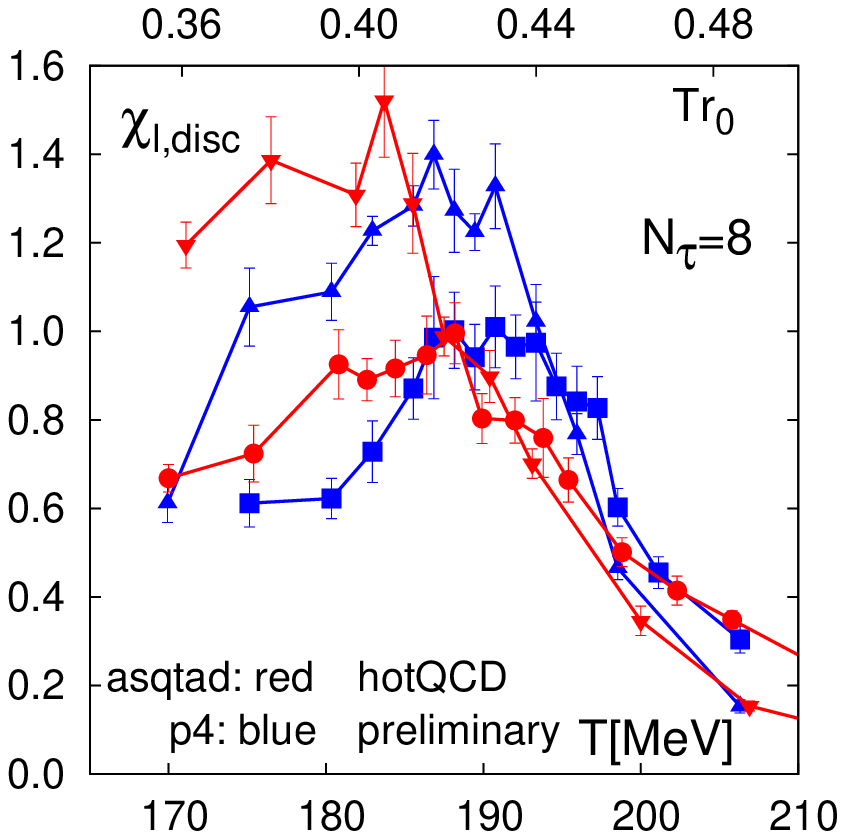}\\
   \end{tabular}
\caption{Left panel: Disconnected light quark susceptibility \vs
  temperature in MeV ($r_0$) units (bottom scale).  Right panel:
  closeup of the peak region. Lines merely connect the points. Red
  circles and downward-pointing triangles, asqtad fermions.  Blue
  squares and upward-pointing triangles, p4fat3.  Squares and circles
  are along a line of constant physics with $m_{ud} = 0.1 m_s$, and
  triangles, with $m_{ud} = 0.05 m_s$.  All results are HotQCD
  preliminary
  \protect\cite{Gupta:Lat2008,Soeldner:Lat2008,DeTar:2007as}.}
\label{fig:asqtad_p4_chiral_sus}
\end{figure}

Figure \ref{fig:asqtad_p4_chiral_sus} gives an example of the peak in the
disconnected chiral susceptibility at the crossover.

Since the chiral susceptibility is the derivative of the chiral
condensate with respect to quark mass, one can immediately derive its
singularities from the expressions for the condensate
in Eq.~(\ref{eq:pbp_sing}).
\bea
\chi_{\rm sing}(a,m,T) \sim \left\{
\begin{array}{ll}
  c_{1/2}(a,T)/(2\sqrt{m}) + c_1/a^2 + {\rm analytic}
      & \mbox{$T < T_c$} , \\
  c_1/a^2 + (c_\delta/\delta) m^{1/\delta-1} + {\rm analytic} & \mbox{$T = T_c$} ,
  \label{eq:pbp2_singularities} \\
  c_1/a^2 + {\rm analytic} 
      & \mbox{$T > T_c$} . 
\end{array}
\right. 
\eea
The RBC-Bielefeld group discusses numerical evidence for the expected
mass dependence \cite{Karsch:2008ch}.  Trends in
Fig.~\ref{fig:asqtad_p4_chiral_sus} are consistent with these
expectations.  In the limit of zero quark mass, this quantity is
infinite below the transition and finite above.  In the continuum
limit it has a temperature-independent ultraviolet divergence.  Thus
the Budapest/Wuppertal group propose subtracting the zero temperature
value, multiplying by the square of the bare quark mass, and dividing by
the fourth power of the temperature \cite{Aoki:2006br}:
\be
  m^2 \Delta \chi_{\rm disc}(a,m,T)/T^4.
\label{eq:BWrenorm_suscept}
\ee
where $\Delta \chi_{\rm disc}(a,m,T) = \chi(a,m,T) - \chi(a,m,0)$.
The $m^2$ cancels the scalar density renormalization factor.  Of
course, this quantity vanishes in the zero mass limit.

\subsection{Setting the temperature scale}

In order to quote dimensionful lattice results in physical units, it
is necessary to determine the lattice spacing in physical units.  The
calibration must be based on a quantity that is reliably determined in
zero-temperature lattice simulations.  Recent favorites are the
splitting of $\Upsilon$ levels, the mass of the $\Omega^-$ baryon, and
the light meson decay constants, such as $f_\pi$ or $f_K$.  These
scale determinations are not guaranteed to agree at nonzero lattice
spacing and at unphysical values of the quark masses.  Indeed, there
can be substantial differences.  For example, for the asqtad action
with a nearly physical strange quark mass, a light quark mass one
tenth as heavy, and a lattice spacing of approximately 0.12 fm, the
$f_K$ scale gives a 15\% lower temperature than the $\Upsilon$
splitting scale.  For the same quark masses, at approximately 0.09 fm
the discrepancy has decreased to 8\%, consistent with an approximately
${\cal O}(a^2)$ scaling error.  Of course, for any quantity of
interest, thermodynamic or not, if possible, we would like to choose a
scale according to which that quantity has only a small variation as
the lattice spacing approaches zero.

\begin{figure}
\begin{tabular}{cc}
\includegraphics[width=0.55\textwidth,bb=30 430 580 700]{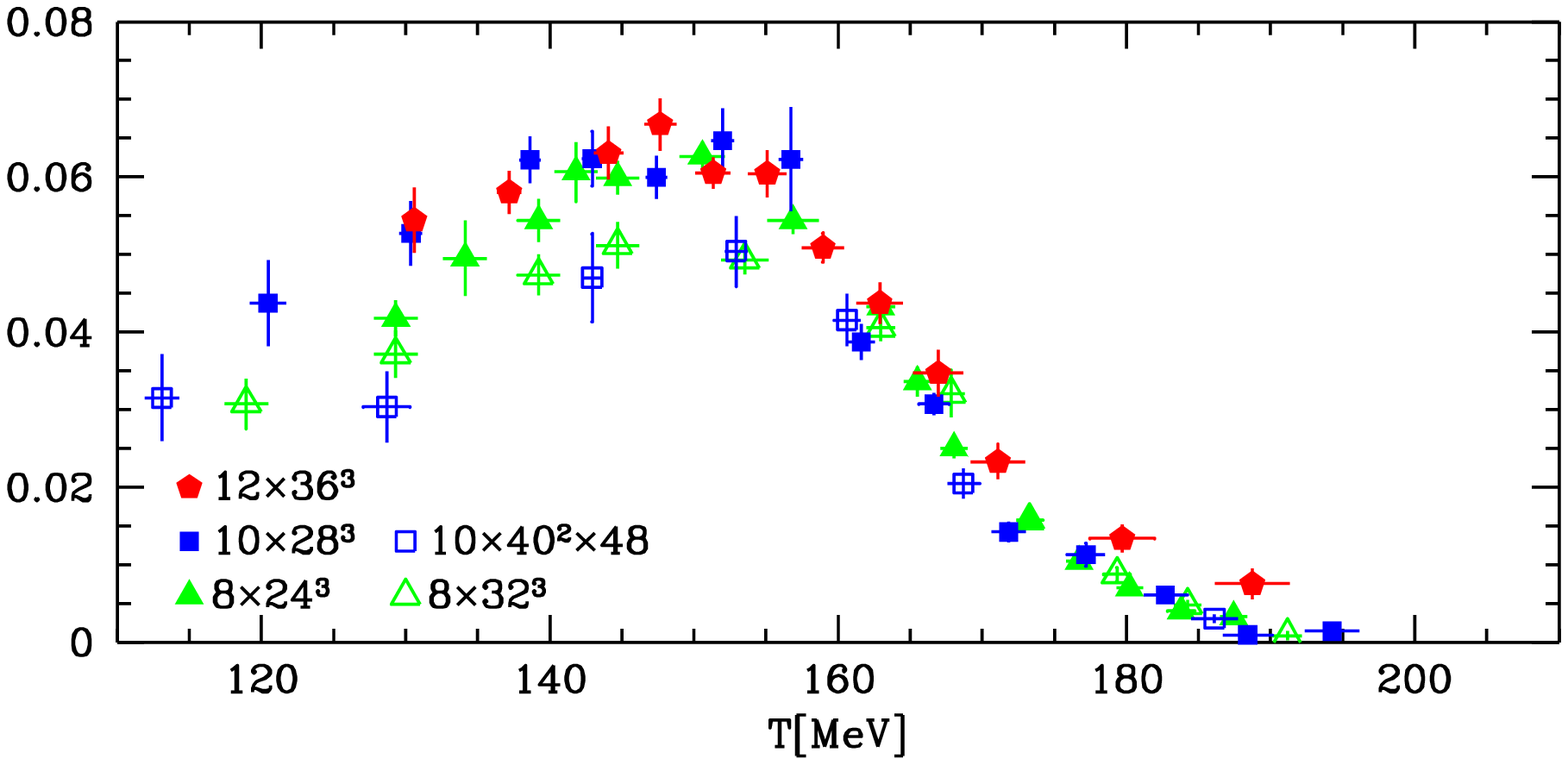}
&
\includegraphics[width=0.40\textwidth]{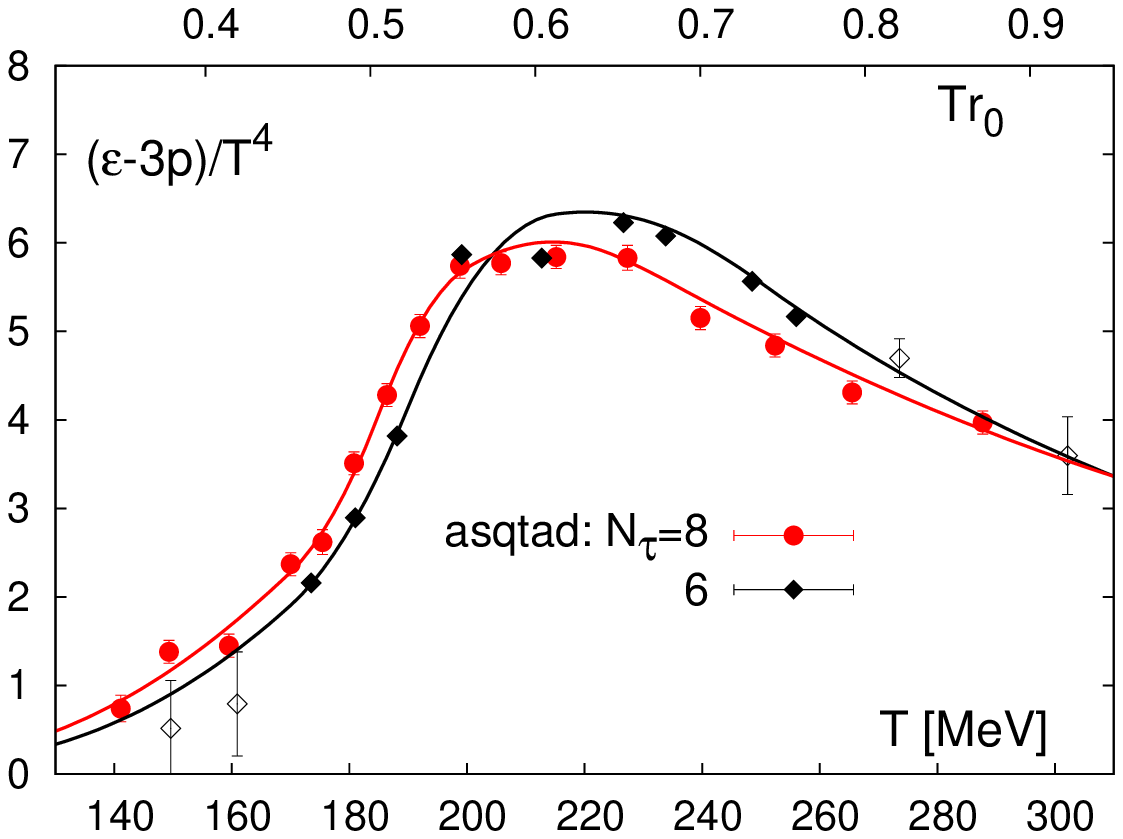}
\end{tabular}
\caption{Left panel: renormalized chiral susceptibility {\it vs}
  temperature ($f_K$ scale) from \protect\cite{Aoki:2009sc}.  Right
  panel: interaction measure {\it vs} temperature ($r_0$ scale) from
  \protect\cite{Bazavov:2009zn}.  (See the definition of this quantity in
  Sec.~\protect\ref{sec:integral_method}.) Note that the interaction
  measure peaks at about 20 MeV above the crossover.  }
\label{fig:BW_HotQCD}
\end{figure}

Recent results from Aoki \etal\ \cite{Aoki:2009sc} give a rather
different temperature $T_c$ for the crossover than the HotQCD
collaboration \cite{Bazavov:2009zn}.  Aoki \etal\ locate the peak in
their renormalized chiral susceptibility at around 150 MeV ($f_K$) for
$N_\tau = 8$, 10, and 12.  The HotQCD collaboration puts the crossover
closer to 190 MeV ($r_0$) for $N_\tau = 8$ and $m_{ud}/m_s =
0.1$. Here are possible reasons for the discrepancy:
\begin{itemize}
\item Much of the difference comes from the different choice of scale.
  The Budapest-Wuppertal collaboration uses $f_K$ to set the scale, and
  the HotQCD collaboration uses the Sommer parameter $r_0$, calibrated
  ultimately from $\Upsilon$ splittings \cite{Aubin:2004wf}.  The
  scale discrepancy alone could explain about 30 MeV of the
  difference.
\item Some of the discrepancy also comes from differences in lattice
  parameters.  The Budapest-Wuppertal collaboration uses a smaller
  lattice spacing and lighter light quark mass.  The HotQCD
  collaboration estimates an approximately 10 MeV ($r_0$ scale)
  downward shift in curves related to the equation of state in the
  continuum limit with physical quark masses.  Some of that shift is
  visible in the right panel of Fig.~\ref{fig:BW_HotQCD}.
\item Some may also come from differences in the fermion formulations.
  The Budapest-Wuppertal group use standard staggered fermions with
  stout gauge links.  This approach reduces effects of taste
  splitting, but does not improve the quark dispersion relation as do
  the actions used by the HotQCD collaboration.  We don't know whether
  such differences would result in a shift in a peak position,
  however.
\end{itemize}
Whatever the differences, no matter how one sets the scale, one
expects all methods to give the same results for the same observable
in the continuum limit at physical quark masses.  So for now we are
left guessing the result of taking that limit.  Since most of the
present difference apparently comes from a choice of scale, it would
help our guessing to know which scale is more suitable for
thermodynamic quantities.  We have seen that the chiral susceptibility
suffers from peculiar singularities that may make it less suitable for
locating the crossover temperature.  Still, the left panel of
Fig.~\ref{fig:BW_HotQCD} suggests that it scales reasonably well in
$f_K$ units.  For the phenomenology of heavy ion collisions,
quantities related more directly to deconfinement, such as the
interaction measure (equation of state) and quark number
susceptibility are important.  As we can see from the right panel of
Fig.~\ref{fig:BW_HotQCD} the interaction measure seems to show better
(but still imperfect) scaling in the $r_0$ scale.  (Preliminary HotQCD
results for the chiral susceptibility are shown in the $r_0$ scale in
Fig.~\ref{fig:asqtad_p4_chiral_sus}.)

\section{QCD Phase Diagram at Zero Density}
\label{sec:phase_diag_zero_dens}

\subsection{General outline of the phase diagram}

\begin{figure}[thb]
\begin{tabular}{cc}
\begin{minipage}[b]{80mm}
\includegraphics[width=80mm]{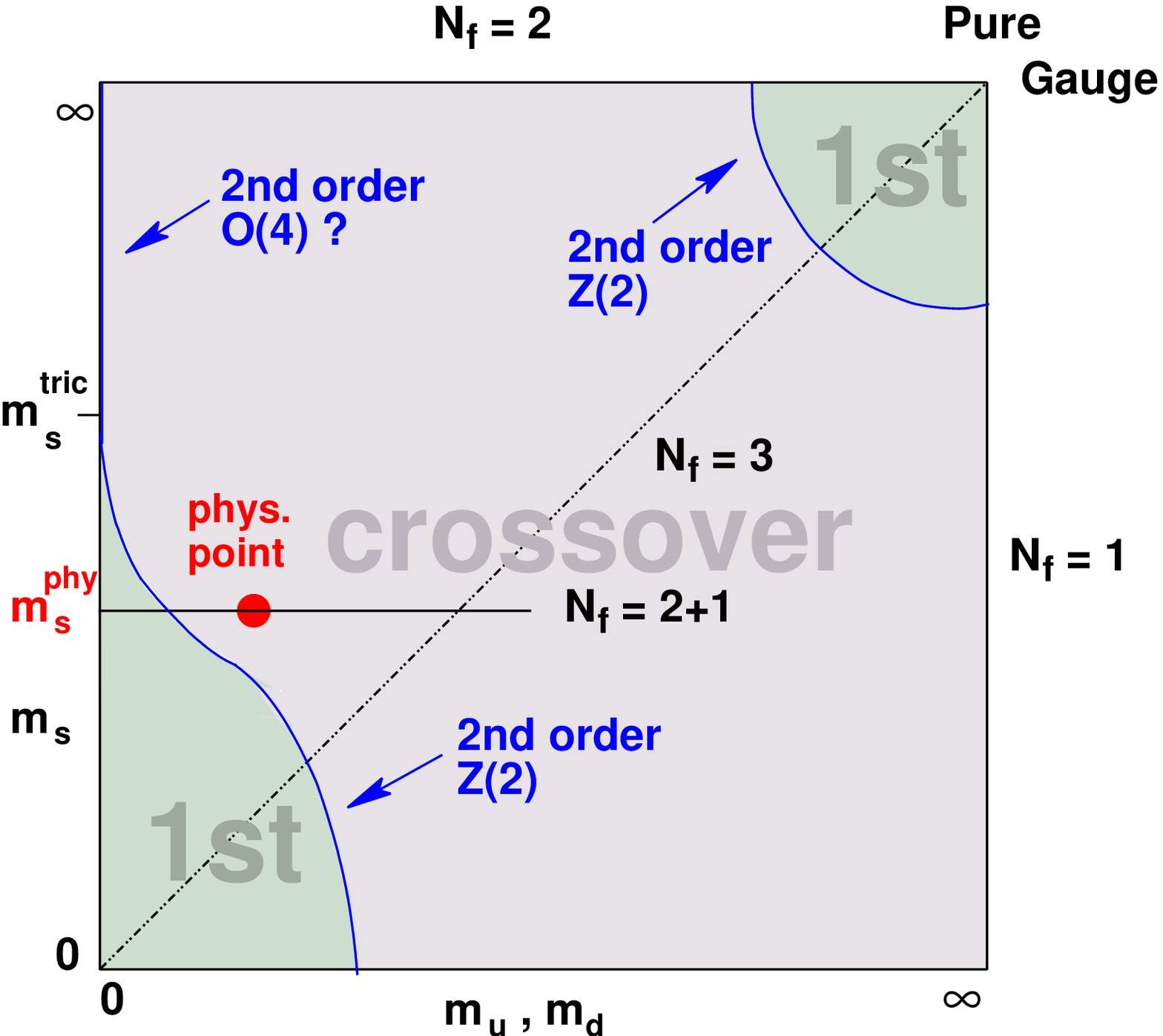} \\
\end{minipage}
&
\begin{minipage}[b]{80mm}
\includegraphics[width=80mm]{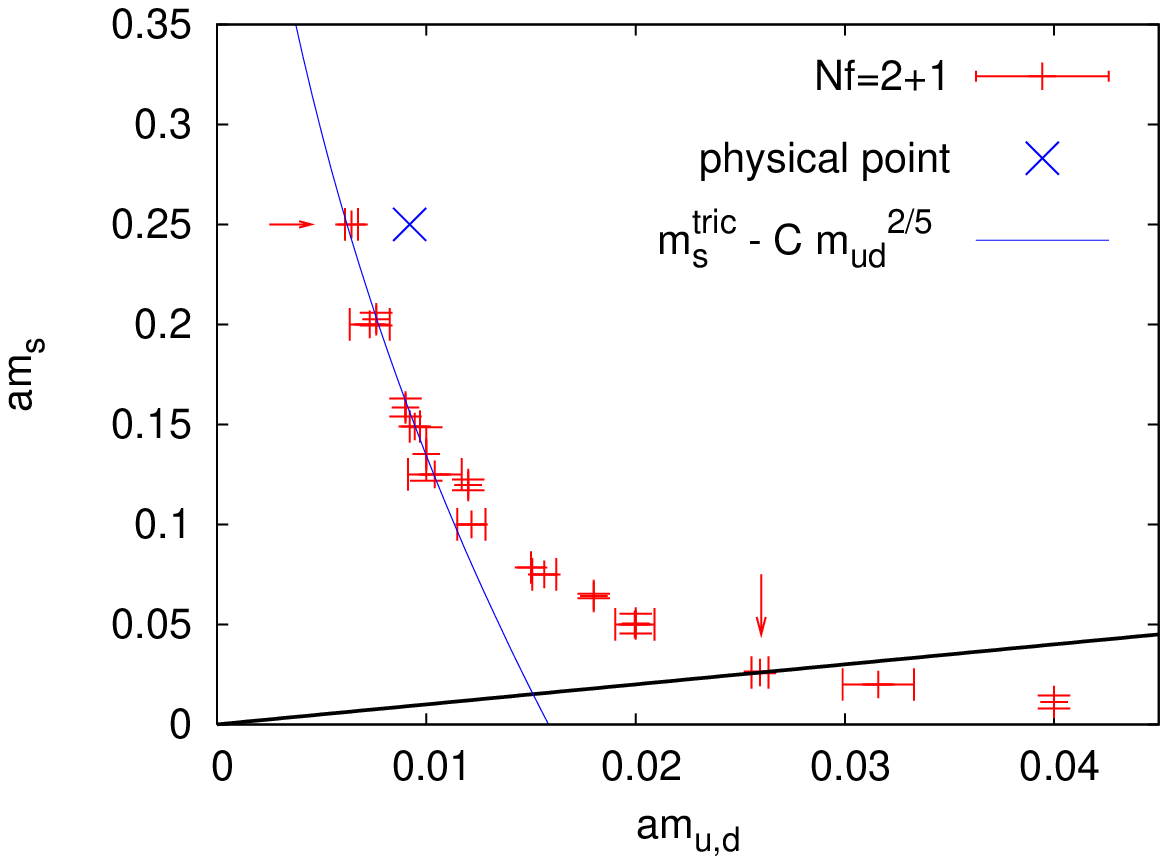} \\
\end{minipage}
\end{tabular}
\caption{Left panel: sketch of the phase diagram for QCD at zero baryon
  density in $2+1$ flavor QCD as a function of the light quark masses
  showing regions where a high-temperature phase transition or
  crossover is expected. For a second-order phase transition, the
  universality class is shown.  The physical point is plotted as a dot
  in the crossover region.  Whether the expected tricritical strange
  quark mass $m_s^{\rm tric}$ is higher or lower than the physical
  strange quark mass $m_s^{\rm phys}$ is not yet firmly established.
  (Similar versions of this figure have appeared in the literature
  including \protect\cite{Laermann:2003cv}.)
  Right panel: result of an actual measurement of a portion of the 2nd
  order $Z(2)$ phase boundary at $N_\tau = 4$ from
  Ref.~\protect\cite{deForcrand:2006pv}.  The axes give bare quark
  masses in lattice units and the blue cross marks the physical point.
}  
\label{fig:mu_ms_phase_diag}
\end{figure}

At infinite quark mass QCD becomes a pure Yang-Mills theory, which has
a well-studied, weak, first-order deconfining phase transition
\cite{Boyd:1995zg}.  As the quark masses are decreased, the first
order transition weakens further and devolves into a crossover, as
indicated in Fig.~\ref{fig:mu_ms_phase_diag}, which summarizes in
{\em qualitative} terms the generally accepted phase structure at zero
chemical potential in the flavors $u$, $d$, and $s$.

Close to zero quark mass, chiral perturbation theory applies, and
quite general arguments can be made about the qualitative nature of
the phase transition \cite{Pisarski:1983ms}, depending on the number
of quark flavors with zero mass and depending on what happens to the
anomaly at the transition.  With a nonzero anomaly and only two quark
flavors the transition certainly occurs at zero $u$ and $d$ quark
masses, and it is in the 3-d $O(4)$ universality class, because of the
$O(4)$ two-flavor chiral symmetry.  If the strange quark is also
massless, the chiral transition is first order, and, since first order
transitions are not usually removed by small symmetry-breaking
perturbations, it persists as the quark masses are increased.
Eventually, at sufficiently large $u$, $d$, and $s$ quark masses the
system is too far from chiral and the first-order transition gives way
to a second-order phase transition in the Ising or $Z(2)$ universality
class: Ising, since at nonzero quark masses, there is no remaining
chiral symmetry.  In the $m_u = m_d$ {\it vs} $m_s$ plane a curve of
such second order transitions separates the first-order regime from
the crossover regime as sketched in the left panel of
Fig.~\ref{fig:mu_ms_phase_diag}.

The quantitative determination of the phase boundaries requires
numerical simulation.  What has emerged is that the second-order
critical line occurs at quite small quark masses, where simulations
are particularly challenging and especially sensitive to cutoff
effects \cite{karsch:2001nf,Karsch:2003va}.  The right panel of
Fig.~\ref{fig:mu_ms_phase_diag} shows recent results from de Forcrand
and Philipsen based on a calculation using unimproved staggered
fermions with $N_\tau = 4$.

\subsection{Order of the phase transition for physical quark masses}

\begin{figure}[thb]
\begin{center}
\includegraphics[clip=true,width=160mm]{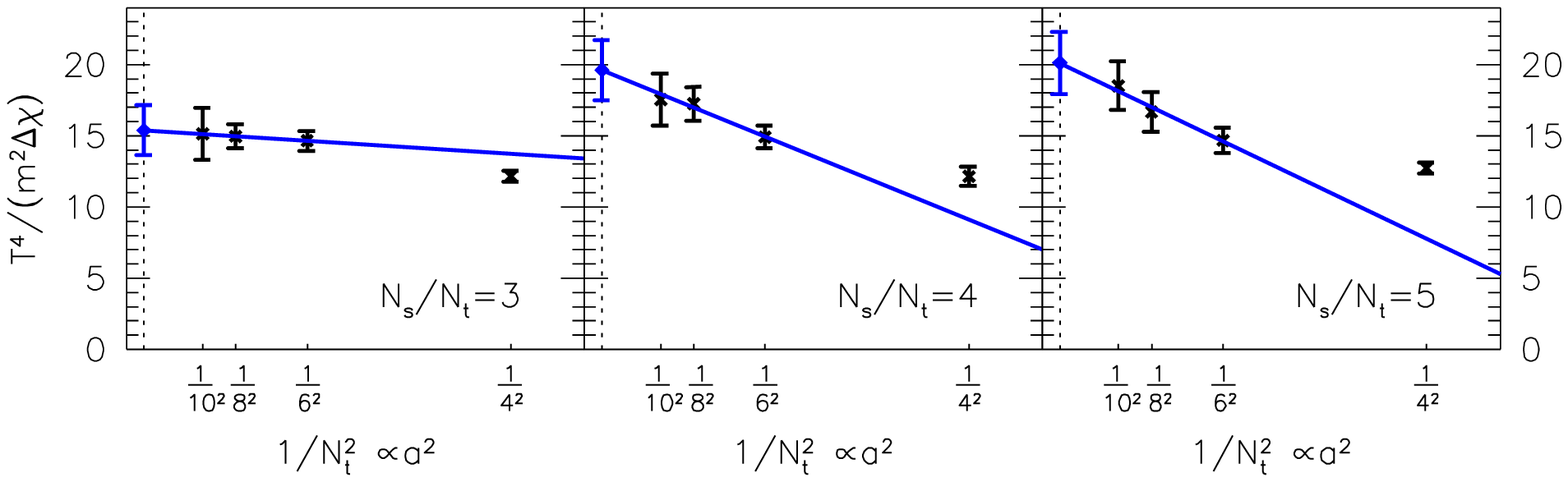} \\
\includegraphics[width=60mm]{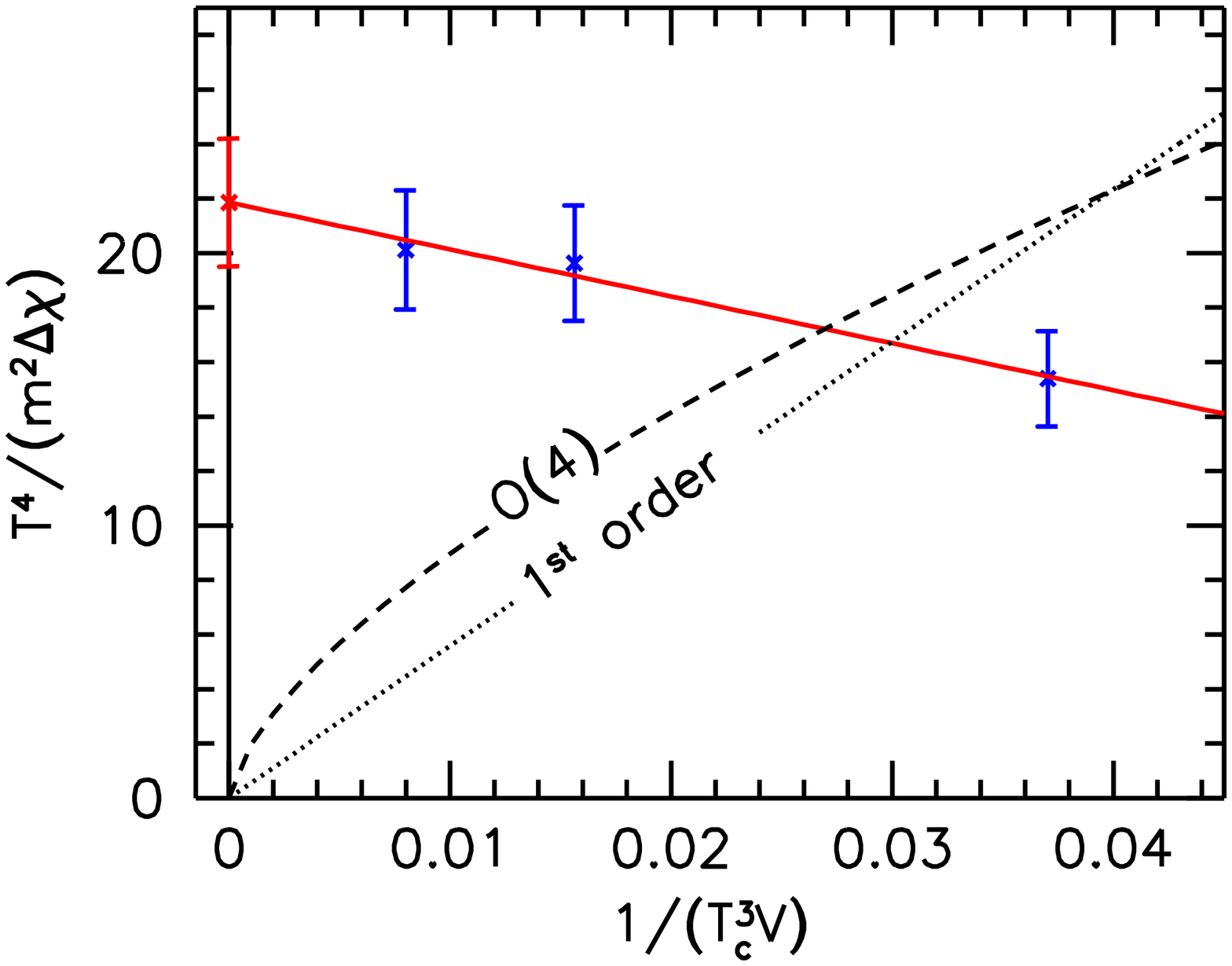}
\end{center}
\caption{Results from 
  \protect\cite{Aoki:2006we}.
  Upper panels: Inverse of the peak height in the renormalized
  disconnected chiral susceptibility \vs squared lattice spacing
  showing the extrapolation to zero lattice spacing.  The lattice
  aspect ratio is varied from left to right.
  Lower panel: Inverse of the peak height in the renormalized disconnected
  chiral susceptibility \vs inverse aspect ratio cubed showing the
  extrapolation to the thermodynamic limit.  Also shown are
  predictions for a first order phase transition and a second order
  transition in the 3-d $O(4)$ universality class.}
\label{fig:aoki_cont_ex_scal}
\end{figure}

A key phenomenological question is whether there is a first order
phase transition at the physical value of the $u$, $d$, and $s$ quark
masses or there is merely a crossover.  All present evidence points to
a crossover at zero chemical potential for these species.  A recent,
thorough investigation has been carried out by the Budapest-Wuppertal
group \cite{Aoki:2006we}.  They examine the conventional signal of the
peak height in the chiral susceptibility, which they renormalize using
Eq.~(\ref{eq:BWrenorm_suscept}).  If there is no phase transition
(\ie, only a crossover), the peak height should be asymptotically
constant in the thermodynamic limit of an infinite lattice volume.  If
there is a first order phase transition, the height is infinite, but
it is limited in a finite volume by finite-size effects.
Asymptotically, it scales linearly with the lattice volume $L^3$.  If
the transition is second order, the volume dependence is weaker, but
the result is still infinite.  The Budapest-Wuppertal group ran a
simulation with conventional staggered fermions on stout links at
$N_\tau = 4$, 6, 8, and 10.  They analyzed their data in two steps.
First they extrapolated the inverse peak height to zero lattice
spacing at fixed lattice aspect ratio $LT$, as shown in the upper
panels of Fig.~\ref{fig:aoki_cont_ex_scal}.  Then they extrapolated
the continuum values to infinite aspect ratio (thermodynamic limit).
The result is compared in the lower panel of
Fig.~\ref{fig:aoki_cont_ex_scal} with predictions for a first order
phase transition and a phase transition in the 3-d $O(4)$ universality
class.  The disagreement is a strong indication that there is no phase
transition.

\subsection{Order of the phase transition for two massless flavors}

There is a related question of significant theoretical interest.  When
all quarks but the $u$ and $d$ are infinitely massive, we have a
two-flavor theory, and, as we have observed above, as long as the
chiral anomaly is not involved, we expect a critical point only at
zero quark mass. Furthermore, since the two-flavor chiral symmetry
$SU(2) \times SU(2) \simeq O(4)$, we expect the high-temperature
deconfining critical point to be in the 3-d $O(4)$ universality class.

\begin{figure}
\begin{center}
\includegraphics[width=0.6\textwidth]{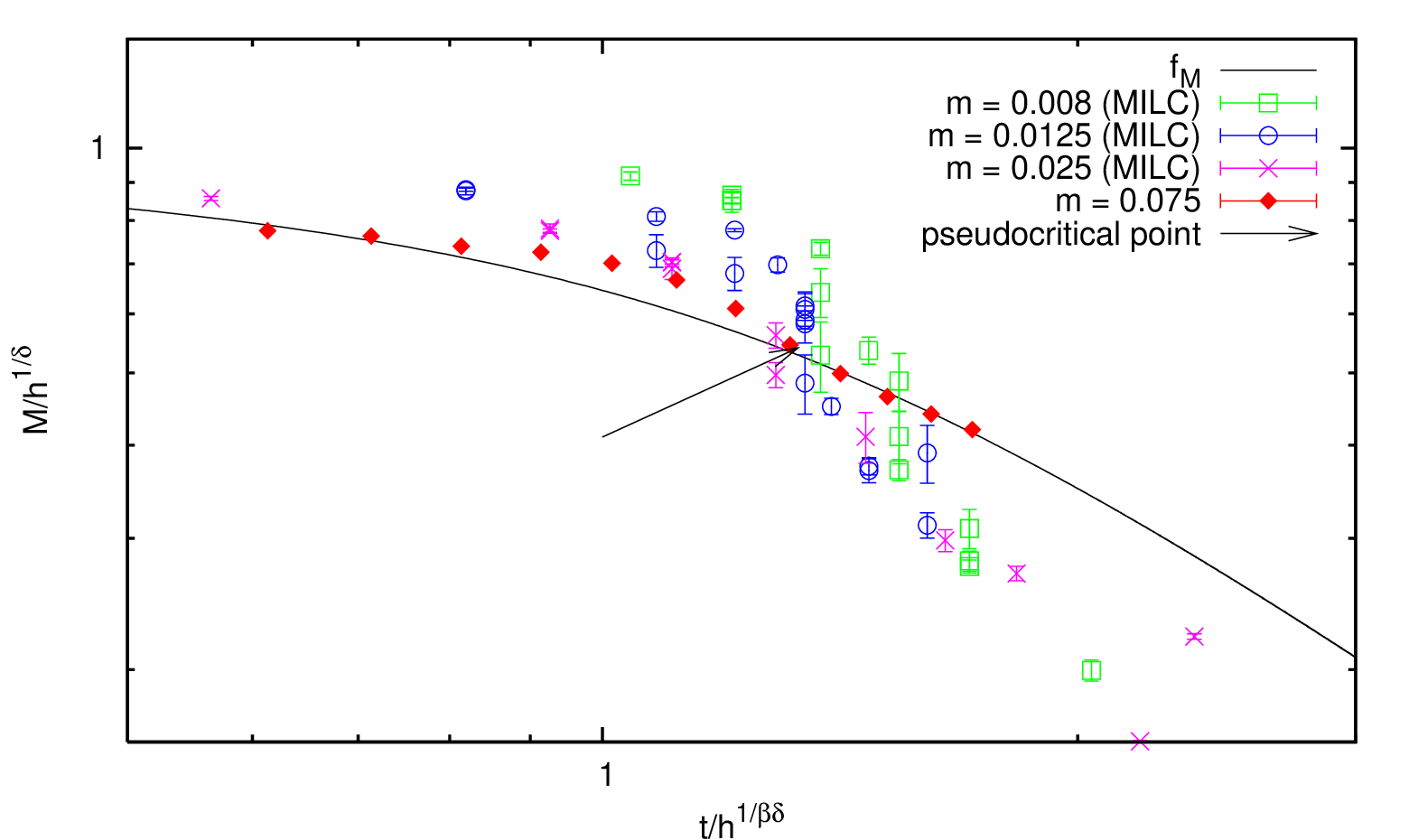}\\
\end{center}
\caption{A double lograrithmic plot showing strong deviations from
  $O(4)$ scaling in this parameter range for two flavors of staggered
  fermions using $N_\tau=4$ lattices, collected in
  \cite{Mendes:2007ve}.  The function $M$ is the chiral condensate
  (magnetization in the spin system), $h$ the quark mass (external
  magnetic field) and $\beta$ and $\delta$ are critical exponents. $t
  = 6/g^2 - 6/g^2_c$ plays the role of the reduced temperature.  }
\label{fig:stag_O4scal}
\end{figure}

\begin{figure}
\begin{center}
\includegraphics[width=0.6\textwidth]{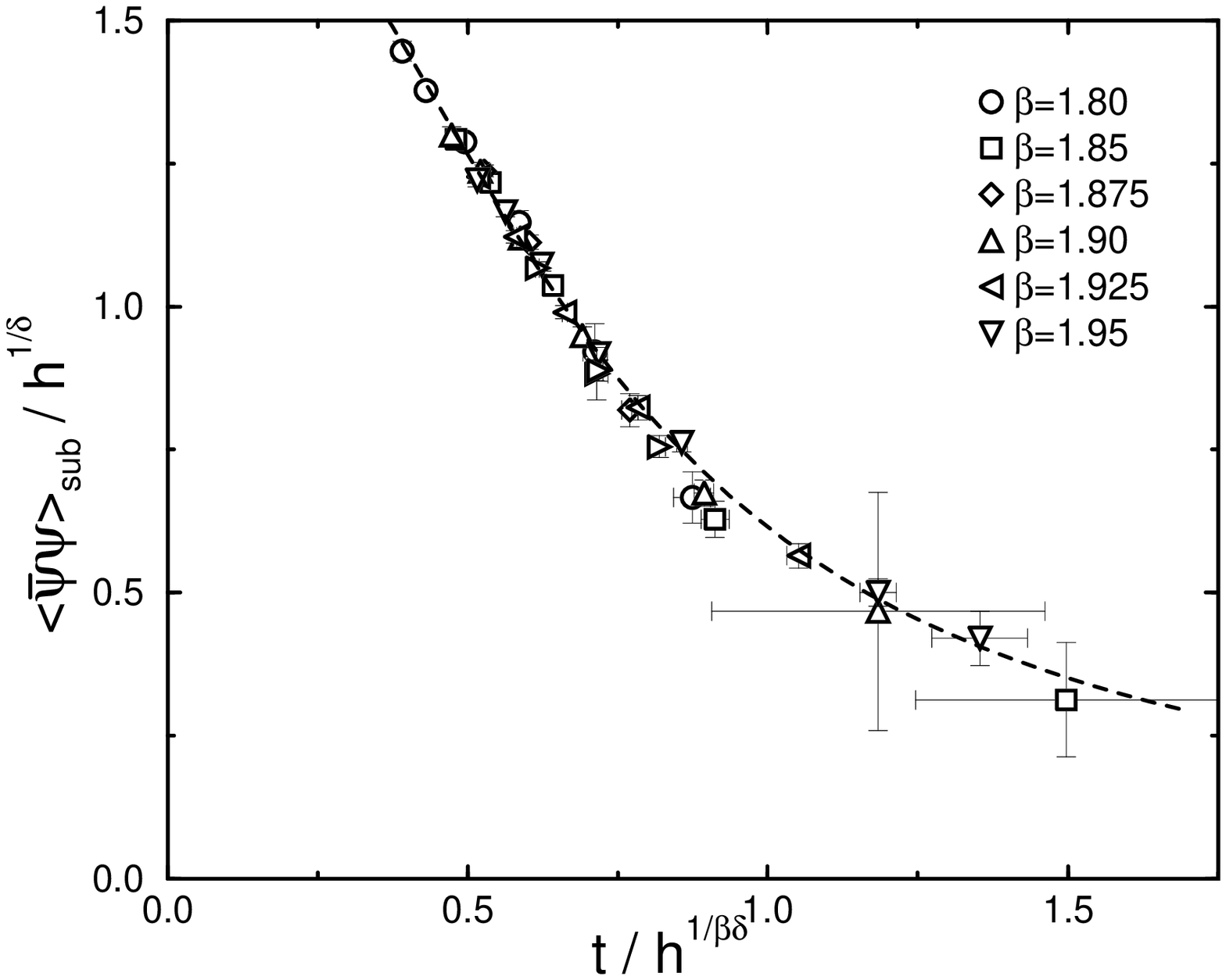}\\
\end{center}
\caption{$O(4)$ scaling, with linear scale, for two flavors of Wilson
fermions using $N_\tau=4$ lattices, from \cite{AliKhan:2000iz}.
}
\label{fig:Wils_O4scal}
\end{figure}

This question has been investigated by several groups with
somewhat contradictory results. Simulations with standard
staggered quarks using $N_\tau=4$ lattices, with large lattice
spacing in the transition region, and hence potentially large
lattice artifacts, as collected by T.~Mendez \cite{Mendes:2007ve},
show some deviations from $O(4)$ scaling as shown in
Fig.~\ref{fig:stag_O4scal}. For $O(4)$ scaling, all data points
should collapse to the curve in the figure.
Two-flavor clover-improved Wilson fermion simulations \cite{AliKhan:2000iz},
on the other hand, indicate good $O(4)$ scaling as seen in
Fig.~\ref{fig:Wils_O4scal}.

\begin{figure}
\begin{center}
\includegraphics[width=0.5\textwidth]{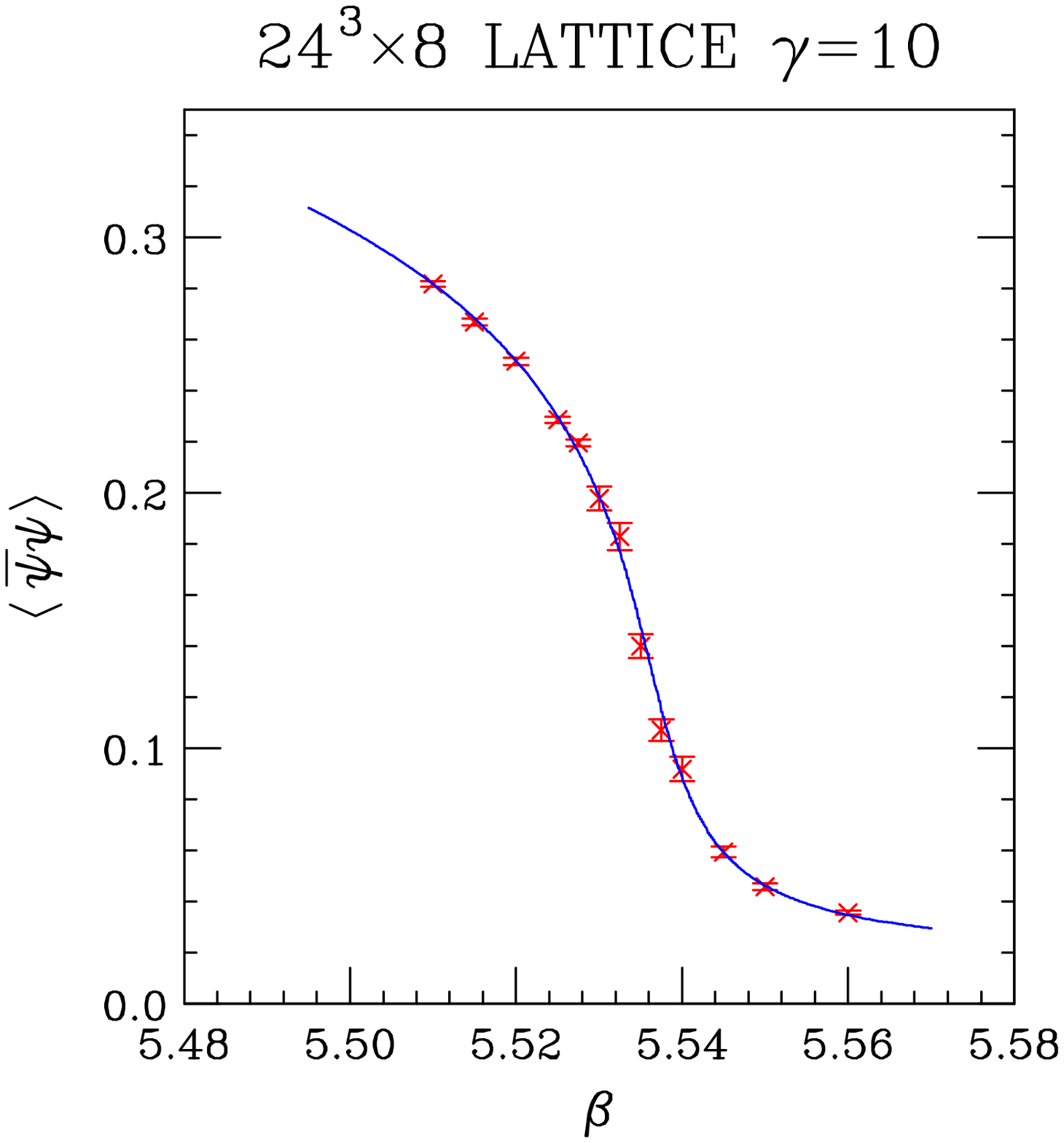}\\
\end{center}
\caption{O(2) scaling in a finite volume for two flavors of massless
staggered fermions with an irrelevant four-fermion interaction,
from \cite{Kogut:2006gt}. The curve comes from an $O(2)$ spin
model simulation with ``matched volume''.
}
\label{fig:O2_finV_scal}
\end{figure}

Since staggered fermions, at the large lattice spacings in the
transition region on high-temperature lattices with small $N_\tau$,
have quite large taste symmetry breaking, one might expect the
transition to be in the $U(1) \times U(1) \simeq O(2)$ universality
class, rather than the $O(4)$ one. More importantly, Kogut and
Sinclair \cite{Kogut:2006gt} argue that finite volume effects
on the fairly small (spatial) lattices used are quite large.
Indeed they found good agreement with $O(2)$ scaling, when taking
the finite volume effects into account as illustrated in
Fig.~\ref{fig:O2_finV_scal}

In contradiction with the theoretical expectations and the above
summarized numerical findings, D'Elia, Di Giacomo, and Pica found
indications of a first-order transition using an unimproved staggered
fermion action and $N_\tau = 4$ \cite{D'Elia:2005bv}. It is important
to check this conclusion with a more refined action. One should
conclude that at present the order of the high-temperature
transition with two massless flavors is still an open question.

\subsection{The phase transition with a physical strange quark}

Suppose, instead, we hold the strange quark mass at its physical value
and then decrease the $u$ and $d$ quark masses toward zero.  According
to the qualitative picture in the left panel of
Fig.~\ref{fig:mu_ms_phase_diag}, depending on where the tricritical
point lies, we could (1) encounter a critical point and enter a first
order regime, or (2) we may have to go to zero quark mass to find a
genuine phase transition.  In the right panel we reproduce a result
from de Forcrand and Philipsen suggesting the first alternative, but
their results were obtained with an unimproved action at $N_\tau = 4$
for which we expect large cutoff effects.

As we have mentioned, cutoff effects complicate the determination of
the phase boundary at small quark mass.  This is especially likely to
be true for simulations based on unimproved staggered fermions (even
improved staggered fermions are not entirely immune), since for them it
is important to take the continuum limit before taking the small quark
mass limit.  Otherwise, one risks being misled by lattice artifacts.

\section{QCD Phase Diagram at Nonzero Densities}
\label{sec:phase_diag_nonzero_dens}

\subsection{Phenomenology}

\begin{figure}
\begin{center}
\includegraphics[width=0.5\textwidth]{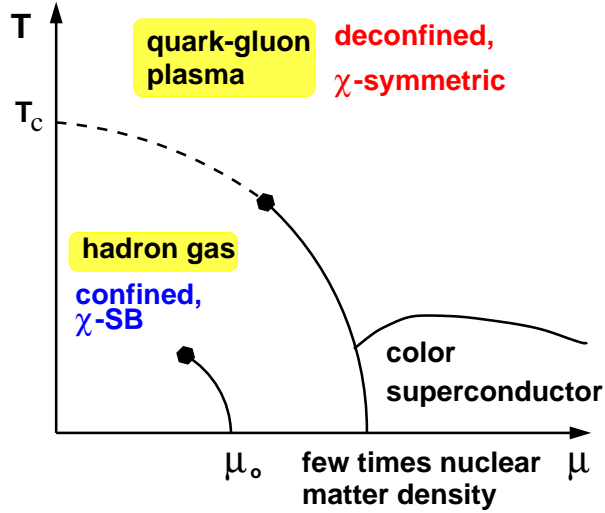}
\end{center}
\caption{Sketch of the expected phase diagram for 2+1 flavor hadronic
  matter as a function of temperature and chemical potential at
  physical quark masses from \cite{Karsch:2006xs}.  The confined,
  chiral symmetry broken phase lies in the lower left, separated from
  the deconfined, chirally symmetric phase by a pseudocritical
  crossover line (dashed) and first order (solid) line of phase
  transitions.  A critical point is indicated by a black hexagon.  A
  nuclear matter phase transition occurs along a line extending from
  $\mu = \mu_0$. At higher densities a color superconducting phase is
  proposed.}
\label{fig:dens_T_phase_diag}
\end{figure}

As the baryon number density is increased (\ie, all the flavor 
chemical potentials are increased from zero), 
according to traditional arguments,
there is a chiral-symmetry restoring phase transition along a line in
the $(\mu, T)$ plane when the $u$ and $d$ quark masses are zero, as
sketched in Fig.~\ref{fig:dens_T_phase_diag} \cite{Halasz:1998qr}.
This tradition is founded on two notions.  The first argues that
asymptotic freedom and consequently deconfinement should reign at very
high temperatures and high chemical potential.  The second argues that
spontaneous chiral symmetry breaking occurs at zero chemical potential
because, when fermions acquire a dynamical mass through symmetry
breaking, the negative energy levels of the Dirac sea are lowered,
lowering the vacuum energy.  With a nonzero chemical potential
the filled positive energy levels rise in energy, counteracting the
advantage of a dynamical chiral mass, and consequently inhibiting
spontaneous symmetry breaking \cite{Kogut:1983ia}.  

At zero $u$ and $d$ quark mass chiral symmetry is exact.  If chiral
symmetry is restored above a critical chemical potential and it is
spontaneously broken below, analyticity requires a phase transition.
There are no such guarantees, however, when quark masses are not zero.
Since we know from numerical simulation that at physical quark masses
there is only a crossover at zero density, the critical line
separating the chirally broken from the chirally restored phase must
move away from the temperature axis as the quark masses are increased.
It then terminates in a critical endpoint $(T_E,\mu_E)$. A crossover
line then fills the gap from there to the temperature axis, as
indicated by the dashed line in Fig.~\ref{fig:dens_T_phase_diag}.  A
key phenomenological question is whether the critical endpoint is
experimentally accessible.

At still higher densities exotic phases have been proposed, including
diquark condensates and color-flavor locked and superconducting phases
\cite{Alford:1998mk,Rapp:1997zu}.  These phases are, thus far,
completely beyond the reach of current lattice simulations.

\subsection{Lattice methods for nonzero densities}

To confirm or refute these traditional arguments requires numerical simulation.
Unfortunately, simulations at nonzero chemical potential are very
difficult, since standard lattice methodology requires that the
Feynman path integrand be treated as a positive probability measure.
In $SU(3)$ gauge theory, the integrand becomes complex at a nonzero
(real) chemical potential.  This creates a fermion ``sign problem''
analogous to the fermion sign problem in condensed matter physics in
strongly-coupled electron systems away from half-filling.  A solution
to either problem would be beneficial to the other.

To see why the problem arises, consider the naive fermion Dirac matrix
$M(U) = \gamma_\nu \nabla_\nu + m$.  The lattice version of the
gauge-covariant derivative $\nabla_\nu$ is given by Eq.~(\ref{eq:cov_deriv}).
The terms in $\nabla_\nu$ in the action allow the quark to hop to next neighbor
sites in the positive and negative $\nu$ direction.  Normally, hopping
in all directions must have equal weight to preserve the discrete
lattice symmetries of axis interchange, parity, time reversal, and
charge conjugation.  The fermion determinant is then real
because taking its complex conjugate corresponds to reversing the
direction of hopping, which has the same weight.  But a positive
nonzero chemical potential promotes quark hopping in the positive
(imaginary) time direction and suppresses it in the negative time
direction.  This is naturally implemented by changing the covariant
time derivative as follows:
\be
   \nabla_0 \psi(x) \to \frac{1}{2a}[U_0(x) e^{a\mu}\psi(x + \hat 0 a) - 
       e^{-a\mu} U_0^\dagger(x-\hat 0 a)\psi(x - \hat 0 a)].
\ee
If a quark hops along a worldline that wraps completely around the
lattice in the imaginary time direction, it accumulates $N_\tau$
factors of $\exp(a\mu)$, and the partition function receives a net
enhancement $\exp(a \mu N_\tau) = \exp(\mu/T)$, the appropriate
statistical weight for the addition of one quark to the grand
canonical ensemble.  A quark hopping backwards is interpreted as an
antiquark, and its contribution is correspondingly suppressed, as it
should be.  With this imbalance the determinant is no longer
guaranteed to be real.  Instead it acquires a complex phase $\phi
\propto \mu V$, \ie, roughly proportional to the lattice volume and
the chemical potential.

A complex determinant creates additional problems for staggered fermions.  With
$2+1$ flavors of staggered fermions at nonzero densities, one requires
the square root and fourth root of the fermion determinants.  When the
determinant is real, there is no phase ambiguity in the root.  But
when the determinant is complex, one has to choose the correct Riemann
sheet.  The ambiguities and an expensive remedy are discussed in
\cite{Golterman:2006rw}.  To be safe, one is limited to small $\mu$
and volumes.

Over the years a number of methods have been proposed for treating a
complex determinant.  We give a brief account of the attempts.  For
recent reviews, see \cite{Philipsen:2005mj,Ejiri:Lat2008}.

\subsubsection{Reweighting the fermion determinant}

As a standard lattice Monte Carlo method, reweighting involves
sampling the Feynman path integral according to one measure and then
making adjustments to achieve the effect of simulating with a
slightly different measure~\cite{Ferrenberg:1988yz,Ferrenberg:1989ui}.

Let us see how this idea is applied to a simulation at nonzero
chemical potentials $\mu_i$, one for each flavor $i$.  (To be precise,
we are speaking of a quark-number chemical potential.  The
baryon-number potential is three times as large ($\mu_{Bi} = 3\mu_i$).
The expectation value of an operator ${\cal O}$ is given by
\be
   \VEV{\cal O}_\mu = 
    \int [dU]\,{\cal O}(U)\exp[-S_G(U)]\prod_i \det[M_i(U,\mu_i)]/Z(\mu) ,
\ee
where $\mu = (\mu_1, \mu_2, \ldots{})$ and 
\be
    Z(\mu) = \int [dU]\,\exp[-S_G(U)]\prod_i \det[M_i(U,\mu_i)].
\ee
Since we can't do importance sampling with the unsuitably complex
determinant $\det[M(U,\mu)]$ in the measure, we can try to do it with
the real determinant $\det[M(U,\mu=0)]$.  That is, we write
\be
   \VEV{\cal O}_\mu = \VEV{{\cal O} R(U,\mu)}_0/\VEV{R(U,\mu)}_0 ,
\label{eq:reweight}
\ee
where $R(U,\mu)$ is the ratio of determinants that reweights the
contributions to the integrand to compensate for the incorrect
sampling measure:
\be
  R(U,\mu) = \det[M(U,\mu)]/\det[M(U,0)] .
\ee
Similarly, we can reweight to imitate a change in any of the
parameters of the action including the quark masses and gauge
coupling.  The reweighting factor $R$ is simply the ratio of the
intended and actual measures.

This procedure, often called the Glasgow method, is mathematically
correct but numerically unstable.  As the chemical potential moves
away from zero, one is no longer doing importance sampling.  In
complex analysis this approach is similar to attempting to estimate a
contour integral in the stationary phase approximation without going
through the saddle point.  The variance in the sampled values of the
numerator and denominator in Eq.~(\ref{eq:reweight}) grows
exponentially as the lattice volume increases, \ie, in the
thermodynamic limit.  The inevitable breakdown is forestalled by
keeping the shift in parameters small, so by working at small $\mu$.

A variant of this method uses the absolute value of the determinant
for the sample weighting.  The reweighting factor is then the phase
\cite{Toussaint:1989fn}.  This method has been applied only to small
lattice volumes.

Fodor and Katz propose reweighting simultaneously in the gauge
coupling $g^2$ and $\mu$~\cite{Fodor:2001au}.  They argue that one
achieves a better overlap with this method.  For example, one might
expect that if one moves along the crossover line in the $\mu-T$
plane, the important integration domain might not change as rapidly as
it would if one moves in some other direction.  To stay on this line
requires changing the gauge coupling along with the chemical
potential.  To locate the critical line, they follow Lee-Yang zeros of
the partition function.  (These zeros lie in the complex temperature
or complex gauge-coupling plane.  If there is a genuine phase
transition, as the lattice volume is increased, they impinge on the
real temperature axis and give rise to a singularity.  If there is
only a crossover, they stay harmlessly away from the real axis.) From
this method they estimate the critical endpoint at $T = 160(3.5)$ MeV
and $\mu_B = 3\mu = 360(40)$ MeV at physical quark masses using
conventional staggered fermions~\cite{Fodor:2004nz}.  This critical
chemical potential is nearly a factor of two smaller than an earlier
estimate at higher quark masses and smaller volumes
\cite{Fodor:2001pe}.  Such sensitivity to the simulation parameters
warrants further study.

\subsubsection{Approximating the determinant with phase quenching}

With degenerate up and down quarks, simulating with the
``phase-quenched'' or absolute value of the determinant and ignoring
the phase completely is equivalent to giving the up quark a positive
chemical potential and the down quark a negative chemical potential,
so it is equivalent to simulating with an isospin chemical potential
\cite{Kogut:2002zg}.  This procedure is numerically tractable, but to
draw conclusions regarding the phase diagram with the standard
chemical potential requires some justification.  Kogut and Sinclair
present the case in \cite{Kogut:2007mz}. See also
\cite{deForcrand:2007uz}.

\subsubsection{Simulating in the canonical ensemble}

Another approach is to simulate in the canonical ensemble of fixed
quark (baryon) number
\cite{Barbour:1997ej,Engels:1999tz,deForcrand:2006ec,Alexandru:2005ix}.
For simplicity, consider a single quark species.  The canonical ensemble
with quark number $q$ is then obtained from the Fourier transform
\be
    Z_q = \int_0^{2\pi}d\phi\,e^{-iq\phi}
    \int [dU]\,\exp[-S_G(U)]\det[M(U,\mu)]|_{\mu/T = i\phi}.
\ee
The sign problem arises in the Fourier transform.  As the quark number
is increased for a given lattice volume and configuration, the Fourier
component decreases rapidly and the sensitivity to oscillations
worsens, so that any discrete approximation to the Fourier transform
develops a severely large variance.  

Meng \etal\ have recently proposed a new ``winding number expansion''
method that starts from the Fourier transform of the logarithm of the
determinant, $\log(\det[M(U,\mu)])=\Tr\log[M(U,\mu)]$ and proceeds via
a Taylor expansion to generate the canonical partition function
\cite{Meng:2008hj,Li:2008fm}.  The method converges much better, but
so far results are reported only for fairly large quark masses.

\subsubsection{Simulating with an imaginary chemical potential}

\begin{figure}
\begin{center}
\includegraphics[angle=270,width=0.5\textwidth]{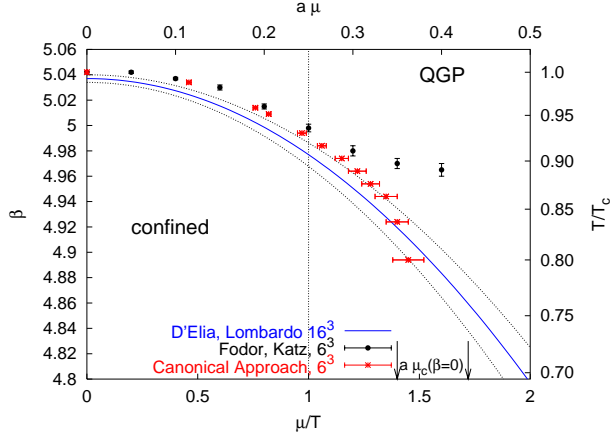}
\end{center}
\caption{From \protect\cite{deForcrand:2006ec}.  Critical line
  as a function of quark chemical potential and temperature for four
  degenerate flavors of unimproved staggered fermions at $N_\tau = 4$,
  bare quark mass $am = 0.05$, and for the spatial lattice volumes
  shown.  Results from three methods are compared:  the imaginary
  chemical potential approach of \protect\cite{D'Elia:2002gd}, the
  canonical ensemble approach of \protect\cite{deForcrand:2006ec}, and
  the multiparameter reweighting approach of \cite{Fodor:2001au}.  A
  range of strong coupling values of the critical chemical potential
  $\mu_c(\beta = 0) $ is also indicated.}
\label{fig:compare_crit_line}
\end{figure}

If we make the chemical potential purely imaginary, the fermion
determinant becomes real, and a direct simulation is
possible~\cite{deForcrand:2002ci}.  To recover results at a physical,
real chemical potential, we must do an analytic continuation.  The
success of such a continuation depends on knowing the analytic form of
the observable as a function of chemical potential.  We do if the
chemical potential is small enough that a Taylor expansion is
plausible.  So in the end, the imaginary potential method provides
essentially the same information as an explicit Taylor expansion about
zero chemical potential. Figure~\ref{fig:compare_crit_line} from de
Forcrand and Kratochvila \cite{deForcrand:2006ec} compares three
methods for determining the critical line.  Each result shown is based
on the same unimproved $N_f = 4$ staggered fermion action.  The
methods agree reasonably well for $\mu/T < 1$. Note that this is a
four-flavor simulation with a first-order phase transition, unlike the
2+1-flavor case of Fig.~\ref{fig:dens_T_phase_diag}.

\subsubsection{Taylor expansion method}

For small chemical potential, we may carry out a Taylor expansion of
the required observables in terms of the flavor chemical potentials at
zero chemical potential \cite{Allton:2002zi,Gavai:2003mf}.  Since all Taylor
coefficients are evaluated at zero chemical potential, determining
them is straightforward.  However, the observables that give the
coefficients are nontrivial.  They involve products of various traces
of the inverse fermion matrix.  The traces are usually evaluated using
stochastic methods.  Furthermore, as the order of the expansion grows,
the number of required terms grows factorially.  Thus it is rare
to find calculations as high as eighth order \cite{Gavai:2004sd,Gavai:2008zr}.

\subsubsection{Probability distribution function method}

The ``probability-distribution-function'' or ``density-of-states''
method is new and promising~\cite{Fodor:2007vv, Ejiri:2007ga}.  It is
related to the reweighting method.  A recent variant by Ejiri combines
reweighting with a Taylor expansion.  To explain the method we start
with a simple case, defining the ``density of states'' or
``probability distribution function'' of the plaquette $P$ with the
Wilson gauge action and arbitrary fermion action:
\be
  w(P^\prime) = \int [dU]\, \delta(P^\prime - P(U))\det[M(U,0)] \exp[-S_G(U)],
\label{eq:dist_func}
\ee
where $\delta(P^\prime - P)$ is the Dirac delta function.  It is
defined like the partition function, but at a fixed value of the
plaquette.  The expectation value for an observable ${\cal O}(P)$
that depends only on $P$ is then
\be
  \VEV{\cal O} = \int dP^\prime\, w(P^\prime) {\cal O}(P^\prime)/
      \int dP^\prime\, w(P^\prime) .
\ee
At nonzero $\mu$ we use reweighting to calculate the partition
function:
\be
  Z(\mu) = \int dP\, R(P,\mu) w(P) ,
\ee
where the plaquette-restricted reweighting function $R(P,\mu)$ is
\be
  R(P,\mu) = \frac{\int [dU]\, \delta(P^\prime - P(U))\det[M(U,\mu)]}
                  {\int [dU]\, \delta(P^\prime - P(U))\det[M(U,0)]},
\ee
\ie, the ratio at nonzero and zero $\mu$.  For the Wilson
action, the gauge weight $\exp[-S_G(U)]$ depends only on $P$, so it
cancels between numerator and denominator in $R(P,\mu)$.  The
distribution function $w(P)$ is still calculated at $\mu = 0$
according to (\ref{eq:dist_func}).  

The sign problem appears in the numeric evaluation of $R(P,\mu)$.
Ejiri offers a way to overcome it~\cite{Ejiri:2007ga}.  His method
begins with a generalization of the distribution function, making it
depend on three variables: the plaquette $P$, the magnitude of the
ratio of determinants $F(\mu) = \det M(\mu)/\det M(0)$, and the phase
$\theta \equiv \Im\log\det M(\mu)$:
\be
  w(P^\prime, |F^\prime|, \theta^\prime)  =
    \int [dU]\, \delta(P^\prime - P(U)) \delta(|F^\prime| - |F|) 
               \delta(\theta^\prime - \theta) 
                \det[M(U,0)] \exp[-S_G(U)],
\ee
Note that the real, positive weight factor in the integrand comes from
the $\mu = 0$ action.  For any value of $\mu$ the partition function
is then
\be
    Z(\mu) = \int dP d|F| d\theta\, F(\mu) w(P,|F|,\theta) ,
\ee
where in place of the reweighting function $R$ we now have simply
$F(\mu)$ itself.

The next step relies on the key assumption that the distribution
function $w(P^\prime, |F^\prime|, \theta^\prime)$ is Gaussian in
$\theta$.  Ejiri argues that this is plausible, at least for large
volume.  A further assumption for rooted staggered fermions is that
the effect on the phase of taking the fourth root is simply to replace
$\theta$ by $\theta/4$ in the Gaussian distribution.  With these
assumptions one can do the $\theta$ integration directly, eliminating
the sign problem.  The result depends only on the width of the
Gaussian, which must be determined numerically.  Finally, to make the
calculation of the ratio of determinants tractable, Ejiri expands
$\log[\det M(\mu)]$ in a Taylor series in $\mu$ about $\mu = 0$.
The same Taylor coefficients appear in an intermediate step in the
Taylor expansion of the pressure or thermodynamic potential.  Since
one is expanding the action instead of the thermodynamic potential,
the convergence properties are different --- possibly more favorable.

Applying this method to p4fat3 staggered fermions with the Wilson
gauge action, a rather coarse lattice with $N_\tau = 4$, and a rather
large quark mass, Ejiri locates the critical chemical potential at
$\mu/T > 2.5$, approximately.  This is an interesting result, which
awaits reconciliation with the questions raised by Golterman, Shamir,
and Svetitsky concerning phase ambiguities of the fourth root of the
staggered fermion determinant~\cite{Golterman:2006rw}.

Thus we see that all of the methods, save, perhaps, the probability
distribution function method, are limited to quite small chemical
potentials.

\subsubsection{Stochastic quantization method}

All of the above lattice methods for simulating at nonzero chemical
potential evaluate the Feynman path integral using Monte-Carlo
importance sampling, a technique that is inherently unstable when the
path integrand is not positive definite.  At nonzero chemical
potential, the $SU(3)$ fermion determinant is complex, and the wide
variety of methods outlined above deal with the complex phase with
limited success.  Instead of quantizing via the Feynman path integral
method, Aarts and Stamatescu \cite{Aarts:2008rr} have recently
proposed using the stochastic quantization method
\cite{Parisi:1980ys}.  In the early days of lattice calculations,
stochastic quantization through the Langevin equation
\cite{Parisi:1984cs} was, in fact, one of the competing numerical
methods for nonperturbative calculations in quantum field theory, and
it met with mixed success \cite{Damgaard:1987rr}.

For purposes of this review, we give just a brief sketch of stochastic
quantization.  For a theory with a scalar field $\phi(x)$ and action
$S$, we generate an ensemble of fields $\phi(x,\tau)$ where $\tau$ is
a fictitious Langevin time (analogous to molecular-dynamics or
Markov-chain time in the standard importance sampling approach.)  The
ensemble satisfies the stochastic equation
\be
   \frac{\partial \phi(x,\tau)}{\partial\tau} = 
   -\frac{\delta S}{\delta \phi(x)} + \eta(x,\tau)
\ee
where $\eta(x,\tau)$ is a Gaussian random field (source), uncorrelated
in $x$.  As long as $S$ has a well-defined minimum and we start with a
solution near that minimum, without the random source the field
relaxes to the classical solution where the action is stationary, \ie,
the variational derivative $\delta S/\delta \phi(x)$ vanishes.  The
random source then induces ``quantum fluctuations'' about the
classical solution.  Quantum observables are estimated in the usual
way as expectation values on the equilibrium ensemble.

When the action $S$ is complex, we get a complex solution and a
complex stationary point, a region that is not reached with
conventional importance sampling.  The hope is that the solution is
still attracted to the appropriate stationary point, \ie, the Langevin
method is stable.  Aarts and Stamatescu have done some preliminary
tests with simplified models that imitate the characteristics of QCD
at nonzero chemical potential.  Their results are promising
\cite{Aarts:2008wh,Aarts:2009hn}.

\subsection{Curvature of the critical surface}

\begin{figure}[thb]
\begin{tabular}{cc}
\begin{minipage}[b]{80mm}
\includegraphics[width=80mm]{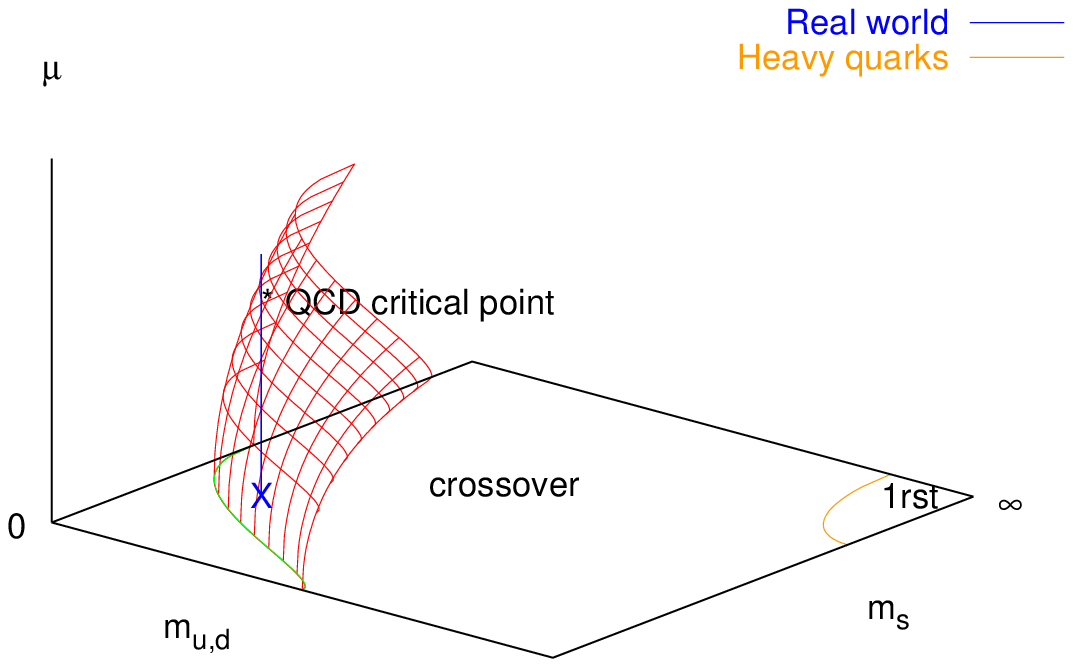} \\
\end{minipage}
&
\begin{minipage}[b]{80mm}
\includegraphics[width=80mm]{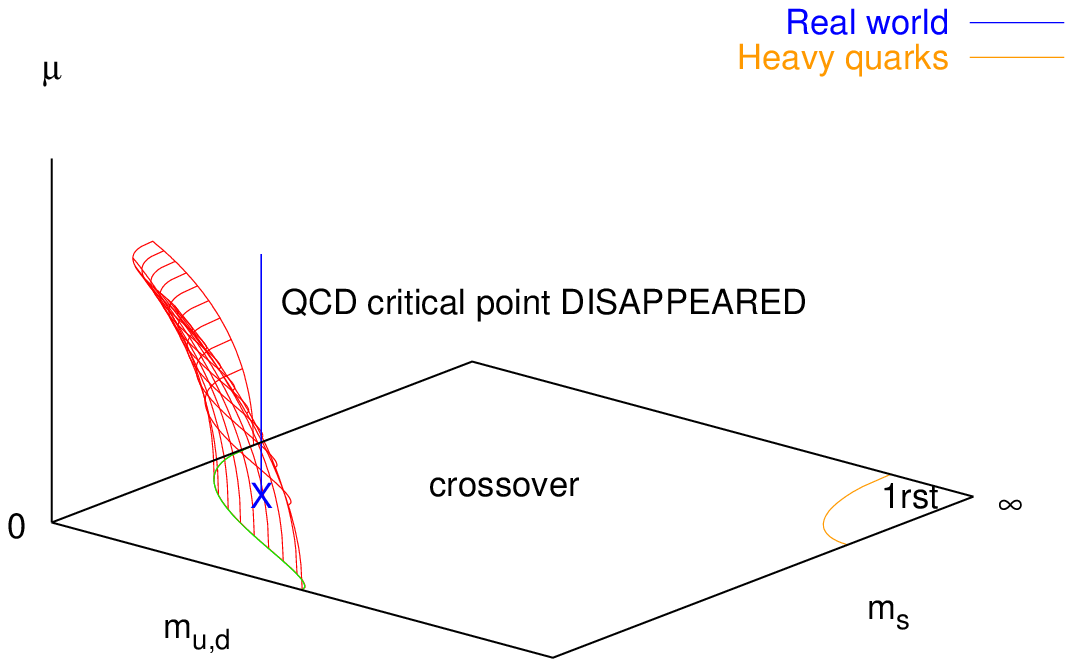} \\
\end{minipage}
\end{tabular}
\caption{Two possible alignments of the chiral critical surface at low
  chemical potential from \protect\cite{deForcrand:2006pv}. Left: the scenario
  permitting a first order phase transition at high densities and
  temperatures.  Right: the scenario allowing only a crossover.}
\label{fig:3dphasediag}
\end{figure}

One question of considerable phenomenological importance can be
addressed with simulations at small chemical potential.  That is
whether the $Z(2)$ critical line sketched in
Fig~\ref{fig:mu_ms_phase_diag} moves closer to the physical quark
masses as the chemical potential is increased or it moves farther
away.  If it moves closer, as shown in the left panel of
Fig.~\ref{fig:3dphasediag}, one may expect a true phase transition in
a suitably baryon-rich environment, such as may occur in a moderately
low-energy heavy-ion collision.  If it moves away, as shown in the
right panel, there would be no such expectation.  De Forcrand and
Philipsen set out to address this question using the imaginary
chemical potential method.  Their results at $N_\tau = 4$ 
suggest that the critical line moves away
\cite{deForcrand:2006pv,deForcrand:2007rq,deForcrand:2008zi,deForcrand:2008vr},
at least
when all three quark flavors are close to having equal masses.

\section{Equation of State}
\label{sec:eos}

The equation of state gives the energy density, pressure, and/or
entropy of the thermal QCD ensemble as a function of temperature at
constant volume.  All quantities are renormalized by subtracting their
values at zero temperature.  The subtraction eliminates an ultraviolet
divergence, but the cancellation of this divergence makes the
computation costly in the continuum limit, since one must compute the
${\cal O}(a^{-4})$ divergent high-temperature and zero-temperature
quantities independently and subtract them to get a finite result.

There are two traditional methods for computing the equation of state
and one recently introduced method.  

\subsection{Derivative method}

The first method is based on the identity
\be
  \varepsilon = \frac{T^2}{V}\left.\frac{\partial \ln Z}{\partial T}\right|_V .
\ee
On the lattice the derivative with respect to temperature at fixed
volume in the first identity translates to a derivative with respect
to $1/(N_\tau a_t)$ at fixed $a_s$, where $a_t$ is the lattice spacing
in the imaginary time direction and $a_s$ is the lattice spacing in
the spatial direction.  At fixed $N_\tau$, we differentiate with
respect to $a_t$ itself.

For example, for the original Wilson plaquette gauge action of
Eq.~(\ref{eq:Wilson_1plaq}) the explicit dependence on $a_t$ and $a_s$ goes
as follows:
\be
   S_G(a_s,a_t,g^2) = 2/g^2(a_s,a_t) \left[
     \frac{a_s}{a_t}\sum_{x} P_t(x) +
     \frac{a_t}{a_s}\sum_{x} P_s(x) \right],
\ee
where we have distinguished the timelike and spacelike plaquettes
\bea
  P_t(x) &=& \sum_i \Re \Tr [1 - U_{P,i,0}(x)] , \\
  P_s(x) &=& \sum_{i<j} \Re \Tr [1 - U_{P,i,j}(x)].
\eea

In the gauge action above, we have indicated the dependence of the
gauge coupling on the lattice constants $a_s$ and $a_t$.  That
dependence is defined through a standard renormalization procedure for
an anisotropic lattice: at a fixed ratio $a_t/a_s$ and gauge coupling
$g$, we compute an experimentally accessible, dimensionful quantity,
such as the splitting of a quarkonium system.  From the experimental
value of the splitting, we can then determine the lattice constants in
physical units.  We repeat the procedure, varying $g$ and $a_t/a_s$ to
get the full dependence of $g$ on the lattice constants.

So from Eq.~(\ref{eq:partitionQCD}) with only the gauge action in this
example, we have \cite{Engels:1990vr} (after setting $a_t = a_s = a$)
\be
  \varepsilon = 
  -T \left.\frac{\partial \ln g^2}{\partial \ln a_t}\right|_{a_s}\VEV{S_G/V} 
  + (6/g^2) T \VEV{P_t - P_s}.
\ee
The partial derivative of the gauge coupling with respect to $a_t$ is
called the Karsch coefficient. It is known up to 1-loop order in lattice
perturbation theory, but a nonperturbative calculation described above
is necessary at experimentally accessible temperatures.  As we
indicated above, that calculation is rather involved.

\subsection{Standard integral method}
\label{sec:integral_method}

A second thermodynamic identity gives the pressure as the volume
derivative of the thermodynamic potential,
\be
  p = T \left.\frac{\partial}{\partial V}\ln Z\right|_T.
\ee
By itself, this identity leads to an expression similar to the energy
density above, but in this case we need the derivative of the gauge
coupling with respect to the spatial lattice spacing $a_s$ at fixed
$a_t$.  We have the same difficulty as before in requiring a
nonperturbative calculation of an unconventional quantity.  

But if we combine the two identities to form the interaction measure
$I$,
\be
  I = \varepsilon - 3p,
\ee
then we get a total derivative of the gauge coupling with respect to
$a = a_s = a_t$ and the lattice thermodynamic identity
\be
  I = -\frac{T}{V} \frac{d \ln Z}{d \ln a}.
\label{eq:int_meas}
\ee
The isotropic derivative of the coupling with respect to the cutoff is
just the commonly computed renormalization group beta function $\beta
= d g^2/d\ln a$.  For the Wilson plaquette gauge action we get
\be
  I = - T/V (d \ln g^2/d\ln a) \VEV{S_G} .
\ee
So the lattice derivative is readily calculated in terms of the
conventional plaquette observable and the beta function.  With
fermions present we require also the chiral condensate and the
derivative of the quark masses with respect to the lattice spacing.
These are also easily accessible in lattice calculations.

We must bear in mind that the physical quantities require subtracting
the zero-temperature values, so in the end we need the difference
\be
  \Delta I = I(T) - I(0).
\ee
We will often drop the $\Delta$ in the following discussion and
figures.

\begin{figure}[t]
  \begin{tabular}{ccc}
   \hspace*{-18mm}
  \includegraphics[width=0.45\textwidth]{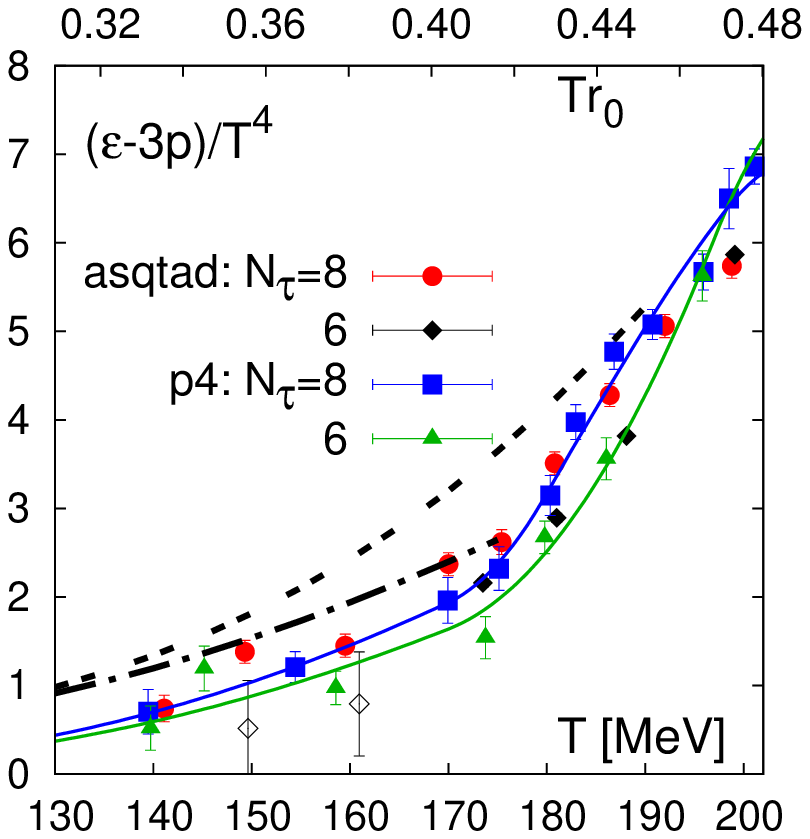}
 &
  \hspace*{-27mm}
  \includegraphics[width=0.45\textwidth]{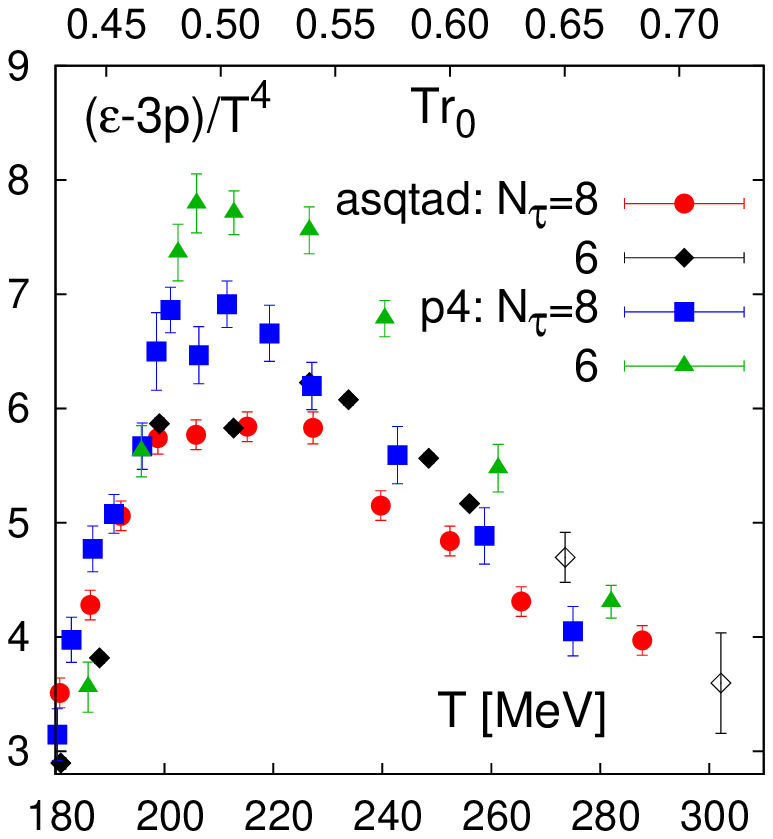}
 &
  \hspace*{-18mm}
  \includegraphics[width=0.45\textwidth]{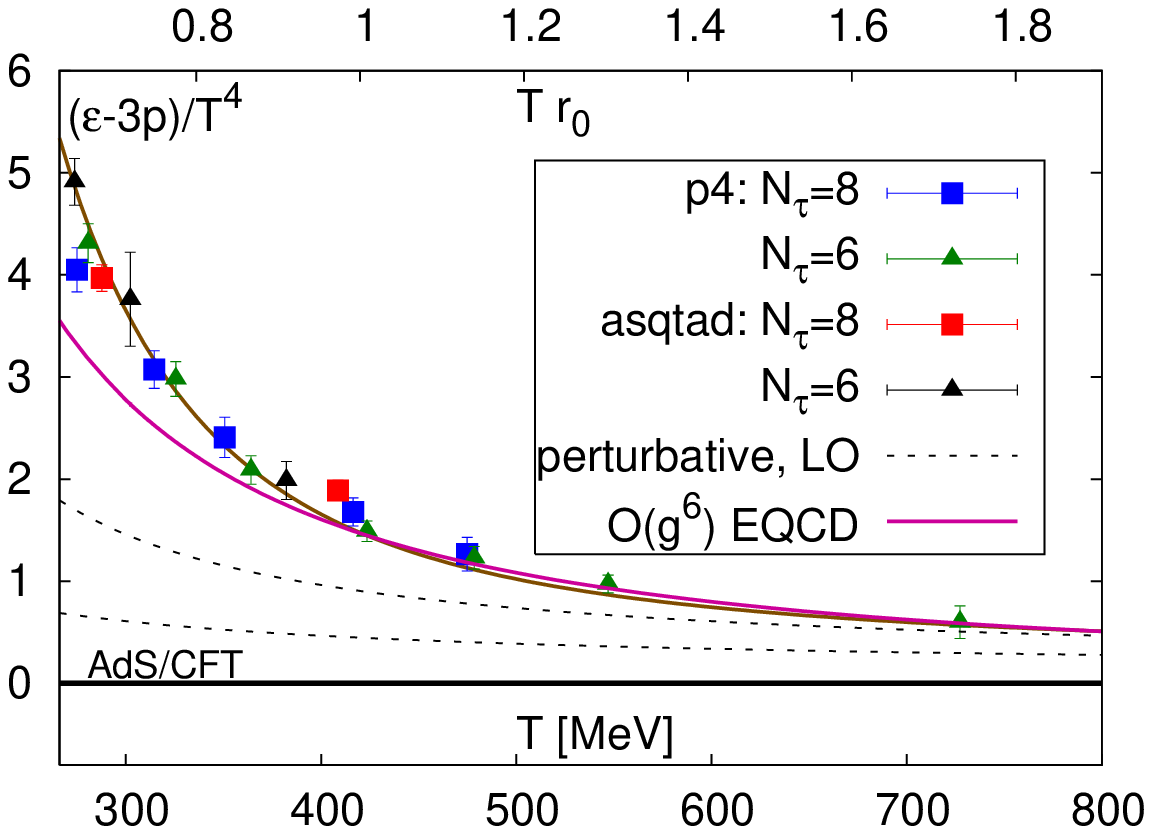} \\
 \end{tabular}
\caption{Details of the dependence of the interaction measure on
  temperature in MeV units (bottom scale) and $r_0$ units (top scale)
  for three temperature ranges left to right: low, middle, and high,
  for $N_\tau = 6$ and $8$ from a HotQCD study comparing p4fat3 and
  asqtad staggered fermion formulations
  \protect\cite{Bazavov:2009zn,Petreczky:QM09}. Measurements in most
  cases are taken along a line of constant physics with $m_{ud} = 0.1
  m_s$. In the low temperature range the dashed and dash-dotted curves
  are predictions of a hadron resonance gas model with different high
  mass cutoffs.  The other curves in that range are spline fits to the
  data.  In the high temperature range the dashed lines are the
  leading order perturbative prediction for $\mu_{\overline {MS}} =
  2\pi T$ and $\mu_{\overline {MS}} = \pi T$. The brown line (the line
  passing through the points) is a fit to leading order perturbation
  theory plus a bag constant, and the magenta line (the line passing
  mostly below the points) is an ${\cal O}(g^6)$ EQCD prediction from
  \protect\cite{Laine:2006cp}.  For a brief mention of EQCD, see
  Sec.~\protect\ref{subsec:spatial_tension}. }
\label{fig:e-3p}
\end{figure}

Figure~\ref{fig:e-3p} shows the interaction measure difference
obtained in a recent $N_\tau = 8$ calculation with equal-mass up and
down quarks and a strange quark.  The mass of the strange quark was
held fixed at approximately its physical value, and the masses of the
up and down quarks were set to a fixed fraction of the strange quark
mass.  Thus the temperature was varied roughly along parameter-space
lines of constant physics, meaning light pseudoscalar mesons (at zero
temperature) had approximately constant masses.

To complete the determination of the equation of state, we need the
energy density and pressure separately.  The pressure is easily
computed in the thermodynamic limit, in which $\ln Z$ is simply
proportional to the volume:
\be
  \ln Z = -pV/T,
\label{eq:thermolimit}
\ee
So the expression (\ref{eq:int_meas}) can also be written as
\be
  I = \frac{T}{V} \frac{d(pV/T)}{d \ln a},
\label{eq:int_meas1}
\ee
or, if we fix $VT^3$ in the derivative, as
\be
  I/T^4 = \frac{d(p/T^4)}{d \ln T} \ .
\label{eq:int_meas2}
\ee
We can then use the identity (\ref{eq:int_meas1}) at fixed $N_\tau$ to
integrate with respect to $\ln a$ (equivalently $\ln T$) to get the
pressure:
\be
   p(a)a^4  - p(a_0) a_0^4 = 
    -\int_{\ln a_0}^{\ln a} \Delta I(a^\prime)(a^\prime)^4 \, d\ln a^\prime.
\label{eq:int_method}
\ee
Here the lower endpoint of integration $a_0$ is a large lattice
spacing, corresponding to a low temperature.  If it is sufficiently
low, we may take $p(a_0) = 0$ and the expression then yields the
pressure at temperature $T = 1/(N_\tau a)$.

The integration is carried out numerically, since the integrand is
determined in a series of simulations done at fixed lattice spacing.
However, the spacing of the points can be set arbitrarily close as
needed. The energy density is then obtained from $\varepsilon = I +
3p$ and the entropy density from $s = \varepsilon + p$.

\begin{figure}
    \hspace*{-35mm}
    \includegraphics[width=0.6\textwidth]{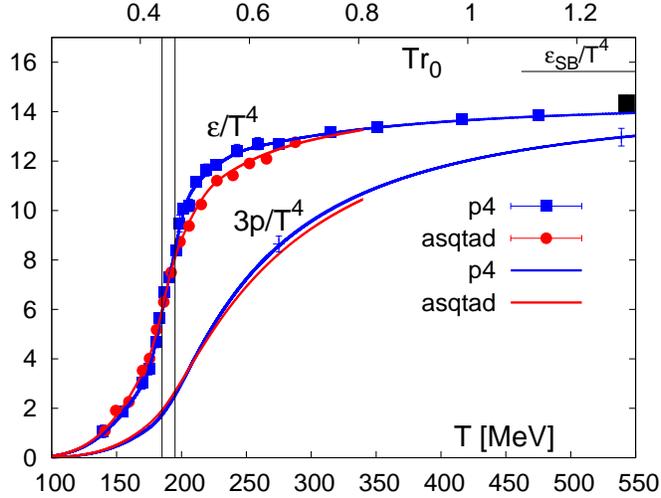}\\
\caption{Equation of state showing energy density and three times the
  pressure, both divided by the fourth power of the temperature \vs
  temperature for $N_\tau = 8$. Measurements are taken along a line of
  constant physics with $m_{ud} = 0.1 m_s$. Results
  are from a HotQCD study comparing p4fat3 and asqtad staggered
  fermion formulations \protect\cite{Bazavov:2009zn}.  The
  blue error bars on the pressure curve indicate the size of the
  error.  The black bar shows a systematic error from setting the
  lower limit of the pressure integration. }
\label{fig:e+3p}
\end{figure}

\begin{figure}
\begin{center}
\includegraphics[width=0.6\textwidth]{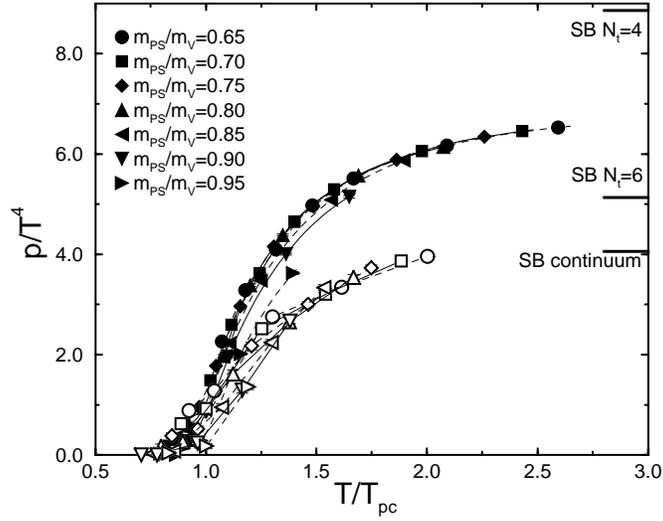}\\
\end{center}
\caption{The pressure as function of $T/T_{pc}$, with $T_{pc}$ the
pseudocritical or crossover temperature for two-flavors of clover
improved Wilson fermions on $16^3 \times 4$ lattices (filled symbols)
and $16^3 \times 6$ lattices (open symbols), from \cite{AliKhan:2001ek}.
The simulation was done for a variety of rather heavy quark masses,
indicated by the vector to pseudoscalar mass ratios $m_{PS}/m_V$.
The lattice artifacts are larger than with improved staggered quarks,
as expected from Table~\ref{tab:SB}.
}
\label{fig:Wils_EoS}
\end{figure}

This integral method was used to complete the construction of the
equation of state with improved staggered quarks shown in
Fig.~\ref{fig:e+3p}. The same method has also been used in a study
with two flavors of clover improved Wilson fermions \cite{AliKhan:2001ek},
as shown in Fig.~\ref{fig:Wils_EoS}.

\subsection{Temperature integral method}

In the standard integral method above we fixed $N_\tau$ and integrated
Eq.~(\ref{eq:int_meas}) with respect to lattice spacing to get the
pressure.  The temperature integral method of \cite{Umeda:2008bd}
instead fixes the lattice spacing and ``integrates''
Eq.~(\ref{eq:int_meas2}) over $N_\tau$ at fixed $N_s$.

The advantage of working at a fixed lattice spacing (so fixed gauge
coupling, quark masses, and Hamiltonian) is that the zero temperature
subtraction is the same for all $N_\tau$, and we are assured of
following lines of constant physics \cite{Levkova:2006gn}.  With the
standard integral method, to carry out the necessary subtraction, we
need a separate zero temperature simulation for each high temperature
point.  Thus one may hope for a savings in computational effort.

The disadvantage of the temperature integral method is that the
integrand is known only at the discrete temperatures $1/(N_\tau a_t)$
for integer $N_\tau$.  To decrease the sample interval at a given
temperature, one must start with a smaller $a_t$, which increases the
cost substantially.  Simulating on an anisotropic lattice helps.

\begin{figure}
\begin{center}
\includegraphics[width=0.5\textwidth]{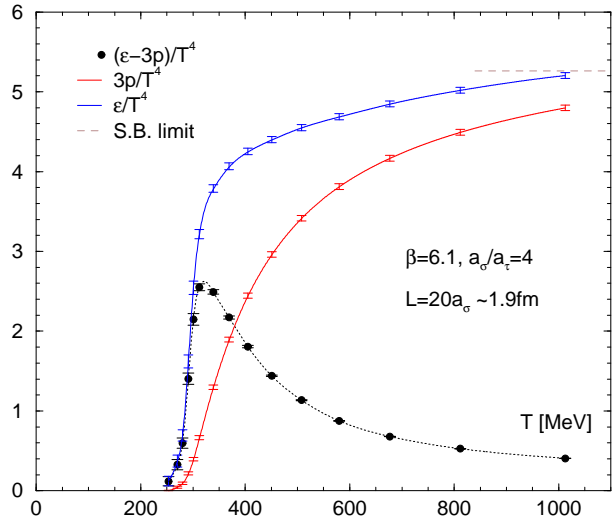} \\
\end{center}
\caption{Equation of state (interaction measure, energy density and
  pressure) for pure Yang-Mills theory, obtained using the $T$
  integral method at fixed lattice spacing $a_\sigma = 0.097$ fm and
  aspect ratio $a_\sigma/a_\tau = 4$
  \cite{Umeda:Lat2008,Umeda:2008bd}. }
\label{fig:tumeda-fig1}
\end{figure}

So far, the method has been tested on a pure Yang-Mills ensemble with
the pleasing result shown in Fig.~\ref{fig:tumeda-fig1}.

\subsection{Step scaling method}

The standard integral method of Eq.~(\ref{eq:int_method}) has the
disadvantage that it requires computing the difference between the
high and zero temperature values of the interaction measure at each
value of the gauge coupling (\ie, each high temperature point).  At
increasingly high temperature we get closer to the continuum limit and
the matching zero temperature calculation becomes very expensive.
Endr\"odi \etal\ propose a step scaling method that alleviates this
problem to some degree \cite{Endrodi:2007tq}.  Their idea is to
compute the pressure at a given temperature as a series of
differences:
\be
  p(T) - p(0) = [p(T) - p(T/2)] + [p(T/2) - p(T/4)] + \ldots{} ~.
\ee
The increment
\be
  \bar p(T) = p(T) - p(T/2)
\ee
must be calculated at the same cutoff $a$ to renormalize properly the
ultraviolet divergence.  In practice, this means matching a
calculation at a given $N_\tau = N$ with a calculation at $N_\tau =
2N$ for the same bare action parameters. (The step factor 1/2 can be
replaced by any factor less than 1.)  The differences
$[p(T)-p(T/2)]/T^4$ are bounded from above, so the series
\be
  [p(T) - p(0)]/T^4 = \bar p(T)/T^4 + \frac{1}{16}\bar p(T)/T^4|_{T/2} 
   + \frac{1}{256}\bar p(T)/T^4|_{T/4} + \ldots{}
\ee
converges rapidly.  

Endr\"odi \etal\ suggest two ways to calculate $\bar p(T)$.  One uses
a modified form of Eq.~(\ref{eq:int_method}).
\be
   p(a,N_\tau=N)a^4  - p(a,N_\tau=2N) a^4 = 
    -\int_{\ln a_0}^{\ln a} 
   [I(a^\prime,N_\tau=N)-I(a^\prime,N_\tau=2N)](a^\prime)^4 \, d\ln a^\prime.
\ee
Here, we have shown the $N_\tau$ dependence explicitly.  We assume that
$a_0$ is large enough that the integration constants $p(a_0,N_\tau)$
are essentially zero.

\begin{figure}
\begin{center}
\includegraphics[width=0.5\textwidth]{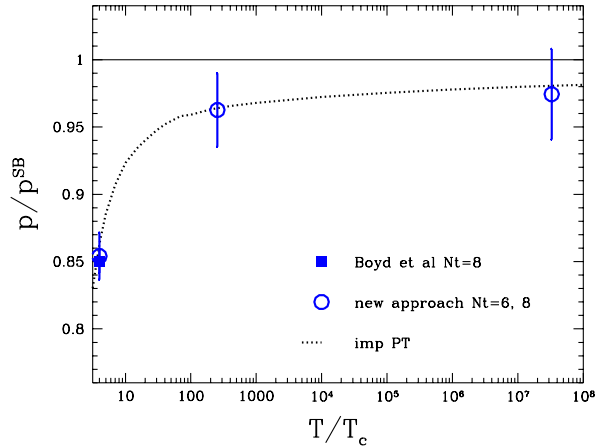} \\
\end{center}
\caption{Circles: pressure from \protect\cite{Endrodi:2007tq} for pure
  Yang-Mills theory at ultra-high temperatures compared with
  predictions of EQCD perturbation theory (dotted line
  \protect\cite{Braaten:1995jr,Kajantie:2002wa,Laine:2006cp}). The
  pressure is given in units of the Stefan-Boltzmann value and the
  temperature in units of the temperature at the phase transition
  $T_c$.  The square is computed using the standard integral method
  \protect\cite{Boyd:1996bx}. }
\label{fig:step_method}
\end{figure}

The second method uses the identity Eq.~(\ref{eq:thermolimit}) to
write
\be
   \bar p(T=1/(aN)) = p(a,N_\tau=N)  - p(a,N_\tau=2N) = 
   [\ln Z(N_\tau = 2N) - \ln Z^2(N_\tau = N)]/(N_s^3N) .
\label{eq:step_increment}
\ee
The rhs is the difference between the partition functions on two
lattices of size $N_s^3 \times 2N$ in which one lattice is intact and the
other is split in half at the midpoint in imaginary time with periodic
(or fermion-antiperiodic) boundary conditions applied to the two
halves.  To compute this difference, Endr\"odi \etal\ modify the
action at the interface by introducing an interpolating parameter
$\alpha$ such that $\alpha = 1$ corresponds to the fully split lattice
and $\alpha = 0$, to the fully intact lattice.  The simulation
measures the derivative of $\ln Z(\alpha)$ with respect to $\alpha$,
which involves only fields at the interface. The increment
(\ref{eq:step_increment}) is then computed from
\be
  \bar p(T) = \frac{1}{N_s^3N}\int_0^1 d\alpha\, 
  \frac{d\ln Z(\alpha)}{d\alpha} .
\ee
There is still a strong cancellation involved in the integration over
$\alpha$, but it is a bit milder than the cancellation in the standard
integral method.  With their method they are able to reach such high
temperatures that contact with perturbation theory is certainly
expected, as shown in Fig.~\ref{fig:step_method}.
For a lower temperature comparison of the $O(g^6)$ EQCD prediction of
Laine \etal\ \cite{Laine:2006cp} with the interaction measure computed
using standard methods, see Fig.~\ref{fig:e-3p}.  For a brief mention
of EQCD, see Sec.~\protect\ref{subsec:spatial_tension}.

\subsection{Equation of state at nonzero densities}
\label{subsec:eos_nzmu}

Heavy ion collisions involve interacting hadronic matter at relatively
low baryon densities and high temperatures.  At the other extreme,
high baryon densities and low temperatures may occur in the cores of
dense stars.  In both cases we would like to know the equation of
state.  For the low density environment of heavy ion collisions the
Taylor series method is effective for lattice simulations.
Unfortunately, thus far we have no reliable lattice method to simulate
the conditions of dense stars.

Consider the $2+1$ flavor case of equal nonzero up- and down-quark
chemical potentials $\mu_u = \mu_d = \mu_{ud}$ and a nonzero strange
chemical potential $\mu_s$.  The pressure can be expanded as follows:
\begin{equation}
\frac{p}{T^4} =
\sum_{n,m=0}^\infty c_{nm}(T) \left(\frac{\mu_{ud}}{T}\right)^n
\left(\frac{\mu_s}{T}\right)^m,
\label{eq:p}
\end{equation}
The coefficients $c_{nm}$ are evaluated at zero chemical potential
\begin{equation}
c_{nm}(T)=
     \left.\frac{1}{n!} \frac{1}{m!} \frac{1}{T^3 V}
     \frac{\partial^{n+m}\ln Z}
      {\partial(\mu_{ud}/T)^n \partial(\mu_s/T)^m}
     \right|_{\mu_{ud,s}=0} \quad.
\label{eq:cn}
\end{equation}
CP symmetry requires that the coefficients vanish for odd $n+m$ at
zero chemical potential.

For increasing $n$ and $m$ the coefficients $c_{nm}$ are increasingly
complicated combinations of traces of the inverse of the lattice Dirac
matrix.  For a simple example, the lowest order mixed coefficient is
\be
  c_{11} = \VEV{\Tr \left(M^{-1}_{ud} 
     \frac{\partial M_{ud}}{\partial(\mu_{ud}/T)}\right)
           \Tr \left(M^{-1}_s \frac{\partial M_s}{\partial(\mu_s/T)}\right)}.
\ee
Such observables are technically difficult to compute because the
trace is over all lattice sites as well as over colors.  Usually such
traces are evaluated by stochastic sampling methods.  As the order $n$
and $m$ increase, not only are the traces more complicated, the
required number of stochastic samples grows rapidly.  In effect, the
computational effort grows factorially in the expansion order.

The quark number densities $\VEV{n_{ud}}$ and $\VEV{n_s}$ can be found
from first derivatives in the same expansion.  For $\VEV{n_{ud}}$ it
is
\be
\VEV{n_{ud}} = \frac{1}{V}\frac{\partial \ln Z}{\partial(\mu_{ud}/T)}
          = T^3
    \sum_{n=1,m=0}^\infty nc_{nm}(T) \left(\frac{\mu_{ud}}{T}\right)^{n-1}
   \left(\frac{\mu_s}{T}\right)^m,
\ee
and for $\VEV{n_s}$,
\be
\VEV{n_s} = \frac{1}{V} \frac{\partial \ln Z}{\partial(\mu_s/T)}
        = T^3
    \sum_{n=0,m=1}^\infty m c_{nm}(T) \left(\frac{\mu_{ud}}{T}\right)^n
   \left(\frac{\mu_s}{T}\right)^{m-1}.
\ee
The leading terms in the expansion are 
\be
\frac{\VEV{n_s}}{T^3} \approx 
     c_{11}(T) \left(\frac{\mu_{ud}}{T}\right) +
     c_{02}(T) \left(\frac{\mu_s}{T}\right).
\ee
The mixed coefficient $c_{11}(T)$ is nonzero (and negative) at low
temperatures, because when we add a strange quark to the ensemble, it
is screened by a light antiquark.  This tendency persists at
temperatures close to, but above the crossover.  So for $\mu_{ud} \ne
0$, the strange quark number density is nonzero for $\mu_s = 0$.  In
heavy ion collisions the mean strange quark number density is zero, so
we need to ``tune'' the strange quark chemical potential to obtain the
experimental conditions.

The quark number susceptibility matrix $\chi_{ab}$ for $a,b \in
{u,d,s}$ is likewise found from second derivatives.  For example, for
the diagonal elements and the equivalent mixed light off-diagonal
elements $\chi_{uu} = \chi_{dd} = \chi_{ud} = \chi_{du}$, we have
\be
\chi_{uu} = \frac{\partial \VEV{n_{ud}/T}}{\partial(\mu_{ud}/T)}
      = T^2
      \sum_{n=2,m=0}^\infty n(n-1)c_{nm}(T) 
       \left(\frac{\mu_{ud}}{T}\right)^{n-2}
       \left(\frac{\mu_s}{T}\right)^m.
\ee
The (diagonal) strange quark number susceptibility $\chi_s = \chi_{ss}$ is
similarly obtained.  The heavy-light mixed quark number susceptibility
$\chi_{us} = \chi_{su}$ is
\be
\chi_{us} = \frac{\partial \VEV{n_{ud}/T}}{\partial(\mu_s /T)}
      = T^2
      \sum_{n=1,m=1}^\infty nm c_{nm}(T) 
      \left(\frac{\mu_{ud}}{T}\right)^{n-1}
      \left(\frac{\mu_s}{T}\right)^{m-1}.
\ee

\begin{figure}
\begin{center}
\includegraphics[width=0.5\textwidth]{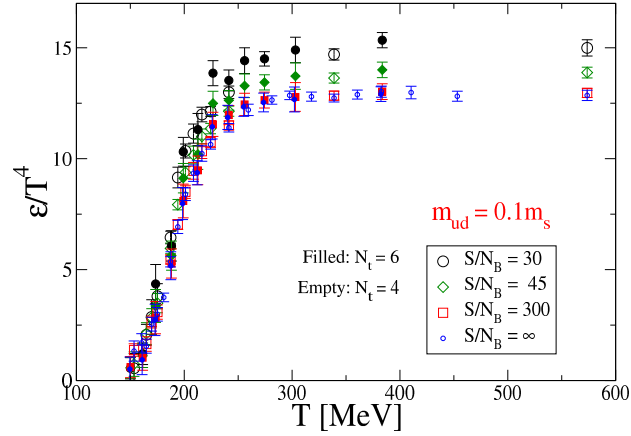} \\
\end{center}
\caption{Energy density vs temperature for constant entropy per baryon
  number, from \cite{Basak:LAT2008}.}
\label{fig:isentropic_eos}
\end{figure}

The interaction measure can also be expanded in this way
\cite{Bernard:2007nm}.  Once we have both pressure and interaction measure,
we can determine the energy density and entropy density for any small
chemical potential.  As an example, we show the equation of state at
constant entropy density per baryon number in
Fig.~\ref{fig:isentropic_eos}.  This is the equation of state
appropriate to an adiabatic expansion or compression of hadronic
matter, conditions that may obtain in a heavy ion collision.

\section{In-Medium Properties of Hadrons}
\label{sec:in_medium}

\subsection{Spatial string tension}
\label{subsec:spatial_tension}

Despite its popular characterization as deconfined, high-temperature
hadronic matter retains vestiges of confinement.  Space-like Wilson
loops still exhibit the area-law behavior associated with confinement.
This is readily seen by considering dimensional reduction, in which
for $T \gg T_c$, the short Euclidean time dimension (of extent $1/T$)
is collapsed, leaving three spatial dimensions
\cite{Ginsparg:1980ef,Appelquist:1981vg}.  Since all dimensions are
Euclidean, any one of them can be interpreted as Euclidean ``time.''
We do a $90^o$ rotation to turn one of the original spatial
coordinates into the Euclidean time coordinate of a 2+1 dimensional
field theory.

The reduction of 4-d QCD to what is sometimes called ``EQCD''
\cite{Braaten:1995jr} has these characteristics:
\bi
\item Quarks acquire a large 3-d mass $\sqrt{(\pi T)^2 + m_q^2}$.  This happens
      because the antiperiodic boundary condition in the small
      dimension requires a minimum momentum component $\pi T$ for that
      coordinate, which then contributes to the energy-momentum
      relation as an additional effective mass.
\item The original fourth component of the color vector potential $A_0$ is
      reinterpreted as a scalar Higgs-like field.  The other three
      vector potential components become the usual vector potential of
      the 2+1 dimensional theory.  We get a confining gauge-Higgs
      theory.
\item The 3-d and 4-d gauge couplings are related through
      $g_3 = g_4 \sqrt{T}$.
\item The spatial Wilson loop of the original 4-d theory is now interpreted 
      as the standard space-time-oriented Wilson loop of the 3-d
      theory.  Because the theory is still confining in 3-d, we get
      a linearly rising potential with a string tension.  
\ei

\begin{figure}
\begin{center}
\includegraphics[width=0.7\textwidth]{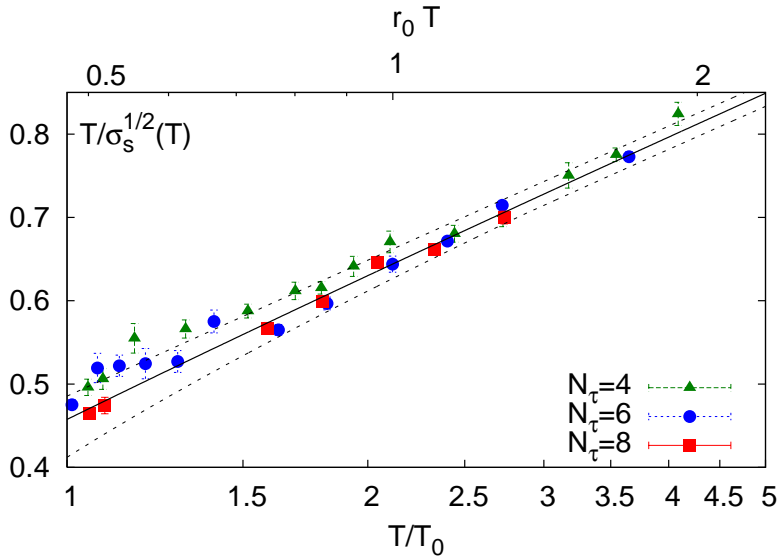} \\
\end{center}
\caption{Temperature divided by the square root of the spatial string
  tension $\sigma_s$ \vs temperature in units of the crossover
  temperature $T_0$ (lower scale) and in $r_0$ units (upper scale) for
  $2+1$ flavors of p4fat3 quarks on lattices with $N_\tau = 4$, 6 and
  8.  The solid curve (with uncertainties indicated by the dashed
  lines) is the prediction of the dimensionally reduced theory
  \protect\cite{Cheng:2008bs}}.
\label{fig:spatial_string}
\end{figure}

In a recent calculation Cheng {\it et al.}~compared the behavior of the
spatial string tension of the full 4-d theory with predictions based on
a perturbative connection between the four- and three-dimensional
coupling and the numerically measured proportionality between string
tension and coupling in three-dimensional $SU(3)$ Yang-Mills theory
\cite{Cheng:2008bs}.  The comparison is shown in
Fig.~\ref{fig:spatial_string}. The good agreement at temperatures as
low as $1.5 T_c$ is unexpected.

\subsection{Screening masses}
\label{subsec:screening}

The Yukawa potential can be thought of as a measure of the spatial
correlation of a pion source and sink (the sources and sinks being
static nucleons).  The important insight here is that the screening
mass $m_\pi$ is the mass of a propagating particle.  In the high
temperature plasma we can consider similar correlations between
interpolating operators of any type.  These spatial correlators are
controlled by confined states, as we indicated in
Sec.~\ref{subsec:spatial_tension}.  Because we no longer have Lorentz
invariance, the spatial screening masses are not expected to be equal
to frequencies of real-time plasma excitations, but one can speculate
that there may be a connection \cite{DeTar:1985kx}.  In any case,
they provide information about the structure of the plasma, they
control the behavior of a variety of susceptibilities, and their
degeneracy patterns provide information about the temperature
dependence of symmetries.

Euclidean thermal hadron propagators (correlators) are defined in the
same way as they are at zero temperature:
\be
  C_{AB}(x) = \VEV{O_A(x) O_B(0)} ,
\ee
where $O_A(x)$ and $O_B(x)$ are interpolating operators for the
desired hadronic state.  

At zero temperature it is typical to project the correlator to zero
spatial momentum, resulting in a time-slice correlator
\be
 C_{AB}(t) = \int d^3{\bf x}\, C_{AB}(t,{\bf x}).
\ee
At large Euclidean time such a correlator has the asymptotic behavior
\be
 C_{AB}(t) \sim Z_A Z_B \exp(-M t) ,
\ee
where $M$ is the mass of the hadron and $Z_A$ and $Z_B$ are overlap
constants.

At nonzero temperatures one cannot explore the asymptotic limit because
of the bound on Euclidean time $0 \le t \le 1/T$, but one can define a
spatial correlator by fixing one of the spatial coordinates and
integrating over the other three, as in
\be
 C_{AB}(z) = \int dt\,dx\,dy\, C_{AB}(t,x,y,z).
\ee
(For fermions, it is necessary to include a Matsubara phase factor
$\exp[i \pi T t]$.)  For large $z$ the asymptotic behavior is
\be
 C_{AB}(z) \sim Z_A(T) Z_B(T) \exp[-\mu(T)z],
\ee
where $\mu(T)$ is the hadronic screening mass.  At zero temperature
$\mu(T=0) = M$.

Even though the high-temperature plasma exhibits deconfining
characteristics in its real time behavior, the spatial correlations
remain confined, so the spectrum of spatial meson and baryon
screening masses retains a gap characteristic of confinement even in
the high-temperature plasma.  However, since the screening mass for
quarks approach $\pi T$ at high temperatures, the
valence-quark-antiquark meson screening masses approach $2\pi T$ and
the valence-three-quark baryon screening masses approach $3\pi T$.
Furthermore, as chiral symmetry is approximately restored at high
temperatures, they must exhibit the approximate degeneracies required
by the chiral multiplets.

\begin{figure}
  \begin{center}
  \includegraphics[width=0.7\textwidth]{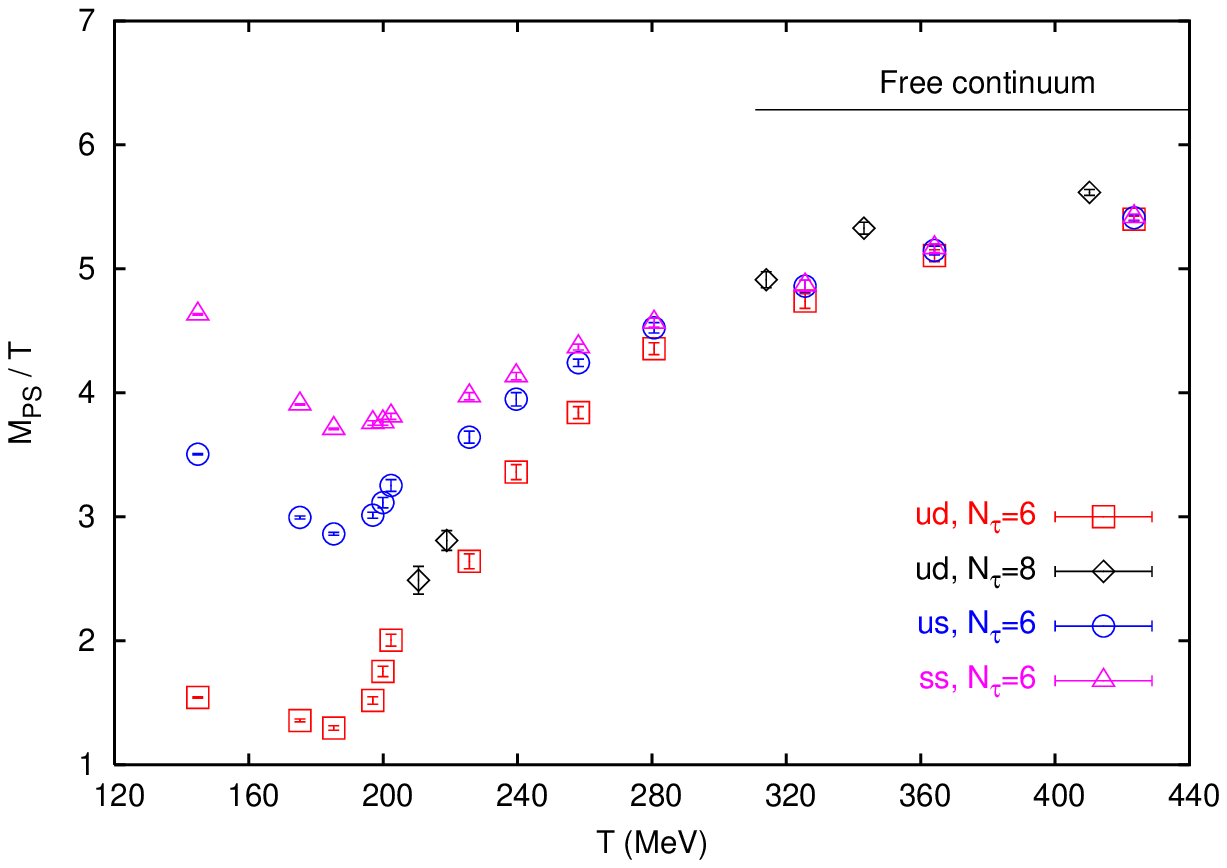} \\
  \includegraphics[width=0.7\textwidth]{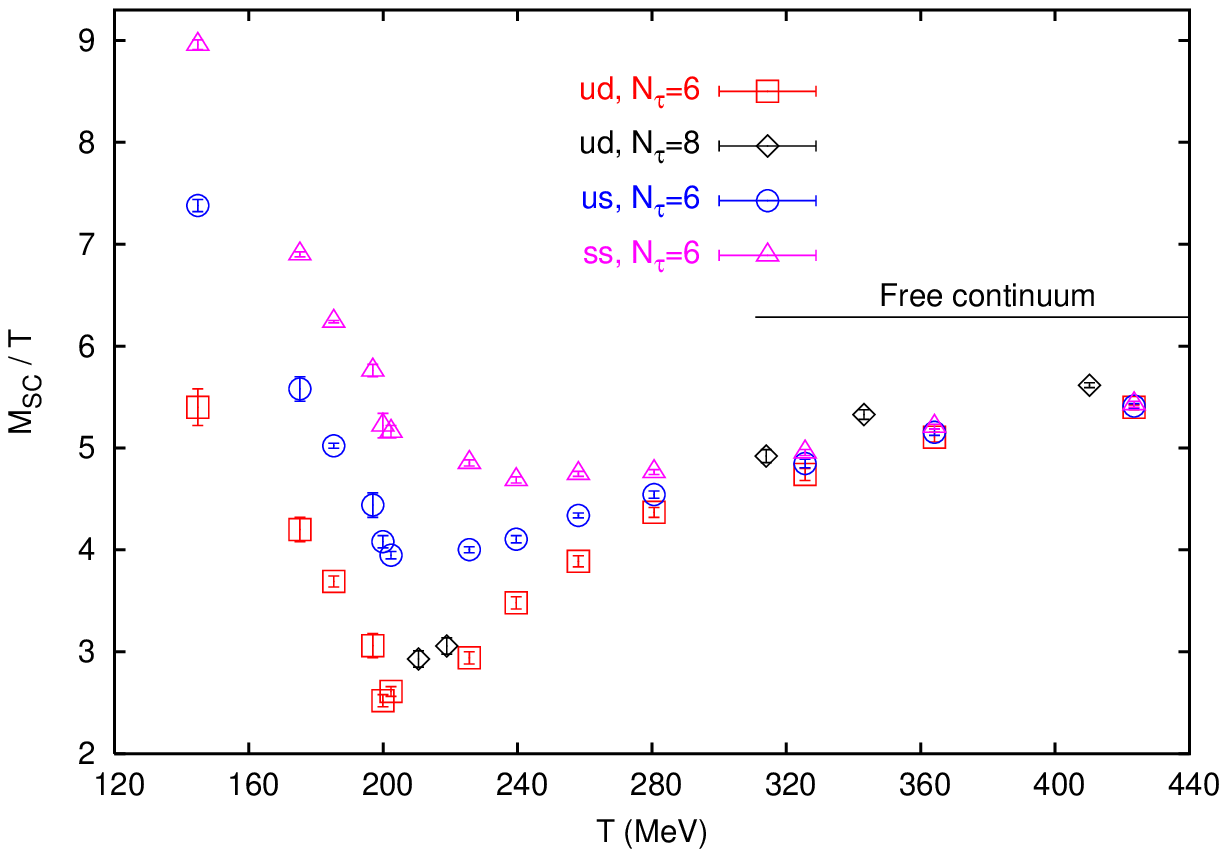} \\
  \end{center}
\caption{Screening masses for the pseudoscalar channel (upper panel)
  and scalar channel (lower panel) \vs temperature in a dynamical
  $2+1$ flavor simulation with p4fat3 staggered fermions
  \cite{Cheng:2007jq}.  Measurements were taken along lines of
  constant physics with $m_\pi \sim 220$ MeV, $m_K = 500$ MeV and $N_\tau
  = 6$ and $8$ \cite{Laermann:Lat2008}. }
\label{fig:RBCBscreen}
\end{figure}

Armed with this background let us consider the temperature behavior of
the screening mass $\mu_\pi(T)$ of the pion.  At low temperatures the
pion is a Goldstone boson, so the screening mass is small.  Above the
transition chiral symmetry is restored.  So the screening mass rises
above the transition temperature, approaching $2\pi T$.  The transition
temperature is marked by the change of slope.  Figure
\ref{fig:RBCBscreen} illustrates this behavior.

The isosinglet scalar $f_0$ ($\sigma$) meson can be generated using
the isosinglet chiral condensate $\pbp_{\rm sing} = \pbp_u + \pbp_d$
as the interpolating operator.  It has a sizable mass at low
temperature, but it joins the pion chiral multiplet at the transition
temperature when the pion screening mass is quite small.  Thus its
mass must dip at the transition temperature and rise again,
approaching $2\pi T$.  Thus a dip in $\mu_{f_0}$ also marks the
transition temperature.

The chiral susceptibilities are related to hadron propagators in
Euclidean space-time.  For example, the isosinglet chiral
susceptibility is
\be
  \chi_{\rm sing}(T) = \int d^4x\, \VEV{\pbp_{\rm sing}(x) \pbp_{\rm sing}(0)} =
     \int dz\, C_{\rm sing}(z,T),
\ee
where $C_{\rm sing}(z)$ is the scalar-isosinglet screening correlator
generated by the isosinglet chiral condensate.  In addition to the
$f_0$, this correlator also contains a two-pion continuum
contribution.  So its asymptotic behavior has terms in
$\exp(-\mu_{f_0}z)$ as well as $\exp(-2 E_\pi z)$ for $E_\pi \ge
\mu_\pi$.  Integration over $z$ of these asymptotic terms yields
contributions to the susceptibility that go as the inverse of the
screening masses.  At low temperatures the two-pion threshold is below
the $f_0$, so the two-pion continuum dominates the susceptibility.  In
the chiral limit this contribution is responsible for the $1/\sqrt{m}$
singularity in the susceptibility.  At high temperatures the pion
screening mass rises, and the $f_0$ screening mass is approximately
degenerate with it.  Thus the two-pion continuum is expected to have a
higher screening mass than the $f_0$, and the susceptibility is finite
in the chiral limit.  Thus this susceptibility should be large at low
temperatures and fall abruptly at the transition temperature.

\subsection{Charmonium}

To the extent the transition to a quark-gluon plasma is a crossover
and not a genuine phase transition, one should not expect low
temperature properties to change abruptly at the crossover
temperature.  Confined hadronic states may persist as plasma
excitations at least for temperatures close to, but above the
crossover temperature.  One of the most studied examples is the
$J/\psi$, since it is readily observed experimentally, and, because of
their large mass, charmed quarks are a good theoretical probe.
Numerical simulation suggests that the $J/\psi$ persists to
temperatures as high as $1.5 T_c$ \cite{Asakawa:2003re,Datta:2003ww}.
(See Sec.~\ref{subsubsec:MEM} below.)  As the temperature increases
beyond $T_c$, it is thought that screening of the heavy-quark
potential eventually prevents the formation of a bound state and
$J/\psi$ production is suppressed \cite{Matsui:1986dk,Datta:2003ww}.

\subsubsection{Static quark/antiquark free energy}

There are two lattice methods for studying thermal effects in
quarkonium.  The first, more model-dependent method, is based on a
Born-Oppenheimer approximation~\cite{Matsui:1986dk}. One measures the
free energy of a static quark-antiquark pair as a function of
separation $r$.  The result is introduced into the Schr\"odinger
equation as a temperature-dependent potential $V(r,T)$ for a given
heavy quark mass.  As the temperature increases, screening effects
weaken the potential, and eventually it does not support a bound state
for quarks of the given mass.  This approximation should be good,
provided the Born-Oppenheimer adiabatic approximation is good, \ie, as
long as the plasma is able to relax to its equilibrium state on the
time scale of the orbital motion of the quarks.

Gauge invariance presents a subtlety in fashionable methods for
extracting the free energy to be used as a Born-Oppenheimer
potential. It is popular to distinguish between color-singlet and
color-octet states of the static quark and antiquark.  Since those
states are supposed to be defined in terms of the colors of only the
spatially separated quarks themselves, the separation is gauge
dependent and probably not phenomenologically significant
\cite{Jahn:2004qr}.

The potential method can be tested entirely in the context of a
lattice calculation.  One starts from the lattice static
potential, derives the spectral function for the thermal quarkonium
propagator (see the next subsection), and compares the result with a
direct determination of the lattice spectral function.  If the static
approximation is correct, the results should agree.  Recent attempts
to follow this approach for $T_c < T < 1.5 T_c$ fail to reproduce any
charmonium states in the spectral function nor any but the 1S state of
bottomonium \cite{Mocsy:2007yj}.  So is the determination of the
lattice spectral function unreliable, or is the static approximation
unreliable for charmonium, or are both unreliable?

Related attempts have been made to derive a heavy-quark potential
suitable for use in the Schr\"odinger equation in real time (as
opposed to lattice imaginary time), but so far the methodology is
developed only in perturbation theory
\cite{Laine:2006ns,Escobedo:2008sy,Brambilla:2008cx}.

\subsubsection{Spectral density}

The second method is model independent, but more difficult.  One
measures the spectral function of a thermal Green's function for the
$J/\psi$ \cite{Umeda:2002vr}.  The correlator is defined for some
suitable local interpolating operator ${\cal O}(x_0, {\bf x})$ as
\be
   C(x_0,{\bf x},T) = \left\langle{\cal O}(x_0, {\bf x}){\cal O}(0, 0)
     \right\rangle.
\ee
The spectral density $\rho(\omega,{\bf q}, T)$ is then obtained by
inverting the Kubo formula for the partial Fourier transform $C(x_0,{\bf
  q},T)$ of the correlator:
\be
  C(x_0,{\bf q},T) = \frac{1}{2\pi}
       \int_0^\infty d\omega \, \rho(\omega,{\bf q}, T) K(\omega,x_0,T) ,
\ee
where 
\be
    K(\omega,x_0,T) = \frac{\cosh \omega(x_0 - 1/2T)}{\sinh(\omega/2T)}.
\label{eq:thermo_kernel}
\ee
Going from the Euclidean correlator $C(x_0,{\bf q},T)$ to the spectral
density $\rho(\omega,{\bf q},T)$ is a very difficult inverse problem.
One would like to extract detailed information about the spectral
density from quite limited information.  Because of time-reflection
symmetry, a simulation at $N_\tau = 8$ has only five, typically noisy,
independent values.

Possible remedies include (1) assuming a functional form for $\rho$
and fitting its parameters (\eg, a delta function for the $J/\psi$ or
a Breit-Wigner shape), (2) decreasing the time interval $a_t$,
allowing a larger $N_\tau$, and (3) adding further constraints on
$\rho$, as in the maximum entropy method.  We outline the last remedy
in the next subsection.

\subsubsection{Maximum entropy method}
\label{subsubsec:MEM}

The maximum entropy method has been used to determine spectral
functions in condensed matter physics for some time \cite{Jarell}.  It
was first applied to lattice QCD by Asakawa, Hatsuda, and
Nakamura~\cite{Nakahara:1999vy,Asakawa:2000tr}.  It is essentially a
Bayesian method with a prior inspired by Occam's razor.  One begins by
defining an unremarkable default prior spectral density
$\rho_0(\omega,T)$.  A typical choice would be the spectral density of a
noninteracting quark-antiquark pair, or at least the density expected
at asymptotically high frequency.  One then requires that the spectral
density $\rho$, inferred from the correlator data, should deviate only
as much from $\rho_0$ as the data seems to require.

The method is applied in the context of a maximum-likelihood fit to
the correlator data. We give a simplified description of the method.
Starting from a parameterization of the spectral density
$\rho(\omega,T)$, one predicts the correlator data and computes the
usual chisquare $\chi^2[\rho]$ difference between prediction and data.
One introduces a Shannon-Jaynes entropy for this $\rho$ as follows
\be
  S[\rho] = \int_0^\infty d\omega \, \left[\rho(\omega) - \rho_0(\omega)
       - \rho(\omega) \ln[\rho(\omega)/\rho_0(\omega)] \right].
\ee
The ``entropy'' vanishes when $\rho = \rho_0$ and for small deviations
from $\rho_0$, it is 
\be 
    S[\rho] \approx -\frac{1}{2} \int_0^\infty d\omega \,
    [\rho(\omega) - \rho_0(\omega)]^2/\rho_0(\omega) .
\ee
So the default prior maximizes the entropy.  One then maximizes the
likelihood $\exp(Q[\alpha,\rho])$ or, equivalently, $Q[\alpha,\rho]$ itself:
\be
    Q[\alpha,\rho] = \alpha S[\rho] - \chi^2[\rho]/2.
\ee
The positive weight $\alpha$ controls the balance between maximum
entropy and minimum chisquare.  In the ``state-of-the-art'' method,
the mean of the best fits $\bar \rho$ is then obtained from the
average:
\be
   \bar \rho = \int d[\rho] d\alpha\, \exp(Q[\alpha,\rho]).
\ee
This is our answer for the spectral density.

\begin{figure}
\begin{center}
\includegraphics[width=0.7\textwidth]{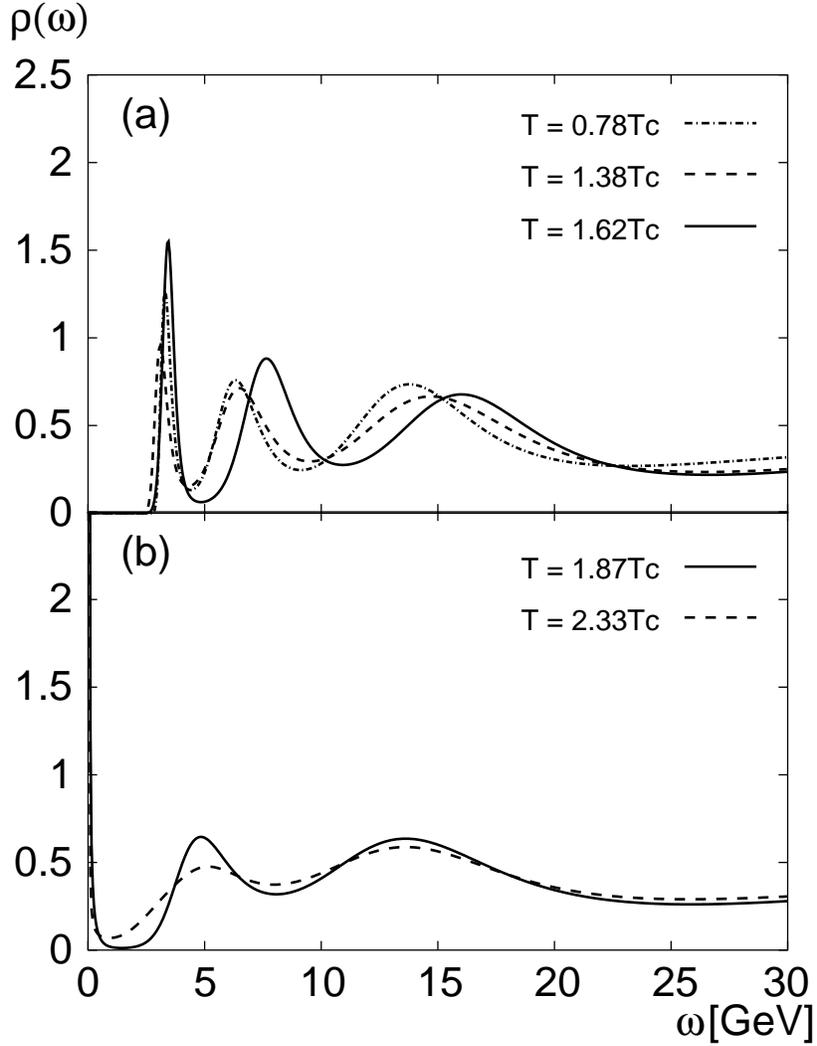} \\
\end{center}
\caption{Spectral density $\rho(\omega)$ for the $J/\psi$ for several
  temperatures shown in units of the crossover temperature $T_c$
  \protect\cite{Asakawa:2003re}.  The ground state peak is visible up
  to $1.62 T_c$.  These results are obtained in a quenched simulation. }
\label{fig:asakawa_hatsuda}
\end{figure}

This method was used by Asakawa and Hatsuda to study the fate of
charmonium in the high-temperature medium \cite{Asakawa:2003re}.  See
also \cite{Datta:2003ww} and, more recently, \cite{Aarts:2007pk}.
Their results for the $J/\psi$ spectral density are shown in
Fig.~\ref{fig:asakawa_hatsuda} and provided some of the first evidence
that the $J/\psi$ exists as a discernible plasma resonance for
temperatures at least as high as $1.62 T_c$ before it ``melts.''

When data are inadequate, results of the MEM method can be quite
sensitive to the choice of the default model.  For example, one may
obtain artifact excited state peaks.  For some examples, see
\cite{Ding:2009se}.

\section{Transport coefficients}
\label{sec:transport}

\subsection{Shear and bulk viscosities}

Among the transport coefficients, the shear  and bulk
viscosities are essential to the hydrodynamical modeling of
the expansion and cooling of the quark-gluon plasma in the aftermath
of a heavy ion collision.  They are obtained from correlators of the
energy-momentum tensor at temperature $T$
\begin{eqnarray}
    C_{\mu\nu,\rho\sigma}(x_0,{\bf x},T) =  
  \left\langle T_{\mu\nu}(x_0,{\bf x})T_{\rho\sigma}(0)\right\rangle .
\end{eqnarray}
We need its spectral function $\rho$, which we obtain from its partial
Fourier transform $C_{\mu\nu,\rho\sigma}(x_0,{\bf q},T)$ and the Kubo formula
\be
   C_{\mu\nu,\rho\sigma}(x_0,{\bf q},T) =  
   \int_0^\infty d\omega \, \rho_{\mu\nu,\rho\sigma}(\omega,{\bf q}, T) 
            K(\omega,x_0,T) ,
\ee
where $K(\omega,x_0,T)$ is given by Eq.~(\ref{eq:thermo_kernel}).
The shear ($\eta$) and bulk ($\zeta$) viscosities are obtained from
the low-frequency behavior of the spectral function $\rho(\omega,{\bf
  q}, T)$:
\be
    \eta(T) = \pi \lim_{\omega\rightarrow 0} 
          \frac{\rho_{12,12}(\omega, 0, T)}{\omega} ,
    \ \ \ \ \ \ \ \ \ \ \ \ 
    \zeta(T) = \frac{\pi}{9} \lim_{\omega\rightarrow 0} 
          \frac{\rho_{ii,jj}(\omega, 0, T)}{\omega} .
\ee
Computing the viscosity has been a well known challenging problem
since it was first attempted by Karsch and Wyld \cite{Karsch:1986cq}.
The correlator is noisy, requiring high statistics. As with the
$J/\psi$ correlator, this is a difficult inverse problem.  A further
complication is that the spectral function has a nasty
$T$-independent, large $\omega$, ultraviolet behavior $\rho \sim
\omega^4$, which tends to overwhelm the low-frequency contribution to
$C(x,\tau)$ for low $x_0$.

Possible remedies include (1) assuming a functional form for $\rho$
and fitting its parameters \cite{Karsch:1986cq}, (2) decreasing the
time interval $a_t$, allowing a larger $N_\tau$ \cite{Burgers:1987mb},
and (3) adding further constraints on $\rho$, such as maximum entropy
\cite{Aarts:2007wj}, and working at small nonzero
momentum\cite{Meyer:2008gt}.

Meyer \cite{Meyer:Lat2008,Meyer:2007dy,Meyer:2007ic} has done a new
high-statistics calculation in pure Yang-Mills theory and uses a
parameterization of the spectral function in terms of an optimized
basis set that folds in appropriate perturbative behavior at large
$\omega$ and then emphasizes deviations from this behavior.  For the
ratio of shear viscosity to entropy density, he finds $\eta/s =
0.134(33)$ at $1.65 T_c$ where perturbation theory gives 0.8, and for
the ratio of bulk viscosity to entropy density, $\zeta/s < 0.15$ at
$1.65 T_c$ and $\zeta/s < 0.015$ at $3.2 T_c$.  These results support
the notion that the plasma is a nearly perfect fluid.

\subsection{Dilepton emission and related quantities}

The dilepton emission rate, the soft photon emissivity, and the
electrical conductivity of the plasma are other important transport
properties.  They are obtained from the thermal correlator of the
electric current
\be
   G_{EM}(x_0,{\mathbf x},T) = \left\langle J_\mu(x_0,{\bf x})J_\mu(0)
           \right\rangle .
\ee
\be
   G_{EM}(x_0,{\mathbf q},T) = \int_0^\infty \frac{d\omega}{2\pi}
       K(\omega, x_0,T) \rho_{EM}(\omega,{\mathbf q},T),
\ee
Again, this is a difficult inverse problem.  The ultraviolet
divergence is milder here than with the spectral function of the
stress-energy tensor.  In this case $\rho \sim \omega^2$.  Otherwise,
the same methods have been applied.

\begin{figure}
\begin{tabular}{cc}
\includegraphics[width=0.4\textwidth]{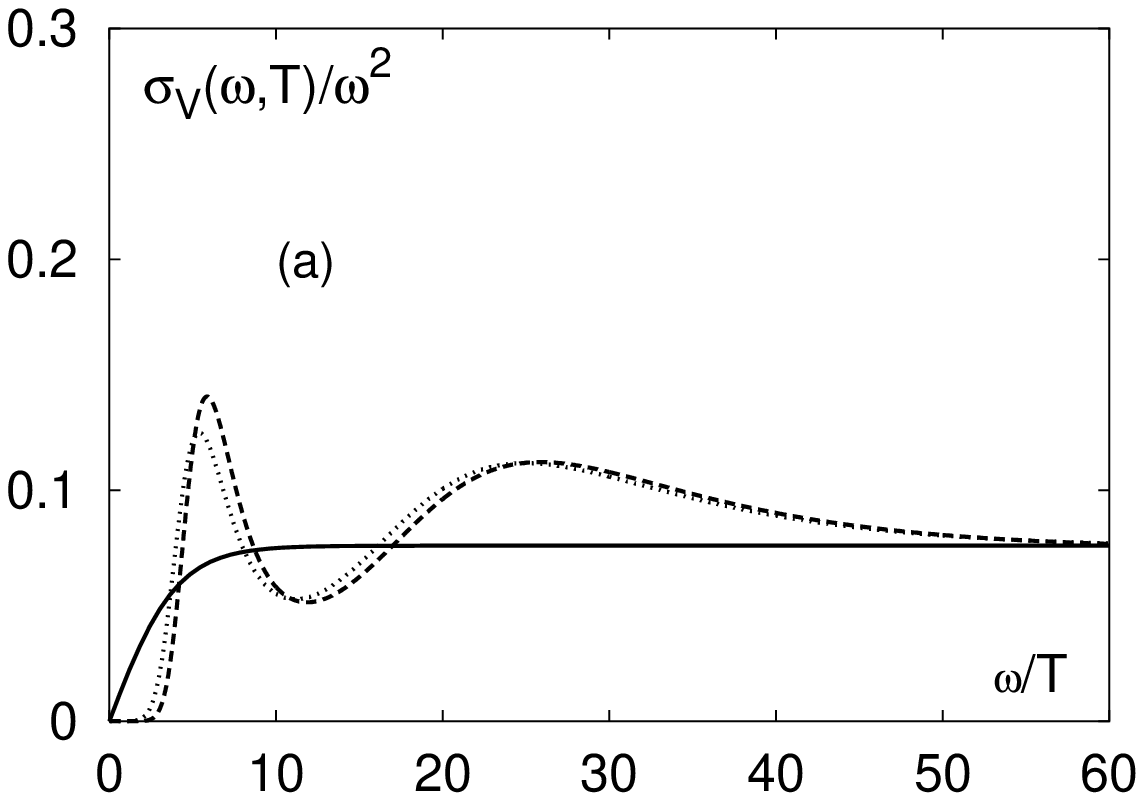} &
\includegraphics[width=0.4\textwidth]{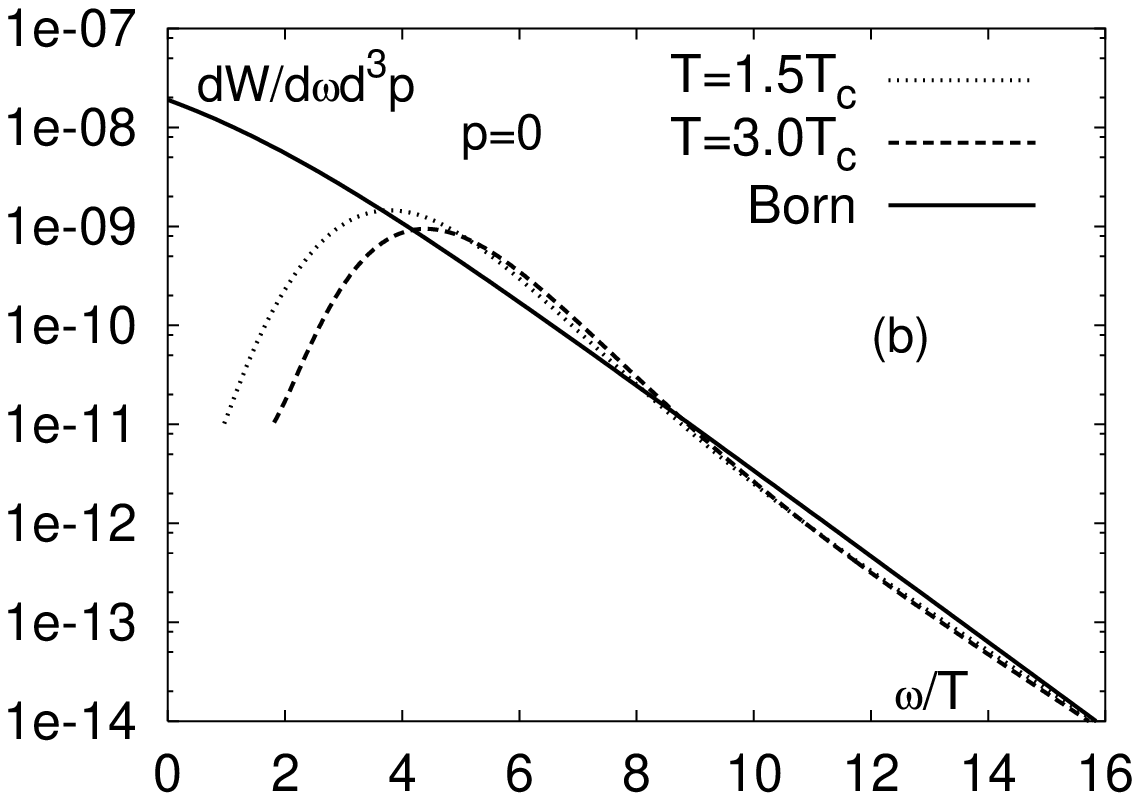} \\
\end{tabular}
\caption{Relationship between (a) the vector meson spectral density
  $\rho_{EM}(\omega,0,T)$ (shown here as $\sigma_V(\omega,T)$ and (b)
  the dilepton differential production rate $dW/d\omega d^3p$ at zero
  three-momentum, plotted as a function of energy $\omega$ in units of
  temperature for two temperatures above the crossover temperature
  $T_c$ \protect\cite{Karsch:2001uw}.  The solid lines represent a
  free quark-antiquark pair.  The dash-dotted lines are lattice MEM
  results that show a peak corresponding to a vector meson
  resonance. Results are obtained in the quenched approximation.}
\label{fig:dilepton}
\end{figure}

The spectral density $\rho_{EM}(\omega,0,T)$ determines the
differential dilepton pair production rate \cite{Braaten:1990wp}:
\be
  \left.\frac{dW^4}{d\omega d^3p}\right|_{\vec p = 0} = 
  \frac{5\alpha_{rm em}^2}{27 \pi^2}
  \frac{1}{\omega^2(e^{\omega/T} - 1)} \rho_{EM}(\omega,0,T).
\ee
An example of the relationship between the MEM determination of the
spectral function and the resulting dilepton rate is given by Karsch
\etal\ \cite{Karsch:2001uw} in Fig.~\ref{fig:dilepton}.  These results
show a strong enhancement over the free quark-antiquark pair
contribution, at least up to three times $T_c$ resulting from a vector
meson resonance.  The hard dilepton rate is obtained from the spectral
function for $\omega/T \gg 1$, and there is rough agreement between
perturbation theory and lattice simulation.

As with the shear and bulk viscosity, the challenge is getting to low
frequency to obtain the soft photon emissivity, and at zero frequency,
the electrical conductivity:
\be
   \sigma(T) = \frac{1}{16}\left.\frac{\partial}{\partial\omega}
        {\rho_{EM}}(\omega,{\mathbf 0},T)\right|_{\omega=0},
\ee
Extracting the spectral function itself is challenging enough.
Extracting its derivative compounds the difficulty. Gupta \etal\
\cite{Gupta:2003zh} tried different Bayesian priors to constrain the
spectral function.

\section{Outlook}
\label{sec:outlook}

Numerical simulations have taught us much about the properties of high
temperature strongly interacting matter.  Here are highlights
discussed in this article:

\begin{itemize}
\item We have a fair understanding of the QCD phase diagram at nonzero
      temperature and zero or small baryon densities and nearly physical
      quark masses.
\item We have a phenomenologically useful determination of the
      equation of state.
\item We have a good understanding of the behavior of the quark number
      susceptibility.
\item We are beginning to understand the small mass limit of
      the chiral condensate and its related susceptibilities.
\item We know the plasma has persistent confining properties that are
      observable in screening masses and the spatial string tension.
\item We have some indications of the persistence of hadronic states
      as resonances in the plasma phase at temperatures close to and above
      $T_c$.
\item We are starting to determine plasma transport coefficients.
\item We are starting to make contact with perturbation theory at high
      temperatures.
\end{itemize}

There are many outstanding questions.  Here are particularly pressing ones:

\begin{itemize}
\item We need a more robust determination of transport coefficients.
\item We don't have a good way to simulate at moderately large or
      higher nonzero baryon number densities.
\item We don't know, yet, whether the critical point in the $\mu/T-T$ plane
      is experimentally accessible.
\item We don't know whether the tricritical point in the $m_s,m_{u,d}$ plane
      lies above or below the physical $m_s$.
\item We would like to understand better the behavior of the equation
      of state in the region where it overlaps with hadron resonance gas
      models.
\item It would be good to develop more confidence in our
      understanding of the continuum limit of phenomenologically important
      quantities.
\item It would be good to have high precision results from fermion
      formulations other than staggered for purposes of corroboration.
\item We would like to develop more confidence in our contact with
      perturbation theory at high temperatures.
\end{itemize}

Work currently underway will help resolve some of these issues.  At
zero or small baryon number densities we expect progress with Wilson
quark formulations, including clover-improved and twisted-mass.
Simulations with domain-wall quarks will help test conclusions about
chiral properties.  Forthcoming simulations with highly improved
staggered quarks (HISQ) will help reduce some of the lattice artifacts
of the staggered fermion formulation, especially at temperatures
leading up to $T_c$, where we suspect they are important.

For simulations at nonzero baryon number densities we really need some
new ideas. Perhaps stochastic quantization will help.  For transport
coefficients and the small-quark-mass region of the phase diagram, we
may expect progress simply by applying more computing power.

Lattice QCD thermodynamics is a very active field.  We expect
continued strong progress in the years to come.

\section*{Acknowledgments}
We thank Ludmila Levkova for a careful reading of the manuscript.
This work is supported by the National Science Foundation under grants
PHY04-56691 and PHY07-57333.

\providecommand{\href}[2]{#2}\begingroup\raggedright\endgroup

\end{document}